\documentclass[11pt,a4paper]{article}
\pdfoutput=1

\usepackage[utf8]{inputenc}
\usepackage{mathtools}
\usepackage{enumitem}
\usepackage{amsmath,amsthm}
\usepackage{amssymb}
\usepackage{mathrsfs}

\allowdisplaybreaks[2]

\usepackage{color}
\usepackage[multiple]{footmisc}
\usepackage{textcomp}
\usepackage{array}
\usepackage{graphicx}
\usepackage{hyperref}
\usepackage[capitalize]{cleveref}

\usepackage{tikz}
\usetikzlibrary{positioning,calc,intersections,through,backgrounds,arrows,decorations.markings,shapes.arrows,patterns,decorations.pathmorphing}

\numberwithin{equation}{section}
\graphicspath{{figures/}}

\def\ie{\begin{equation}\begin{aligned}}
\def\fe{\end{aligned}\end{equation}}

\newcommand{\sch}{\mathscr{H}}
\newcommand{\tA}{\widetilde{A}}

\newcommand{\sfc}{\mathsf c}
\newcommand{\sfB}{\mathsf B}
\newcommand{\sfK}{\mathsf K}
\newcommand{\sfP}{\mathsf P}
\newcommand{\sfh}{\mathsf h}

\newcommand{\cJ}{\mathcal J}

\newcommand{\zA}{z_{+}}
\newcommand{\zB}{z_{-}}

\newcommand{\fk}{\mathsf{K}}
\newcommand{\fp}{\mathsf{P}}

\def\tilde{\widetilde}
\def\bar{\overline}

\newcommand{\C}{\mathbb C}

\newcommand{\R}{\mathbb R}
\renewcommand{\d}{\mathrm{d}}

\newcommand{\del}{\partial}

\newcommand{\de}{d^2}
\newcommand{\dl}{D^2}

\def\Tr{{\mathrm{Tr}}}

\newcommand{\rd}{\textrm{d}}

\newcommand{\wbar}{\overline{w}}
\newcommand{\zbar}{\overline{z}}

\renewcommand{\varkappa}{h}

\begin{document}

\begin{titlepage}
    \enlargethispage{1in}
    \begin{flushright}
        RIKEN-iTHEMS-Report-26
    \end{flushright}
    \vskip 0.5in
    \begin{center}
        {\bf\Large{Renormalization Group Flow \\
        \bigskip
        from 4d Chern--Simons Theory}}
        \vskip 0.5cm
        \vskip 0.05in
        Meer Ashwinkumar$^{a \, \ast}$ and Masahito Yamazaki$^{b,c,d,e\,\dagger}$

        {\small{ $^a$\textit{Institute for Theoretical Physics, Albert Einstein Center for Fundamental Physics,\\  University of Bern,
        Sidlerstrasse 5, CH-3012 Bern, Switzerland}}}\\
        {\small{ $^b$\textit{Department of Physics, Graduate School of Science, University of Tokyo,\\
        Hongo 7-3-1, Bunkyo-ku, Tokyo 113-0033, Japan}}}\\
        {\small{ $^c$\textit{Kavli Institute for the Physics and Mathematics of the Universe,\\
        UTIAS, University of Tokyo, 5-1-5 Kashiwanoha, Chiba 277-8583, Japan}}}\\
        {\small{ $^d$\textit{Trans-Scale Quantum Science Institute, University of Tokyo,\\
        Hongo 7-3-1, Bunkyo-ku, Tokyo 113-0033, Japan}}}\\
        {\small{ $^e$\textit{Center for Interdisciplinary Theoretical and Mathematical Sciences, \\ RIKEN, Saitama 351-0198, Japan}}}

    \end{center}
\begingroup
\renewcommand{\thefootnote}{\fnsymbol{footnote}}
\footnotetext[1]{%
  \href{mailto:meerashwinkumar@gmail.com}%
       {\texttt{meerashwinkumar@gmail.com}}%
}
\footnotetext[2]{%
  \href{mailto:yamazaki@hep-th.phys.s.u-tokyo.ac.jp}%
       {\texttt{yamazaki@hep-th.phys.s.u-tokyo.ac.jp}}%
}
\endgroup

    \vskip 0.5in
    \baselineskip 16pt

    \begin{abstract}
        The four-dimensional Chern--Simons theory coupled with surface defects is known to realize two-dimensional integrable quantum field theories (QFTs). We discuss the renormalization group (RG) flow of such two-dimensional integrable QFTs from first-principles Feynman diagram computations intrinsically in the four-dimensional framework. We first derive the one-loop RG flow of two-dimensional QFTs realized by the four-dimensional Chern--Simons theory on $\Sigma \times C$ coupled with the order surface defects supported on $\Sigma $. Here, the affine Kac--Moody symmetry of the defects underlies the one-loop RG flow. We then derive the one-loop RG flow of the principal chiral model (PCM) with Wess--Zumino term, which is realized using disorder surface defects, within the Batalin--Vilkovisky formalism. 
      We obtain the nontrivial one-loop RG flow of the theory, reproducing the known asymptotically-free beta function in the absence of a Wess--Zumino term, using two complementary methods: by explicitly evaluating one-loop Feynman amplitudes, and by computing an analogue of analytic torsion for four-dimensional Chern--Simons theory.

    \end{abstract}

   \par\vfill
\end{titlepage}
\pagenumbering{arabic} 

\tableofcontents
\newpage

\section{Introduction}

The four-dimensional Chern--Simons (4d CS) theory provides us with a powerful unifying framework for the study of integrable models \cite{Costello:2013zra, Costello:2017dso, Costello:2018gyb, Costello:2019tri} (see \cite{Lacroix:2021iit, Yamazaki:2025yan, Ashwinkumar:2025btr} for reviews). With regard to the study of two-dimensional integrable field theories, Costello and the second-named author \cite{Costello:2019tri} showed that these theories arise from order or disorder surface defects in the 4d CS theory.

A wealth of subsequent works followed, where various integrable field theories (IFTs) were derived from the 4d CS theory. The analysis has been mostly restricted to classical aspects of these integrable field theories, with the exception of works such as \cite{Ashwinkumar:2023zbu, Levine:2023wvt, Schmidtt:2023slc,Lacroix:2025ias,Schmidtt:2025zsb,Bonelli:2026flm}. However, the  quantization of integrable field theories starting from the first-principles quantization of four-dimensional Chern--Simons theory has remained elusive thus far.

A particularly challenging open problem is to derive the \textit{integrable}  quantization of ``nonultralocal" integrable field theories (realized using disorder surface defects) from the 4d CS theory. Nonultralocality here refers to the presence of a term involving the derivative of a Dirac delta function in the Poisson brackets of the Lax connection of these theories. Such a term obstructs a straightforward integrable quantization of these Poisson brackets in terms of a Yang--Baxter equation. Understanding and overcoming this obstruction to integrable quantization would be highly desirable, especially using the approach of 4d CS.

In this work, we shall initiate the study of first-principles quantization of integrable field theories using the 4d CS theory, by using the latter to derive their renormalization group (RG) flow.  While previous works have focused on computations in the effective 2d theory, we shall use Feynman diagram techniques intrinsically in the 4d framework. For ultralocal theories, the 4d propagator identifies the coupling with the separation of the defects, and the RG flow becomes a flow of the defect positions; for the PCM, the RG flow arises from a genuine one-loop computation in the 4d theory. This will make it possible to leverage the full power of the 4d CS theory in the exploration of quantum integrability of a large class of 2d integrable field theories.

Our main goal in this paper is to derive, from the 4d CS theory, the universal formula for the one-loop renormalization group (RG) flow of the effective action of the integrable field theories it realizes. While our consideration has a much wider applicability, for concreteness, let us consider a theory whose interaction is described by an action
\ie
    \label{S_1}
    S_1=
\varkappa
    \int  \operatorname{Tr}\left({\mathcal{J}}_{+} {\mathcal{J}}_{-}\right)
    \;,
\fe
where $\varkappa$ is the coupling constant of the theory,
$\mathcal{J}_{ \pm}$ are model-dependent currents that are conserved and flat (as for the PCM), and take values in the Lie algebra $\mathfrak{g}$ of the underlying Lie group $G$.
This theory is integrable in that it has the Lax connection that
has lightcone components of the form
\ie
    \mathcal{L}_{ \pm}(z)=\frac{\mathcal{J}_{ \pm}}{1 \pm z} \;,
\fe
with $z$ being the spectral parameter,
satisfying the flat connection equation
\ie
    \partial_{+} \mathcal{L}_{-}(z)-
    \partial_{-} \mathcal{L}_{+}(z)
    + [\mathcal{L}_{+}(z), \mathcal{L}_{-}(z)]=0\;,
\fe
for $z$ away from the poles $z=\pm 1$.
For current-current deformations of a conformal field theory, the one-loop running of the coupling constant can be written in the universal formula
\ie
    \label{universal_RG}
    \beta_{\varkappa} =
    \frac{d \varkappa}{d \log \mu} =- \pi \mathsf{h}^{\vee}  \varkappa^2 \;,
\fe
where $\mu$ is the energy scale, and $\mathsf{h}^{\vee}$ is the dual Coxeter number of $\mathfrak{g}$ (normalized as in \cref{app:Lie}). The negative coefficient here implies that the theory is asymptotically free.

This formula was derived by Kutasov \cite{Kutasov:1989dt}. In the nonultralocal case, Levine \cite{Levine:2022hpv} showed that the one-loop divergences take a universal form in terms of the Lax currents, so that the inverse coupling runs linearly (see also Komatsu et al.~\cite{Komatsu:2026cqz} for a recent discussion). For example, consider the principal chiral model with the action
\begin{equation}
    \label{S_PCM}
    S_{\mathrm{PCM}}=-\frac{1}{16\pi^2 \lambda} \int \operatorname{Tr}\left(g^{-1} \partial_\mu g g^{-1} \partial^\mu g\right)\;,
\end{equation}
with a field $g$ valued in a group $G$, where we use the mostly-plus signature and the positive-definite trace of \cref{app:Lie}.
The one-loop beta function of $\lambda$ is
\begin{equation}
    \label{varkappa_RG}
    \beta_{\lambda}=
    \frac{d\lambda}{d\log \mu}=
    -\pi \mathsf{h}^{\vee}\lambda^2+O\left(\lambda^3\right) \;,
\end{equation}
which has the same form as \eqref{universal_RG}. Note, however, that for the PCM the coefficient of \eqref{S_1} is proportional to the inverse coupling $1/\lambda$, which runs linearly.
As another example, consider an IFT involving a current-current deformation such as the non-abelian Thirring model, with action
\begin{equation}
    \label{S_JJ}
    S=S_{\rm CFT}+\varkappa \int J^a \bar{J}^a \;,
\end{equation}
where the currents take values in the adjoint representation of the symmetry group $G$. Here and in \cref{sec:ultralocal}, $\varkappa$ denotes the coefficient of $\int J^a\bar J^a$ in the exponent of the two-dimensional path-integral weight.
The one-loop beta function of the coupling constant $\varkappa$ is given by \eqref{universal_RG}.

One should note that the coefficient of the one-loop beta function depends on the definition of the coupling constant $\varkappa$, and can be changed by rescaling of $\varkappa$. We have a canonical normalization of $\hbar$ in the realization from the 4d CS theory, so that $\varkappa$ can be written as
\begin{align}
    \label{h_pm}
    \varkappa= -2i\,\frac{\hbar}{\zA-\zB}\;,
\end{align}
where $\hbar$ is the coupling constant of the 4d CS theory, while $\zA$ and $\zB$ are positions of the order surface defects for the theory \eqref{S_JJ}, in the language of \cite{Costello:2019tri}, and the integral in \eqref{S_JJ} is taken with respect to the Cartesian measure on $\Sigma$ (see \cref{sec:ultralocal}). The universality of the one-loop RG equation \eqref{universal_RG} holds in this definition of $\varkappa$. For the theory \eqref{S_PCM}, which is realized using disorder surface defects, the coupling is instead determined by the zeroes and poles of the meromorphic one-form, see \cref{sec:nonultralocal}, where we use the inverse coupling $H=1/(2\pi\lambda)$. In terms of $H$, \eqref{varkappa_RG} reads $\rd H/\rd\log\mu=\mathsf h^\vee/2$, which agrees with \cite{Levine:2022hpv} (whose coupling, denoted $h$ there, and dual Coxeter number $c_G$ correspond to our $H$ and $\mathsf h^\vee/2$).
We can moreover identify the underlying symmetry group $G$ with the gauge group of the 4d CS theory.

We note at this point that while there has been work on providing a 4d CS interpretation of the RG flow of 2d nonultralocal IFTs in \cite{Lacroix:2025ias, Levine:2023wvt} (based on previous analysis in two dimensions \cite{Levine:2022hpv, Hassler:2023xwn, Lacroix:2024wrd}), these papers did not derive the RG flow from Feynman diagram computations in the 4d CS theory itself; rather, one first derives the two-dimensional effective field theory,
whose RG flow was then studied by two-dimensional methods.
In addition, the work of Derryberry \cite{Derryberry:2021rne} provides a mathematical setting for the derivation of the RG flow of nonultralocal IFTs from 4d CS theory.

One of the main techniques that we shall use to derive the RG flow is the Batalin--Vilkovisky (BV) formalism \cite{Batalin:1981jr,Batalin:1983ggl}. This formalism has previously been applied to 4d CS in the work of Costello \cite{Costello:2013zra} as well as the work of Khan \cite{Khan:2022vrx}, which featured the computation of the framing anomaly that arises when the topological surface (that supports line operators in the construction of integrable lattice models) of 4d CS theory has nonzero curvature. The recent work of Bittleston and Costello on the renormalization group flow of the self-dual sector of Yang--Mills theory also utilized the BV formalism of 6d holomorphic BF theory on twistor space
\cite{Bittleston:2025jmk}. We shall utilize similar techniques to derive the RG flow of integrable field theories from 4d Chern--Simons theory.

The BV formalism has been studied in the context of classical integrable field theories and 4d CS theory before, notably, in the recent paper \cite{Alfonsi:2026pth} that describes the relation of 4d CS theory and the principal chiral model at the level of cyclic $L_{\infty}$-algebras. The work \cite{Benini:2026pwa} also relates 4d CS theory to integrable field theories at the level of $L_{\infty}$-algebras using homotopy transfer techniques. However, these papers do not  address the quantization and renormalization group flow of these integrable field theories.

The rest of this paper is organized as follows. In \cref{sec:ultralocal} we discuss the quantization of ultralocal field theories from 4d Chern--Simons theory, including the trigonometric and elliptic cases, higher-loop contributions and multiple defects.
In \cref{sec:nonultralocal} we then discuss the quantization of nonultralocal field theories from 4d Chern--Simons theory using the BV formalism, focusing on the principal chiral model (PCM) with Wess--Zumino term. The RG flow of the PCM with Wess--Zumino term is computed using two methods in this formalism, that is, by explicitly evaluating one-loop amplitudes, and by deriving it from an analogue of analytic torsion (\cref{sec:rg-torsion}). We make speculative comments on elliptic integrable field theories in \cref{sec:comments}. We also include several appendices for technical materials: Lie algebra conventions (\cref{app:Lie}), conventions for elliptic IFTs (\cref{ellipticcon}), the adjoint of the rescaled BRST differential and the associated Laplacians (\cref{app:Qadjoint,app:Qlaplacian}), the short-time asymptotics of the one-loop amplitudes (\cref{shorttimeheat}), and the background-coupled Laplacians (\cref{app:bg}); in particular, \cref{app:2dPCM} reviews the two-dimensional derivation of the RG flow of the PCM, with which our results are compared.

\noindent
\textbf{Note: }
During the final stages of this work, a preprint by Costello and Liniado~\cite{Costello:2026gve}
on holography for integrable field theories arising from 4d Chern--Simons theory
appeared on the arXiv.  While there is some overlap in the physical questions
addressed, the methods are distinct. They derive the
RG flow of integrable field theories from a holographic description in terms of a topological string and
planar spectral curves, while here we derive the one-loop RG flow directly and intrinsically from
the perturbative quantum field theory of 4d Chern--Simons theory itself.

\section{RG Flow for IFTs Arising from Order Surface Defects}\label{sec:ultralocal}

\subsection{Tree Level Action from Feynman Diagrams}

Consider the 4d CS theory on a product of two 2d surfaces $\Sigma\times C$, with local complex
coordinates $(w,\bar w)$ on $\Sigma =\C$ and $(z,\bar z)$ on the Riemann surface $C$, governed by the action
\ie
    S_{\textrm{4d-CS}}=\frac{1}{2\pi \hbar}\int_{\Sigma \times C}\omega \wedge \textrm{Tr}\left( A\wedge\textrm{d} A +\frac{2}{3} A\wedge A \wedge A\right) \;,
\fe
where the overall constant $\hbar$ controls the perturbative expansion of the theory.\footnote{In our convention, the expectation value of an operator $\mathcal{O}$ is given by the path integral
\begin{align}
    \displaystyle
    \langle \mathcal{O} \rangle
    = {
    \int \mathcal{D} A \, \mathcal{O} \exp(i S)
    }/{\int \mathcal{D} A \, \exp(i S) }\;.
\end{align}
}
The connection $A = A_{w} dw + A_{\wbar} d\wbar+A_{\zbar} d\zbar$ has components along $w, \wbar, \zbar$-directions, and takes values in the Lie algebra $\mathfrak{g}$ associated with the gauge group $G$.

In the following we will denote the generators of $\mathfrak{g}$ as $T^a$ with adjoint indices $a, b, \dots$, satisfying commutation relations
$[T^a, T^b] =  f^{ab}{}_c T^c$ with real structure constants $f^{ab}{}_c$ (the $T^a$ are anti-Hermitian generators).
We will choose the normalization of the generators as
\begin{align}
    \textrm{Tr}(T^a T^b) = \delta^{ab}\;,
\end{align}
and the dual Coxeter number $\mathsf{h}^{\vee}$ is defined by (in this normalization, $\mathsf h^\vee$ is twice the dual Coxeter number in the standard normalization, see \cref{app:Lie})
\begin{align}
    \label{Coxeter_def}
    f^{ab}{}_c f^{ab}{}_d = \mathsf{h}^{\vee}\delta_{cd}\;.
\end{align}

In order to realize two-dimensional integrable field theories,
we first consider the 4d CS theory coupled with order surface defects.
We then integrate out the fields along the curve $C$,
to obtain a two-dimensional effective theory on $\Sigma$.
The perturbative expansion of the resulting two-dimensional effective action is
organized with respect to the 2d coupling constant $\varkappa$, which is determined by the combination of $\hbar$ and $\omega$. The terms of order $\varkappa$ give the classical interaction, while the terms of order $\varkappa^2$ give the first quantum correction, i.e., the one-loop running of the theory.

The perturbative computation of the effective action requires the propagator for the 4d CS theory, which depends on the gauge choice.
Let us discuss two concrete choices of the gauge,
the Lorentz(-like) gauge
\ie
    \label{eq:Lorentz}
    \partial_z A_{\zbar} + \partial_w A_{\wbar} +\partial_{\wbar} A_w =0 \;,
\fe
and the holomorphic gauge
(which may also be called the axial(-like) gauge):
\ie
    A_{\zbar}=0 \;.
\fe
While the final result is independent of the gauge choices, the intermediate computations are quite different, and hence we discuss both choices below.

Let $\sfP^{ab}(X_1, X_2)$ denote the propagator two-form
of the 4d CS theory with flat metrics on both surfaces and $\omega=\textrm{d}z$, where $a$ and $b$ are Lie algebra indices, and
we have collectively denoted the coordinates as $X_i=\{w_i,\bar{w}_i,z_i, \bar{z}_i\}$ ($i=1, 2$).
By definition, the propagator
depends only on the differences of the coordinates as $\sfP^{ab}(X_1, X_2)=\sfP^{ab}(X_1-X_2)$, where the latter satisfies the defining relation
\begin{equation}
    \label{P_def}
    \frac{i}{2\pi  } \omega \wedge \d\, \sfP^{ab}=\delta^{ab}\delta^{(4)}(w,\bar{w},z,\bar{z})\,  \d w \wedge \d\bar{w} \wedge \d z \wedge \d\bar{z} \;,
\end{equation}
where $\delta^{(4)}(w,\bar{w},z,\bar{z}) \d w \wedge \d\bar{w} \wedge \d z \wedge \d\bar{z}$
on the right-hand side is the four-form current localized at $X=0$, i.e., $w=\bar w=z=\bar{z}=0$. This is the normalization of \cite[eq.~(4.7)]{Costello:2017dso}, with the same action and path-integral weight. More explicitly, in Lorentz-like gauge,
\begin{equation}
    \label{eq:bulkprop}
    \begin{aligned}
        \sfP^{ab}(x)
        &=\delta^{ab}\frac{i}{4\pi }
        \frac{
        w\,\d\bar w\wedge \d\bar z
        + \bar w\,\d\bar z\wedge \d w
        + 2
        \bar z\,\d w\wedge \d\bar w
        }{
        \big(|w|^2+|z|^2\big)^2
        }\;.
    \end{aligned}
\end{equation}
Note that in the Cartesian coordinates $x$ and $y$ on $\Sigma$, where $w=x+i y$ and $\bar{w} =x-i y$, we have
\begin{equation}
    \sfP^{ab}(x)
    =\delta^{ab}\frac{1}{2 \pi }
    \frac{
    x\,\d y\wedge \d\bar z
    + y\,\d\bar z\wedge \d x
    + 2
    \bar z\,\d x\wedge \d y
    }{
    \big(x^2+y^2+|z|^2\big)^2
    }\;,
\end{equation}
which is the same expression as in \cite{Costello:2017dso}.

We shall firstly be interested in the case where $C=\mathbb{CP}^1$, corresponding to rational integrable field theories.

\definecolor{figcyan}{RGB}{0,255,255}
\begin{figure}[htbp]
    \centering
    \begin{tikzpicture}[scale=1.0]
        \draw[fill=figcyan, draw=black, line width=0.4pt]
            (-2.9,-1.1) -- (-1.3,-1.35) -- (-1.3,1.15) -- (-2.9,1.4) -- cycle;
        \draw[fill=figcyan, draw=black, line width=0.4pt]
            (1.3,-1.1) -- (2.9,-1.35) -- (2.9,1.15) -- (1.3,1.4) -- cycle;
        \draw[decorate, decoration={snake, amplitude=2.5pt, segment length=6.9pt},
            line width=1.3pt]
            (-2.25,0.0) -- (2.25,0.0);

        \node at (-2.15,0.33) {$A_{\bar w}$};
        \node at (-2.18,-0.32) {$\mathcal{J}_w$};
        \node at (2.15,0.33) {$A_w$};
        \node at (2.18,-0.32) {$\mathcal{J}_{\bar w}$};
        \node at (0,0.45) {$\langle A_{\bar w} A_w\rangle$};

    \end{tikzpicture}
    \caption{Tree-level Feynman diagram generating the leading classical piece of the two-dimensional effective action.}
    \label{figure1}
\end{figure}
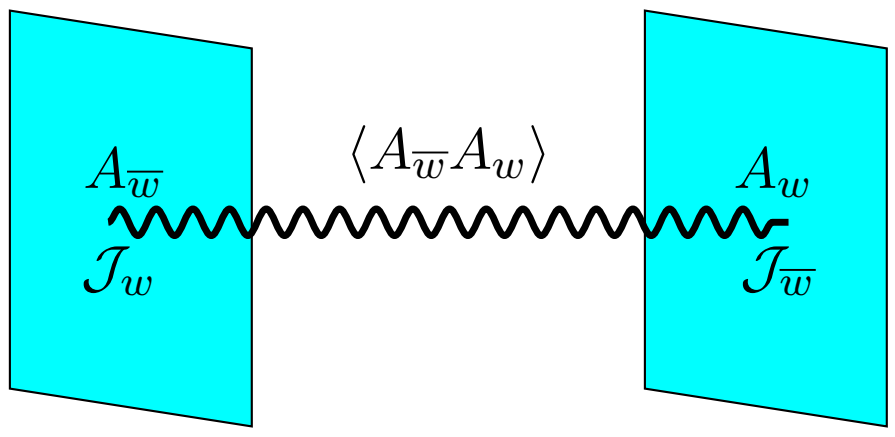

We shall first compute the interaction term of ultralocal IFTs using this Lorentz gauge propagator. This arises from the tree-level Feynman diagram depicted in Figure \ref{figure1}, with defects at positions $\zA$ and $\zB$ respectively,
where the vertices represent the coupling of the surface defects to the gauge field of the 4d CS theory:\footnote{We do not include factors of $i$ in this coupling. Correspondingly, in this section we expand the exponential of the defect couplings without factors of $i$ and identify the connected correlators directly with terms in the exponent of the two-dimensional weight. Thus $h$ is the coefficient of $\int\cJ_w\cJ_{\bar w}$ in this exponent, and the sign of the beta function \eqref{rationalbeta} refers to this convention.}
\begin{align}
    \int_{\Sigma\times \{z_+ \}} \d^2 w \, A_{\bar w} \mathcal{J}_w\;,
    \quad
    \int_{\Sigma\times \{z_- \}} \d^2 w\, A_{w} \mathcal{J}_{\bar w}\;
\end{align}
where $d^2 w =   \d w \wedge \d\wbar$.
The Feynman diagram evaluates to the integral
\ie
    \hbar\int_{\Sigma} \d^2 w \int_{\Sigma} \d^2 w' \, \cJ_{w}^a(w,\bar{w})
    \sfP_{ab}(w-w',\bar{w}-\bar{w}')
    \cJ_{\bar{w}}^b(w',\bar{w}')
    \;,
\fe
where the power of $\hbar$ is associated with the propagator.
Substituting the expression \eqref{eq:bulkprop} of the propagator, we obtain
\ie
    \frac{i}{4 \pi }\hbar\int_{\Sigma} \d^2 w \int_{\Sigma} \d^2 w' \, \frac{\cJ_{w}^a(w,\bar{w}) \cJ_{\bar{w}}^a(w',\bar{w}')\, 2(\bar{z}_{+}-\bar{z}_{-})}{(|w-w'|^2 +|\zA-\zB|^2)^2}
    \;.
\fe
Shifting the integration variables as $w' \rightarrow w'+w$, the integral becomes
\ie
    \frac{i}{4 \pi }\hbar\int_{\Sigma} \d^2 w \int_{\Sigma} \d^2 w' \, \frac{\cJ_{w}^a(w,\bar{w}) \cJ_{\bar{w}}^a(w'+w,\bar{w}'+\bar{w})\, 2(\bar{z}_{+}-\bar{z}_{-})}{(|w'|^2 +|\zA-\zB|^2)^2}
    \;.
\fe
Taylor expanding $\cJ_{\bar{w}}^a(w'+w,\bar{w}'+\bar{w})$ as $\sum_{m,n=0}^{\infty} \frac{1}{m!}\frac{1}{n!}(w')^m(\bar{w}')^n\partial^m_{w}\partial^n_{\bar{w}}\cJ_{\bar{w}}^a(w,\bar{w})$, we find that the only terms in the Taylor expansion that contribute to the integral are those where $m=n$, due to the angular integral on the $w'$-plane. We can then compute the integral over the $w'$-plane, using polar coordinates, whereby $\d^2w' = -i 2 r_{w'}\d r_{w'}\d\theta_{w'}$. The $m=0$, $n=0$ term gives rise to the expected interaction term,
\ie \label{standard interaction}
    \frac{\hbar}{z_{+}-z_{-}}
    \int_{\Sigma} \d^2 w  \, {\cJ_{w}^a(w,\bar{w}) \cJ_{\bar{w}}^a(w,\bar{w})}
    \;.
\fe
Here and below, the tree-level amplitude is obtained from the coefficient of $\d w\wedge\d\bar w$ in the propagator two-form $\sfP(X_1-X_2)$, with $X_1$ on the defect at $\zA$ (which couples to $A_{\bar w}$) and $X_2$ on the defect at $\zB$ (which couples to $A_w$).
After converting $\d^2w$ to the Cartesian measure, the overall coefficient ${\hbar}/{(\zA-\zB)}$ gives the two-dimensional coupling constant $h$ in \eqref{hdef} below, in agreement with \eqref{h_pm}.
The integrals for the remaining terms corresponding to $m=n$, when performed with suitable IR cutoffs, give rise to higher-derivative interactions. However, these interactions vanish upon imposing the equations of motion of the surface defect, which requires that $\mathcal{J}_w$ and $\mathcal{J}_{\bar{w}}$ are holomorphic and antiholomorphic, respectively. In the language of BV quantization that we shall use in the next section, such interactions are trivial in the cohomology of the BV differential.
In the following subsection, we shall identify counterterms that give rise to the running of the coupling entering this term.

In the computation with the Lorentz gauge, the intermediate steps (including the propagator) involved
the anti-holomorphic spectral parameters $\zbar$, and we recovered the holomorphicity in $z$ only at the end of the computation. The holomorphicity, in contrast, is manifest throughout
if we choose the holomorphic gauge, in which the propagator satisfying \eqref{P_def} is given by
\ie \label{holprop}
    \sfP^{ab}(x_1,x_2)
    =
    \langle A_{\bar w}^a(w_1,z_1)A_{w}^b(w_2,z_2)\rangle \,
    d^2 w
    =
    \delta^{ab}\delta^{(2)}(w_1-w_2)\frac{1}{z_1-z_2}
    d^2 w
    \;,
\fe
as can be checked with the help of the formula
$\partial_{\bar{z}}(z^{-1}) = -2\pi i \delta(z, \bar z)$, where the delta function is normalized as $\int \d z\wedge\d\bar z\,\delta(z,\bar z)=1$ (and similarly $\int\d^2w\,\delta^{(2)}(w)=1$ with $\d^2w=\d w\wedge\d\bar w$).
In this gauge, the Feynman diagram of \cref{figure1} evaluates to
\ie \label{inntterm}
    \hbar
    \int_{\Sigma} d^2 w \int_{\Sigma} d^2 w' \, \delta^{(2)}(w-w') \frac{\cJ_{w}^a(w,\bar{w}) \cJ_{\bar{w}}^a(w',\bar{w}')}{\zA-\zB}
    \;,
\fe
which again gives \eqref{standard interaction}.
In the following, we shall write the measure on $\Sigma$ as $\d^2w= -2i \d^2x $, where $\d^2x$ is the measure for Cartesian coordinates. Then, \eqref{inntterm} can be written as
\ie
    h
    \int_{\Sigma} \d^2 x  \,  \cJ_{w}^a(w,\bar{w}) \cJ_{\bar{w}}^a(w,\bar{w})
    \;,
\fe
where
\ie \label{hdef}
    h= -2i \frac{\hbar}{z_+-z_-} \;.
\fe
This shows that computations are simplified in holomorphic gauge, which we will use in the loop computations below.

\subsection{One-Loop RG Flow from Feynman Diagrams}

In the following, we shall describe how the one-loop RG flow for ultralocal IFTs can be derived from a Feynman diagram expansion. We shall consider the following Feynman diagrams at order $\hbar^{2}$ that arise when computing the correlation function of order surface defects in 4d CS. From the four-dimensional point of view these diagrams contain two propagators and no loop of the gauge field; we refer to them as one-loop since they give the first quantum correction, of order $h^2$, to the two-dimensional coupling.

\definecolor{figcyan}{RGB}{0,255,255}

\begin{figure}[htbp]
    \centering
    \begin{tikzpicture}[scale=1.0, baseline=(current bounding box.center)]

        \begin{scope}[xshift=0cm]

            \draw[fill=figcyan, draw=black, line width=0.4pt]
                (-2.9,-1.1) -- (-1.3,-1.35) -- (-1.3,1.15) -- (-2.9,1.4) -- cycle;

            \draw[fill=figcyan, draw=black, line width=0.4pt]
                (1.3,-1.1) -- (2.9,-1.35) -- (2.9,1.15) -- (1.3,1.4) -- cycle;

            \draw[decorate, decoration={snake, amplitude=2.5pt, segment length=6.9pt},
                line width=1.3pt]
                (-2.25,0.25) -- (2.25,0.25);

            \draw[decorate, decoration={snake, amplitude=2.5pt, segment length=6.9pt},
                line width=1.3pt]
                (-2.25,-0.25) -- (2.25,-0.25);

            \node at (-2.47,0.33) {$\mathcal{J}_{ w}$};
            \node at (-2.47,-0.32) {$\mathcal{J}_w$};

            \node at (2.47,0.33) {$\mathcal{J}_{\bar w}$};
            \node at (2.47,-0.32) {$\mathcal{J}_{\bar w}$};

        \end{scope}

        \node[font=\Large] at (3.8,0) {$+$};

        \begin{scope}[xshift=7.4cm]

            \draw[fill=figcyan, draw=black, line width=0.4pt]
                (-2.9,-1.1) -- (-1.3,-1.35) -- (-1.3,1.15) -- (-2.9,1.4) -- cycle;

            \draw[fill=figcyan, draw=black, line width=0.4pt]
                (1.3,-1.1) -- (2.9,-1.35) -- (2.9,1.15) -- (1.3,1.4) -- cycle;

            \draw[decorate, decoration={snake, amplitude=2.5pt, segment length=6.9pt},
                line width=1.3pt]
                (-2.25,0.25) -- (2.25,-0.25);

            \draw[decorate, decoration={snake, amplitude=2.5pt, segment length=6.9pt},
                line width=1.3pt]
                (-2.25,-0.25) -- (2.25,0.25);

            \node at (-2.47,0.33) {$\mathcal{J}_{ w}$};
            \node at (-2.47,-0.32) {$\mathcal{J}_w$};

            \node at (2.47,0.33) {$\mathcal{J}_{\bar w}$};
            \node at (2.47,-0.32) {$\mathcal{J}_{\bar w}$};

        \end{scope}
    \end{tikzpicture}

    \vspace{0.4cm}

    \begin{tikzpicture}[scale=1.0, baseline=(current bounding box.center)]

        \node[font=\Large] at (-3.8,0) {$=$};
        \node[font=\Large] at (-2.3,0) {$2\times$};

        \begin{scope}[xshift=1cm]

            \draw[fill=figcyan, draw=black, line width=0.4pt]
                (-2.9,-1.1) -- (-1.3,-1.35) -- (-1.3,1.15) -- (-2.9,1.4) -- cycle;

            \draw[fill=figcyan, draw=black, line width=0.4pt]
                (1.3,-1.1) -- (2.9,-1.35) -- (2.9,1.15) -- (1.3,1.4) -- cycle;

            \draw[decorate, decoration={snake, amplitude=2.5pt, segment length=6.9pt},
                line width=1.3pt]
                (-2.25,0) -- (-0.9,0);

            \draw[decorate, decoration={snake, amplitude=2.5pt, segment length=6.9pt},
                line width=1.3pt]
                (-0.9,0) .. controls (-0.35,0.85) and (0.35,0.85) .. (0.9,0);

            \draw[decorate, decoration={snake, amplitude=2.5pt, segment length=6.9pt},
                line width=1.3pt]
                (-0.9,0) .. controls (-0.35,-0.85) and (0.35,-0.85) .. (0.9,0);

            \draw[decorate, decoration={snake, amplitude=2.5pt, segment length=6.9pt},
                line width=1.3pt]
                (0.9,0) -- (2.25,0);

            \node at (-2.48,0) {$\mathcal{J}_w$};
            \node at (2.48,0) {$\mathcal{J}_{\bar w}$};

        \end{scope}

    \end{tikzpicture}
    \caption{Summation of two Feynman diagrams via OPE of currents. The lower diagram represents the OPE of the currents on the defects, not a loop of four-dimensional gauge fields.}
    \label{fig:combined}
\end{figure}
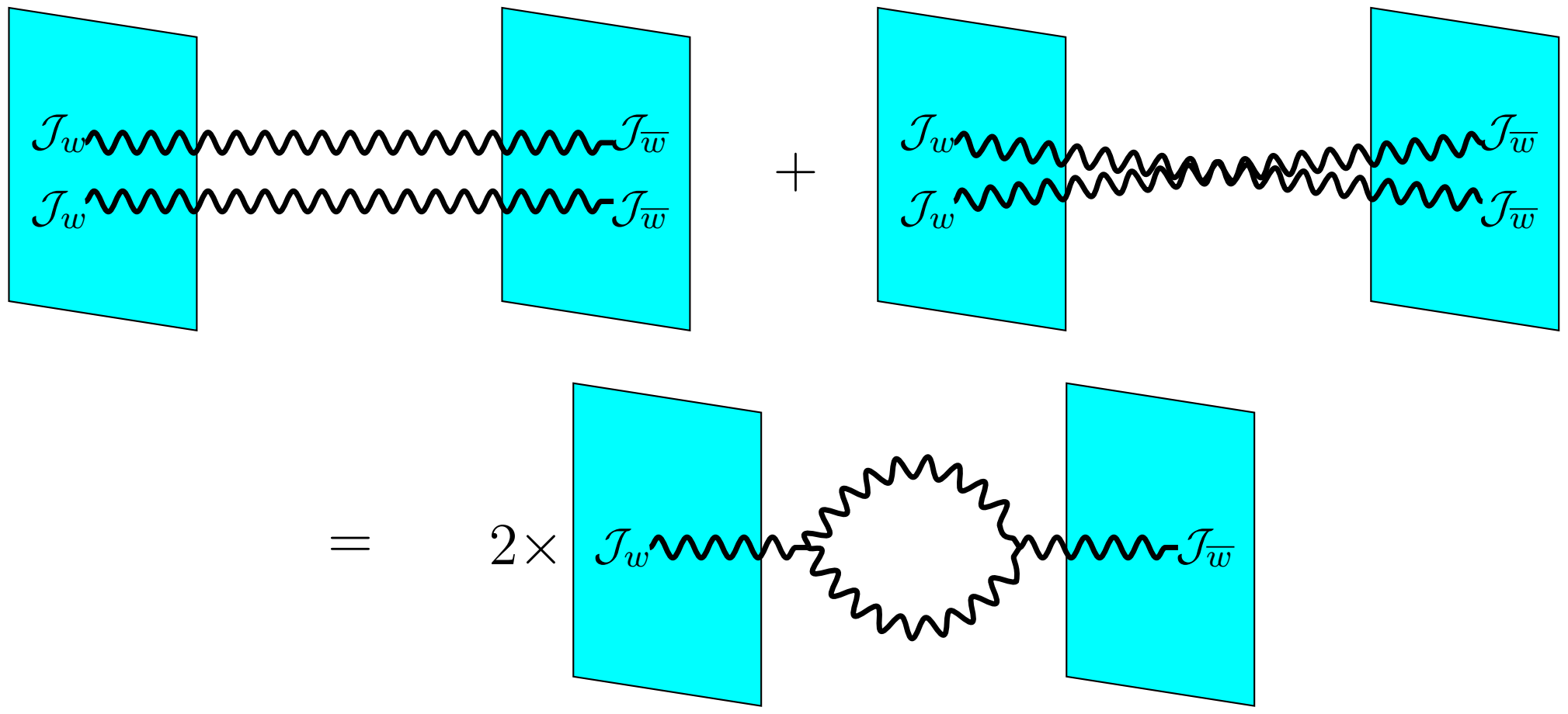

\subsubsection{Computations in Holomorphic Gauge}\label{onelooprational}

Let us consider the following correlation function that contributes to the effective action of 4d CS with the chiral and antichiral surface defects at second order in $\hbar$
\ie \label{2ndordcor}
    \left(\frac{1}{2!} \right)^2
    \hbar^2 \Big\langle &\int_{\{\zA\}} d^2w_1 \, \mathcal{J}_w^a(w_1,\bar{w}_1)A_{a\bar{w}}(w_1,\zA)\int_{ \{\zA\}} d^2w_2\,\mathcal{J}_w^b(w_2,\bar{w}_2)A_{b\bar{w}}(w_2,\zA)\\& \times
    \int_{\{\zB\}}d^2w_3 \, \mathcal{J}_{\bar{w}}^c(w_3,\bar{w}_3)A_{c{w}}(w_3,\zB)\int_{\{\zB\}} d^2w_4 \,\mathcal{J}_{\bar{w}}^d(w_4,\bar{w}_4)A_{d{w}}(w_4,\zB) \;
    \Big\rangle \;,
\fe
where the factors of $(1/2!)^2$
originate from the expansion of the exponential, and the factor of
$\hbar^2$ originates from the propagators.

Let us consider the contractions between the first and third gauge field, as well as the second and fourth gauge field in the correlation function.
Substituting in the propagator \eqref{holprop}, and integrating over the delta functions, we find
\ie
    \frac{1}{4}
    \varkappa^2
    \left\langle
    \int_{\{\zA\}} d^2x_1\int_{ \{\zA\}} d^2x_2\, \mathcal{J}_w^a(w_1,\bar{w}_1) \mathcal{J}_w^b(w_2,\bar{w}_2)\mathcal{J}_{\bar{w}a}(w_1,\bar{w}_1)\mathcal{J}_{\bar{w}b}(w_2,\bar{w}_2)
    \right\rangle \;,
\fe
where $h$ is defined in \eqref{hdef}.
Now, we can use the holomorphic OPE of currents that corresponds to the affine Kac--Moody algebra
\ie
    \mathcal{J}_w^a(w_1,\bar{w}_1) \mathcal{J}_w^b(w_2,\bar{w}_2)\sim \frac{f^{ab}{}_c\mathcal{J}_w^c(w_2,\bar{w}_2)}{w_1-w_2} + \frac{k\delta^{ab}}{(w_1-w_2)^2}\;,
\fe
as well as its anti-holomorphic counterpart, 
to obtain (the cross terms between the simple and double poles vanish since $f^{aa}{}_c=0$, and the product of the double poles only renormalizes the identity operator)
\ie
    \frac{1}{4}
    \varkappa^2 f^{ab}{}_c f_{ab}{}^d
    \left\langle
    \int_{\{\zA\}} d^2x_1\int_{ \{\zA\}} d^2x_2 \,\frac{\mathcal{J}_w^c(w_2,\bar{w}_2) \mathcal{J}_{\bar{w}d}(w_2,\bar{w}_2)}{|w_1-w_2|^2}
    \right\rangle \;.
\fe
We can now shift variables as $w_1 \rightarrow w_1+ w_2 $, and perform the integral over the $w_1$ plane with short distance and long distance cutoffs denoted by $\Lambda^{-1}$ and $\mu^{-1}$, respectively.
This integral gives us
\ie
    \int d^2x_1 \frac{1}{|w_1|^2} = \int_{\Lambda^{-1}}^{\mu^{-1}} r dr  \int_0^{2\pi} d\theta \frac{1}{r^2}= 2\pi\log \left(\frac{\Lambda}{\mu}\right) \;,
\fe
where we used $d^2 x=   r dr d\theta$.
Using the definition of the dual Coxeter number in \eqref{Coxeter_def}, we find
\ie
    \frac{1}{4}
    2\pi \varkappa^2 \mathsf{h}^{\vee}  \log \left(\frac{\Lambda}{\mu}\right)\int_{ \{\zA\}} d^2x_2 \,\mathcal{J}_w^c(w_2,\bar{w}_2) \mathcal{J}_{\bar{w}c}(w_2,\bar{w}_2)
    .
\fe

Taking into account the other possible contraction of gauge fields, we find that the total amplitude is twice this value
\ie
    \pi \varkappa^2 
    \mathsf{h}^{\vee}  \log \left(\frac{\Lambda}{\mu}\right) \int_{ \{\zA\}} d^2x_2 \,\mathcal{J}_w^c(w_2,\bar{w}_2) \mathcal{J}_{\bar{w}c}(w_2,\bar{w}_2)
    .
\fe
This result is represented diagrammatically in \cref{fig:combined}.
We thus get the running coupling constant
\ie
    \varkappa(\mu)=
    \varkappa+{\pi \varkappa^2}\mathsf{h}^{\vee}  \log \left(\frac{\Lambda}{\mu}\right) \;,
\fe
where $\varkappa=\varkappa(\Lambda)$ on the right-hand side denotes the bare coupling at the UV cutoff,
leading to the beta function
\ie
\label{rationalbeta}    \frac{d\varkappa(\mu)}{d\log \mu
    } = -{\pi }\mathsf{h}^{\vee} \varkappa^2 \;.
\fe
In particular, for real positive $\varkappa$, the negative sign indicates that the integrable quantum field theory becomes weakly coupled at short distances/high energies and is thus asymptotically free. This result agrees with the two-dimensional derivation of the RG flow of Kutasov \cite{Kutasov:1989dt}.\footnote{A common convention is to have the coupling constant multiplied by a factor of ${1}/{(2\pi)}$. Defining $\varkappa'$ by extracting a factor of ${1}/{(2\pi)}$ from $\varkappa$, i.e., $\varkappa=\varkappa'/(2\pi)$, we find that the beta function has the familiar form
\ie
    \frac{d\varkappa'(\mu)}{d\log \mu
    } = -\frac{\mathsf{h}^{\vee}}{2}{ } \varkappa'^2 \;,
\fe
where $\mathsf h^\vee/2$ is the dual Coxeter number in the standard normalization.}
For real and positive $h$, the geometric interpretation of the RG flow is that the separation of the order surface operators on $\mathbb{CP}^1$ measured by $|\zA-\zB|$ becomes larger as the energy scale becomes larger.
In this derivation, the four-dimensional input is the tree-level propagator, which identifies the coupling with the separation of the defects on $C$ and leads to the geometric interpretation above, while the logarithmic divergence itself arises from the OPE of the currents on the defects, as in the two-dimensional computation. A loop computation in the four-dimensional theory itself is the subject of \cref{sec:nonultralocal}.

From \cite{Ashwinkumar:2023zbu}, we know that for the case of order surface defects, their gauge anomalies lead to the emergence of disorder surface defects as quantum corrections to $\omega =\textrm{d}z$, involving simple poles and zeroes of the meromorphic one-form $\omega$. However, these are order $\hbar$ corrections to $\textrm{d}z$, so these effects can only contribute to our analysis in the ultralocal case at two loops or higher.
\subsubsection{Trigonometric and Elliptic Integrable Field Theories}

It is straightforward to generalize the derivation of the one-loop effective action and renormalization group flow to the cases of trigonometric and elliptic integrable field theories, where we have $C=\mathbb{C}^{\times}$ and $C=E$ (the elliptic curve). This is easiest to achieve in holomorphic gauge, where the propagator now takes the generic form
\ie \label{holgenprop}
    \langle A_{\bar w}^a(w_1,z_1)A_{w}^b(w_2,z_2)\rangle =\delta^{(2)}(w_1-w_2)\hbar \, r^{ab}(z_1,z_2) \;,
\fe
where $r^{ab}(z_1,z_2)$ denote the elements of either a trigonometric or elliptic classical $r$-matrix. Unlike in \eqref{holprop}, we have included here the factor of $\hbar$ associated with the propagator.

We can then directly generalize the derivation in \cref{onelooprational}, by computing the amplitude corresponding to the 4d Chern--Simons Feynman diagrams in Figure \ref{fig:combined} in the case where $C$ is an elliptic curve. Once again, we find that the two possible contractions lead to the same amplitude, and summing these gives the total amplitude
\ie
    \label{ellampl}
    \Delta S_{\rm eff}&=-2i \pi \hbar^2 r_{ac}(\zA,\zB)r_{bd}(\zA,\zB) f^{ab}{}_f f^{cd}{}_e \log \left(\frac{\Lambda}{\mu} \right)\\
    &\qquad\times\int_{ \{\zA\}} d^2w_2 \, \mathcal{J}_w^f(w_2,\bar{w}_2) \mathcal{J}^e_{\bar{w}}(w_2,\bar{w}_2)
    .
\fe
We consider this expression in the elliptic case, since the trigonometric case can be understood as a limit of the former. In particular, we would like to understand the factor of $r_{ac}(\zA,\zB)r_{bd}(\zA,\zB) f^{ab}{}_f f^{cd}{}_e$.

Let us focus on $\mathfrak{g}=\mathfrak{sl}(N)$, which is the only case with nondegenerate elliptic solutions of the classical Yang--Baxter equation \cite{bd}, and work in the Belavin basis described in \cref{ellipticcon},
where the current-current interaction term of the IFT is written as
\begin{equation}
   \hbar \sum_{\alpha \in \mathbb{A}} \int_{\Sigma} \d^2 w\, r^\alpha\left(z_+-z_-\right)
    \cJ_{w, \alpha}(w,\wbar) \cJ_{\wbar}^{ \alpha}(w,\wbar) \;.
\end{equation}
Writing $r=\sum_i r_i'\otimes r_i''$, the group-theory factor
$r_{ac}r_{bd}f^{ab}{}_f f^{cd}{}_e$ in \eqref{ellampl} is the component form of
$\sum_{i,j}[r_i',r_j']\otimes[r_i'',r_j'']$. For $r=\sum_\alpha r^\alpha\, T^\alpha\otimes T_\alpha$, the commutation relations of \cref{ellipticcon} give
\ie
    \sum_{\alpha,\beta\in\mathbb A} r^\alpha r^\beta\,[T^\alpha,T^\beta]\otimes[T_\alpha,T_\beta]
    =\sum_{\gamma\in\mathbb A}\Bigg(\sum_{\substack{\alpha, \beta \in \mathbb{A} \\ \alpha+\beta=\gamma}} f_{\alpha, \beta} f^{\alpha, \beta} r^\alpha r^\beta\Bigg) T^\gamma\otimes T_\gamma \;.
\fe
Hence \eqref{ellampl} has the same form as the tree-level interaction term, and comparing the two,
the one-loop flow of the coupling constants is
\begin{equation}
    \frac{{d} r^\gamma (z_{+-})}{{d} \log \mu}=2 \pi i \hbar \sum_{\substack{\alpha, \beta \in \mathbb{A} \\ \alpha+\beta=\gamma}}  f_{\alpha, \beta} f^{\alpha, \beta} r^\alpha(z_{+-}) r^\beta(z_{+-}) \;,
    \label{rgammaflow}
\end{equation}
where $z_{+-}=z_+-z_-$.
Moreover, using the identity
\ie \mathsf{h}^{\vee} \partial_z r^\gamma(z)=-\sum_{\substack{\alpha, \beta \in \mathbb{A} \\ \alpha+\beta=\gamma}}  f_{\alpha, \beta} f^{\alpha, \beta} r^\alpha(z) r^\beta(z) \;,
\label{rmatrixidentity}\fe
which follows from the classical Yang--Baxter equation (see \cref{ellipticcon}),
and the chain rule, we find from \eqref{rgammaflow} that
\begin{equation}
    \frac{{d} z_{+-}}{{d} \log \mu }=-2\pi i\hbar \mathsf{h}^{\vee} \;.
    \label{zflow}
\end{equation}
The flow \eqref{zflow} agrees with the RG flow in the rational case, which can be checked by replacing the definition $h=-2i{\hbar}/{z_{+-}}$ in \eqref{rationalbeta}.

We thus find from 4d Chern--Simons theory that the divergent contribution to the effective action for an elliptic ultralocal integrable field theory can be reabsorbed into the running of the coupling.

\subsection{Higher Loop Contributions}\label{sec:higherloops}

We only considered correlation functions that are quartic in the gauge fields so far. It is possible to compute higher loop contributions to the RG flow, at least in a particular scheme, by evaluating correlators that are of sixth and higher order in the gauge field.  As before, we contract the gauge fields in holomorphic gauge. In this gauge, interaction vertices of 4d CS do not play a role in the evaluation, since the correlators do not involve the component $A_{\bar{z}}$. (We do not address here the subtleties of the holomorphic gauge, such as residual gauge symmetries and the infrared behaviour of the propagator.)

Indeed, the chiral defect couples only to $A_{\bar w}$ and the antichiral defect only to $A_w$, and in holomorphic gauge the only nonvanishing propagator is \eqref{holprop}. At order $\hbar^{2n}$, every $A_{\bar w}$ at $z_+$ is therefore contracted with an $A_w$ at $z_-$, and the $n!$ such contractions, together with the factors $(1/n!)^2$ from the two exponentials, give
\ie
    \frac{1}{n!}\Big\langle\Big(h\int_\Sigma\d^2x\,\mathcal J_w^a\mathcal J_{\bar w}^a\Big)^n\Big\rangle \;,
\fe
where the expectation value is taken in the theory on the defects. Summing over $n$, the correlator of the defects in 4d CS theory in holomorphic gauge becomes the partition function of the defect theory deformed by the current-current interaction with coupling $h$, and its perturbative expansion is the one evaluated in \cite{Gerganov:2000mt} using the current algebra.
Thus, in the rational case, the 4d CS computation reproduces the two-dimensional computation of \cite{Gerganov:2000mt}, and we recover the all-order formula proposed in \cite{Gerganov:2000mt},
\ie
    \beta_{\varkappa}=-\pi 
    \frac{\mathsf{h}^{\vee} \varkappa^2}{\left(1-k \pi \varkappa \right)^2} \;,
    \label{GLMbeta}
\fe
where $k$ is the level in the OPE of \cref{onelooprational}.
Here we have converted the formula of \cite{Gerganov:2000mt} to the present conventions: their coupling $g$ and Hermitian currents correspond to $\varkappa=g/2\pi$ and $\mathcal J=i\mathcal J_{\rm GLM}$, so that their level, which is positive for unitary WZW models, is $k_{\rm GLM}=-k$, and the denominator reads $(1+k_{\rm GLM}\pi\varkappa)^2$ in their conventions.
Note that at higher orders in $\varkappa$, the level $k$ of the affine Kac--Moody algebra begins to play a role.
Here, we implicitly assume that both order surface operators support affine Kac--Moody algebras with the same level, that is, $k_{\rm chiral}=k_{\rm antichiral}$, and this is in fact required by the analysis of anomaly inflow of the order surface operators whereby the one-form $\omega=\rd z$ receives quantum corrections involving simple poles and zeroes, see \cite{Ashwinkumar:2023zbu}. As mentioned above, this emergence of zeroes/disorder defects needs to be treated carefully at two loops and higher, and we shall not pursue this in detail. See \cite{Ludwig:2002fu} for a four-loop computation which disagrees with the all-order formula of \cite{Gerganov:2000mt}.

\subsection{Multiple Defects}

A strength of 4d CS is the possibility of constructing ultralocal IFTs from multiple chiral and antichiral surface operators of varying types, such as free-fermion operators, $\beta\gamma$ systems, and coadjoint orbit surface operators \cite{Ashwinkumar:2023zbu}. Each of these defects realizes an affine Kac--Moody algebra at a fixed level, and with the levels of these algebras satisfying the anomaly cancellation condition
\begin{equation}
    \sum_i k^i_{\rm chiral} =\sum_j k^j_{\rm antichiral} \;,
\end{equation}
where $k^i_{\rm (anti)chiral}$ denotes the level at the $i$-th (anti-)chiral defect, as explained in \cite{Ashwinkumar:2023zbu}. The tree-level interactions between each pair of  chiral and antichiral defects give a current-current interaction of an IFT, and the one-loop renormalization of the coupling can then be computed from  diagrams that are order 4 in the gauge field by directly generalizing the derivation above. At one loop, each coupling $h_{ij}=-2i\hbar/(z_i-z_j)$ between the $i$-th chiral and the $j$-th antichiral defect runs according to \eqref{rationalbeta}. (Diagrams involving the currents of two different chiral defects, say, do not contribute to the logarithm: these currents have no singular OPE with each other, and the remaining poles $1/(\bar w_1-\bar w_2)$ and $1/(\bar w_1-\bar w_2)^2$ of the antichiral OPE, multiplied by the regular Taylor expansion in $w_1-w_2$, vanish upon angular integration.) This is consistent with a flow of the defect positions in which all the separations $z_i-z_j$ change at the same rate. At higher loops, the levels $k^i_{\rm (anti)chiral}$ of the individual defects enter, and the formula \eqref{GLMbeta} for equal levels does not apply directly. Higher loop contributions to the  couplings can also be computed, in principle.

\section{One-Loop RG Flow for the Principal Chiral Model in the BV Formalism}\label{sec:nonultralocal}

In this section, we shall compute the one-loop RG flow for a particular nonultralocal integrable field theory that is derived using disorder surface defects in 4d Chern--Simons theory.
Specifically, we shall employ the Batalin--Vilkovisky (BV) formalism to derive the RG flow of the principal chiral model.
Unlike in \cref{sec:ultralocal}, where we expanded the defect couplings without factors of $i$, in this section we use the weight $e^{iS}$ throughout, and we state the result as the running of the coefficient $H$ of the Euclidean action $S_E=-iS_{\rm PCM}$, see \eqref{sec:torsion:classical}.

Compared with the previous derivation, an advantage of this derivation is that one does not need an explicit form of the propagator. In fact, we will see that the propagator can be obtained from a heat kernel, and we find that only the short-distance behavior of the latter, which is known in general contexts, is important for the discussion of the one-loop RG flow.

\subsection{BV Formalism}

We shall compute the quantum effective action of 4d Chern--Simons theory coupled to disorder surface defects and isolate a UV-divergent term that gives rise to the RG flow of IFTs. The 4d CS gauge field shall be expanded as
\ie
    \tA+\mathsf{B}\;,
\fe
where $\tA$ is a classical 4d CS solution, and $\sfB$ is a fluctuation.  The action thus has the form
\ie
    S[\tA+\mathsf{B}]=S[\tA]+\frac{1}{2 \pi \hbar} \int \omega \wedge \textrm{Tr}\left(\mathsf{B} \wedge \rd_{\tA} \mathsf{B}+\frac{1}{3}\mathsf{B}\wedge[\mathsf{B}, \mathsf{B}] \right)
    \;,
\fe
where $\mathrm{d}_{\tA} \mathsf{B}=\mathrm{d} \mathsf{B}+[\tA, \mathsf{B}]$ with $[X,Y]:= X\wedge Y-(-1)^{|X||Y|}Y\wedge X$, where $|X|$ indicates the sum of ghost number and form degree.
Given that this action has a gauge symmetry under the transformation $\mathsf{B} \rightarrow \mathsf{B}+\mathrm{d}_{\tA} \epsilon+[\mathsf{B}, \epsilon]$, we ought to gauge fix the path integral to remove redundant degrees of freedom. We shall follow the BV formalism to achieve this.

The BV formalism requires the inclusion of a ghost field and anti-fields for both the ghost and fluctuation field. The BV field space realizes a differential graded Lie algebra with an odd symplectic pairing. The BV field space is given by
\ie
    \label{BV_field_space}
    \Omega^{\bullet}_{\text{CS}_4}(\Sigma \times C; \mathfrak{g}) = \bigoplus_{m+n= \bullet} \Omega^m_{\text{dR}}(\Sigma; \mathfrak{g}) \otimes \Omega^{(0,n)}_{\bar{\del}}(C; \mathfrak{g})\;,
\fe
where $\Omega^{m}_{\text{dR}}(\Sigma; \mathfrak{g})$ denotes the space of $\mathfrak{g}$-valued $m$-forms on $\Sigma$, and $\Omega^{(0,n)}_{\bar{\del}}(C; \mathfrak{g})$ denotes the space of $\mathfrak{g}$-valued $(0,n)$ forms on $C$. In addition, there is a grading given by the form degree $F$, which is related to the ghost number by
\ie
    \text{gh} = 1-F\;.
\fe
Explicitly, the BV fields consist of
\ie
    \mathsf{c} \,\, &\in \,\, \Omega^0_{\text{CS}_4}(\Sigma \times C ; \mathfrak{g})\;,\\
    \mathsf{B}_i\text{d}x^{i} + \mathsf{B}_{\bar{z}} \text{d}\bar{z} \,\, &\in \,\, \Omega^1_{\text{CS}_4} (\Sigma \times C; \mathfrak{g})\;,\\
    \mathsf{B}^{\vee}_{ij} \text{d}x^{i} \wedge \text{d}x^{j} + \mathsf{B}^{\vee}_{i \bar{z}} \text{d}x^{i} \wedge \text{d} \bar{z} \,\, &\in \,\, \Omega^2_{\text{CS}_4} (\Sigma \times C; \mathfrak{g})\;,\\
    \mathsf{c}^{\vee}_{i j\bar{z}} \text{d}x^{i} \wedge \text{d}x^{j} \wedge \text{d}\bar{z} \,\, &\in \,\, \Omega^3_{\text{CS}_4} (\Sigma \times C; \mathfrak{g}) \;,
\fe
where $\mathsf{c}$ is the ghost field with ghost number $1$, $\mathsf{B}$ is the fluctuation field  with ghost number $0$, and $\mathsf{B}^{\vee}$ and $\mathsf{c}^{\vee}$ are their anti-fields with ghost numbers $-1$ and $-2$, respectively. We have also used the indices $i$ and $j$ to distinguish local coordinates when defining one-forms on $\Sigma$.

The aforementioned Lie algebra structure of the BV field space arises from the combination of the algebra of differential forms defined using the wedge product and the Lie algebra $\mathfrak{g}$. The differential on the BV field space is
\ie
    \mathrm{d}_{\mathrm{CS}_4}=\mathrm{d}_{\tA}^{\Sigma}+\bar{\partial}_{\tA}^{C} \;,
\fe
and it is nilpotent since $\widetilde A$ is a classical gauge field that satisfies the 4d CS equations of motion (for the PCM, in the sense $\omega\wedge F(\tA)=0$; on the rescaled complex of \cref{heatk} this becomes $Q_J^2=0$). The symplectic pairing of odd degree is
\begin{equation}\label{sympair}
    \langle \mathcal{X},\mathcal{Y} \rangle :=\int_{\Sigma \times C} \omega \wedge \textrm{Tr}(\mathcal{X}\wedge \mathcal{Y}) \;,
\end{equation}
where $\mathcal{X}$ and $\mathcal{Y}$ belong to the BV field space. The BV action for 4d CS theory can be written as
\ie S_{\text{BV}}[\mathcal{B}] = \frac{1}{2\pi \hbar} \left\langle \mathcal{B}, \text{d}_{\text{CS}_4}\mathcal{B} + \frac{1}{3}[\mathcal{B},\mathcal{B}]  \right\rangle \;, \fe
where the BV field is
\ie
    \mathcal{B}=\sfc+\sfB+\sfB^{\vee}+\sfc^{\vee}\;.
\fe
 Note that since $\omega$ has zeroes, the symplectic form \eqref{sympair} would naively be degenerate at these zeroes, and in this case the BV formalism would need extra care. However, as we shall explain below, since some gauge field component develops a singularity at one of these zeroes, the BV pairing of the suitably rescaled fields and antifields (see \eqref{rescaledfields}) is nondegenerate.

\subsection{Superdeterminant}

In terms of components of the BV field, the BV action is given by
\begin{align} \begin{split}
        S_{\text{BV}} & = \frac{1}{2\pi\hbar} \int_{\Sigma \times C} \omega \wedge \textrm{Tr}(\mathsf{B}\wedge \,\text{d}_{\tA} \mathsf{B} + \frac{1}{3}\mathsf{B}\wedge [\mathsf{B},\mathsf{B}]) \\
        &+ \frac{1}{\pi\hbar} \int_{\Sigma \times C} \omega \wedge \textrm{Tr} ( \mathsf{B}^{\vee}\wedge ( \text{d}_{\tA} \mathsf{c} + [\mathsf{B},\mathsf{c}] ))
        + \frac{1}{2\pi\hbar} \int_{\Sigma \times C} \omega \wedge \textrm{Tr}( \mathsf{c}^{\vee}[\mathsf{c},\mathsf{c}])\;.
    \end{split}
    \label{thebvaction}
\end{align}

To arrive at this action, we have integrated by parts and dropped a term of the form $\int\omega \wedge  \mathsf{B} \wedge \tilde{F}$ using the equation of motion $\omega\wedge \tilde{F}=0$.
Integration by parts leads to a term proportional to $\int_{\Sigma\times C}\textrm{d}\omega \wedge \mathsf{B}\wedge \widetilde A$. This term vanishes using the boundary conditions of the gauge field, which are imposed on both its background and fluctuation components.
Crucially, there are two types of interaction terms from this action that shall be relevant for our analysis, namely, the $\sf{B} \tA \sfB$ interaction and the $\sfB^{\vee}\tA \sfc$ interaction.

In order to remove redundant degrees of freedom, we ought to fix the gauge symmetry of the theory. In the BV formalism, this corresponds to picking a Lagrangian subspace of the BV field space. We shall achieve this by choosing a metric on $\Sigma \times C$, denoted $g_{\Sigma} \oplus g_C$, and defining
\begin{equation}
    \delta=\left(\d^{\Sigma}\right)^{\dagger}+2\left(\bar{\partial}^C\right)^{\dagger}\;,
\end{equation}
which consists of the sums of the adjoints of $\d^{\Sigma}$ and $\bar{\partial}^C$, given explicitly by
\begin{equation}
\begin{aligned}
\left(\d^{\Sigma}\right)^{\dagger} & =g^{ij} \iota_{\frac{\partial}{\partial x^i}} D_j
\;, \\
\left(\bar{\partial}^C\right)^{\dagger} & =g^{z \bar{z}} \iota_{\frac{\partial}{\partial \bar{z}}} D_z \;,
\end{aligned}
\end{equation}
where $D$ denotes the covariant derivative with respect to the Christoffel connection.
As we shall see, in our application, we shall pick a metric whose principal part is flat (see \eqref{main:hk:principalmetric}); the warp factor of the metric along $\Sigma$ nevertheless gives rise to a connection that mixes $\Sigma$ and $C$.
The gauge-fixing Lagrangian subspace of the BV field space is then
\begin{equation}
    \mathfrak{L}=\left\{X \in \Omega^{\bullet}_{\text{CS}_4}(\Sigma \times C; \mathfrak{g}) \mid \delta X=0\right\}\;.
\end{equation}
This corresponds to an analogue of the Lorentz gauge. Here and in \cref{heatk}, we define the adjoints without the conventional minus sign (see the remark below the definition of $\delta$ in \cref{heatk}). For the PCM, $\delta$ is replaced by the adjoint $Q^\dagger$ of the rescaled BRST differential with respect to the Hermitian metric \eqref{hermitianmetric}, derived in \cref{app:Qadjoint}, and $\mathfrak L=\{X\mid Q^\dagger X=0\}$.

We shall compute the Gaussian BV integral on the gauge-fixing Lagrangian
$\mathfrak L$, keeping this gauge choice fixed as the background varies.
Let $\mathcal H_{\widetilde A}$ denote the Hessian of the quadratic
part of \eqref{thebvaction} restricted to this space, including the
BV pairing \eqref{sympair}, and define
\[
 \mathcal X
 =\mathcal H_0^{-1}
       \bigl(\mathcal H_{\widetilde A}-\mathcal H_0\bigr) \;,
\]
where $\mathcal{H}_0:=\left.\mathcal{H}_{\widetilde A}\right|_{\widetilde A=0} $.
The Hessian perturbation is linear in the background.
Writing $\Gamma^{(1)}=-i\log Z_{\mathrm{1-loop}}$, its
background-dependent contribution is
\ie
 \Gamma^{(1)}[\widetilde A]-\Gamma^{(1)}[0]
 =\frac{i}{2}\operatorname{STr}'\log(1+\mathcal X) \;.
 \label{STr_log}
\fe
The primes indicate omission of zero modes, and the traces are
understood with regularization. Expanding in the background gives
\ie
 \frac{i}{2}\operatorname{STr}'\log(1+\mathcal X)
 =\frac{i}{2}\operatorname{STr}'\mathcal X
  -\frac{i}{4}\operatorname{STr}'\mathcal X^2
  +O(\widetilde A^3) \;.
\fe
We shall compute the quadratic contribution,
\ie
 \Gamma^{(1)}_{(2)}
 =-\frac{i}{4}\left(
   \operatorname{Tr}'_{\rm B}\mathcal X_{\rm B}^2
   -2\operatorname{Tr}'_{\rm gh}\mathcal X_{\rm gh}^2
   \right) \;.
 \label{diffbosghos}
\fe
Here the ghost trace counts one ghost--antifield pair.
The odd Hessian acts on both members of this pair, so its trace
contains two identical contributions. The operators
$\mathcal X_{\rm B}$ and $\mathcal X_{\rm gh}$ compose the
corresponding free field propagator with a background insertion leading to one loop amplitudes that we will compute below.

\subsection{Heat Kernels and Propagators}\label{heatk}

An obstacle to deriving the explicit propagator is the presence of zeroes in $\omega$, which result in the noninvertibility of the symplectic structure on the field space of the theory. However, this does not pose a problem if one suitably restricts the field space.
Consider the 4d CS theory with a meromorphic one-form that has simple zeroes. Following the arguments of Costello and the second-named author \cite{Costello:2019tri}, locally, we can write the action as
\ie \label{zeroaction}
    \int_{\mathbb{C}\times C} z \mathrm{d} z \wedge A \wedge \mathrm{d} A=\int_{\mathbb{C}\times C} z \mathrm{d} z \wedge \mathrm{d} w \wedge \mathrm{d} \bar{w} \wedge \mathrm{d} \bar{z}\,2\left(-A_w \partial_{\bar{z}} A_{\bar{w}}+A_w \partial_{\bar{w}} A_{\bar{z}}-A_{\bar{w}} \partial_w A_{\bar{z}}\right)\;,
\fe
up to total derivatives, where $z$ is a local coordinate around a zero.
To ensure an elliptic propagator, Costello and the second-named author showed that one of the gauge field components along $\Sigma =\mathbb{C}$, for example, $A_w$, must be allowed to have a first-order pole at $z=0$. In this section, we shall discuss the derivation of the propagator with such restrictions.

Feynman diagram computations involving UV-divergent amplitudes require us to pick a regularization scheme. A convenient regularization scheme involves expressing the propagator in terms of a heat kernel. Hence, in what follows, we shall
focus on the heat kernel description of propagators, and we shall first
review the derivation of the propagator of 4d Chern--Simons theory with $\omega=\textrm{d}z$ in terms of a heat kernel.

The heat kernel $\fk_{\tau>0}$ on $\mathbb{R}^2\times \mathbb{C}$ satisfies
\begin{align}
    &\partial_{\tau} \fk_{\tau}(z, w)
    = \Delta_{z, w} \fk_{\tau}(z, w ) \\
    &\lim_{{\tau}\to 0}\fk_{\tau}(z, w )
    = \delta^{(4)}(z, w) \;,
\end{align}
where $\Delta:=\delta \textrm{d}+\textrm{d}\delta$ is the Laplacian on $\R^2 \times \C$. Here, and in the following, the dependence of heat kernels and propagators on the antiholomorphic coordinates $\bar{z}$ and $\bar{w}$ are implicit. 
On the flat space $\mathbb{R}^2\times \mathbb{C}$ the heat kernel takes the form 
\ie 
\fk_{\tau}(z, w ) =\frac{1}{(4\pi {\tau})^{2}}    e^{-|z|^2 / 4 {\tau}} e^{-|w|^2 / 4 {\tau}} \;.
\fe

In the flat case, let us define
\ie
    \delta:=(\tilde{\mathrm{d}}^{\dagger} +2\bar{\partial}^{\dagger} ) \;,
\fe
where
\ie
    \begin{aligned}
        \tilde{\mathrm{d}}^{\dagger}  & = \partial_{i} \iota_{\partial_i} \;, \\
        \bar{\partial}^{\dagger} &=g^{z \bar{z}} \partial_z \iota_{{\partial_{\bar{z}}}} \;.
    \end{aligned}
\fe
Note that we define these adjoints without the conventional minus sign, so that $\Delta=\delta\d+\d\delta$ above is the non-positive Laplacian and the heat equation takes the form $\partial_\tau\fk_\tau=\Delta\fk_\tau$. For the PCM, from \eqref{hermitianmetric} on and in the appendices, we instead use the non-negative Laplacians $\Delta_p=QQ^\dagger+Q^\dagger Q$ with heat kernels $e^{-\tau\Delta_p}$.
The regularized propagator $\fp_{\de}^{\dl}(z', w' ; z'',  w'' )$ is defined by
\ie
    \frac{i}{2\pi}
    \textrm{d}z \wedge \fp_{\de}^{\dl}(z', w' ; z'',  w'' )=
    \int_{\de}^{\dl} \rd {\tau} \, \delta \fk_{\tau}(z', w' ; z'',  w'' ) V \;,
\fe
where
\ie
    V=\rd \bar{z} \wedge \rd z \wedge \rd w \wedge \rd \bar{w}
\fe
is the volume form on $\mathbb{R}^2 \times \C$.
The limits $\de$ and $\dl$ of the ${\tau}$-integral have dimensions of length squared, and can be understood as the inverses of the squares of UV and IR regulators, respectively.

Since $\delta^2=0$, we can show that the regularized propagator obeys the gauge fixing condition
\ie
    \delta \fp_{\de}^{\dl}(z', w' ; z'',  w'' ) =0 \;.
\fe
Moreover, since
\ie
    \frac{i}{2\pi}
    \rd z \wedge \rd \fp_{\de}^{\dl}(z', w' ; z'',  w'' )
    &=-\frac{i}{2\pi}\rd\left(\rd z \wedge \fp_{\de}^{\dl}(z', w' ; z'',  w'' )\right)\\
    &=-\int_{\de}^{\dl} \rd {\tau}\, d\delta \fk_{\tau}(z', w' ; z'',  w'' ) \\
    &=-\int_{\de}^{\dl} \rd {\tau} \, \Delta \fk_{\tau}(z', w' ; z'',  w'' ) \\
    &=-\int_{\de}^{\dl} \rd {\tau} \, \partial_{\tau} \fk_{\tau}(z', w' ; z'',  w'' ) \;,
\fe
where we have used the fact that $\fk_{\tau}V$ is a top-form on $\mathbb{R}^2\times \C$, we recover the defining condition of the propagator
\ie
    \frac{i}{2\pi}
    \rd z \wedge \rd \fp^{\dl}_{\de}=\delta^{(4)} (z', w' ; z'',  w'' ) \;,
\fe
in the limit where $\de \rightarrow 0$ and $\dl\rightarrow \infty$.

In the presence of disorder defects, the defining condition on the propagator is generalized to
\ie
    \frac{i}{2\pi}
    \omega (z') \wedge \rd \fp(z', w' ; z'',  w'' ) =\delta^{(4)} (z', w' ; z'',  w'' )
\fe
or
\ie
    \lim_{\de\rightarrow 0, \dl\rightarrow \infty }
    \frac{i}{2\pi}
    \omega (z') \wedge \rd \fp_{\de}^{\dl}(z', w' ; z'',  w'' ) =\delta^{(4)} (z', w' ; z'',  w'' )
\fe
when UV and IR regulators are introduced.

We must now take into account the zeroes and poles of $\omega$ when deriving the propagator.
Let us first recall the local construction using the metric induced by $\omega$; this motivates the rescalings below, but is not used in the computations of \cref{sec:oneloopPCM}. As noted in \cite{Costello:2019tri}, close to a simple zero of $\omega$, this metric takes the form
\ie \label{conemetric}
    \rd s^2 = 4 \varrho^2 \rd \theta^2 +\rd \varrho^2 \;,\qquad \varrho=\frac{|u|^2}{2} \;,
\fe
where we denote the local holomorphic coordinate around the zero as $u=|u|e^{i\theta}$.  The angle around the conical singularity at  the zero is $4\pi$.
As explained in \cite{Costello:2019tri}, upon rescaling the field that has a simple pole at $u=0$, one finds an elliptic kinetic operator of the form
\ie
    \bar{\partial} + u \partial \;.
\fe
This can be understood as a complex structure deformation on the $w$-plane via a Beltrami differential.
We can then use this kinetic operator to derive the form of the propagator for the rescaled field close to $u=0$.
The defining condition of the propagator is locally
\ie
    \rd u \wedge (\bar{\partial} + u \partial)\fp =\delta^{(4)} (u', w'; u'',  w'' ) \;.
\fe
The rescaled-field propagator defined in terms of a heat kernel is thus locally of the form
\ie
    \fp_{R,\de}^{\dl}(u', w'; u'', w'')=\int_{\de}^{\dl} \rd {\tau} \,  (\bar{\partial} + u \partial)^{\dagger}\fk_{R,{\tau}}(u', w' ; u'',  w'') \;,
\fe
where the heat kernel is defined with respect to the  metric \eqref{conemetric}. The defining condition is obeyed in the
$\de \rightarrow 0$ and $\dl\rightarrow \infty$ limits since we have defined the heat kernel with respect to the Laplacian $\Delta_u = (\bar{\partial} + u \partial)(\bar{\partial} + u \partial)^{\dagger} +(\bar{\partial} + u \partial)^{\dagger}(\bar{\partial} + u \partial)$.
We then find that the components of the ordinary propagator for the unrescaled field can be obtained by multiplying the corresponding component of $\fp^{\dl}_{R,\de}$ by $1/u$.

To derive the explicit propagator, it is helpful to note that the action \eqref{zeroaction} with the prescribed simple pole for $A_w$ can be written as a holomorphic 4d $\beta \gamma$ system perturbed by a term quadratic in components of $\gamma$. Indeed, as noted in \cite{Costello:2019tri},
the action \eqref{zeroaction} is equivalent to
\begin{equation}
    \int \mathrm{d} z \rd w \rd \bar{w} \rd \bar{z}\left(-A^{\rm R}_w \partial_{\bar{z}} A_{\bar{w}}+A^{\rm R}_w \partial_{\bar{w}} A_{\bar{z}}\right)-\int z \rd z \rd w \rd \bar{w} \rd \bar{z} A_{\bar{w}} \partial_w A_{\bar{z}} \;,
\end{equation}
after defining the rescaled field $A^{\rm R}_w= z A_w$. Here, we identify $\textrm{d}z\wedge \textrm{d}w A^{\rm R}_w$ with a two-form field $\beta$, and $A_{\wbar}$ and $A_{\zbar}$ with components of an antiholomorphic one-form field $\gamma$. It is well-known that the propagator for the holomorphic 4d $\beta\gamma$ system can be given in terms of the Bochner--Martinelli kernel \cite{Williams:2018ows}
that satisfies\begin{equation}
    \bar{\partial}_x \omega_{\mathrm{BM}}(x ; y) \wedge d w \wedge d z=\delta_y(x) \;.
\end{equation}
The additional term linear in $z$ can be treated as an interaction term, whereby the correlator of $\beta$ with itself becomes nonzero.  Hence, we may obtain, e.g., the $A_w-A_{\zbar}$ propagator by multiplying the $\zbar$ component of the BM kernel by ${1}/{z}$.

It is desirable for our goal to have a more globally defined propagator. Let us present a derivation in the case of the PCM, where $\omega= \frac{(z-z_0)(z-z_1)}{z^2}\rd z$. In what follows, we shall first restrict ourselves to the case without Wess--Zumino term, and explain the generalization of the derivation to include the Wess--Zumino term in \cref{wzterm}.  In the case without Wess-Zumino term, which corresponds to $z_0=-z_1$, we may rescale $z_0$, $z_1$ and $z$ by a factor of $h=z_0$ such that $\omega$ can be written as
\ie\label{homegapcm}
    \omega = h  \frac{(z-1)(z+1)}{z^2}\rd z \;.
\fe

The kinetic term of the action in this case takes the form, up to the overall factor $1/(2\pi\hbar)$ and the trace,
\ie
    &h \int \frac{(z-1)(z+1)}{z^2} \rd z A \rd A \;.
\fe
We denote a generic rescaled field by $A^{\rm R}$, reserving
$\widetilde A$ for the classical background used in the loop amplitude.
Let us write
\ie \label{rescaledfields}
    A_w& = A^{\rm R}_w\frac{z}{z-1}\\
    A_{\bar{w}}& = A^{\rm R}_{\bar{w}}\frac{z}{z+1}\\
    A_{\bar{z}}& = A^{\rm R}_{\bar{z}}z \;,
\fe
in order to make the respective boundary behaviours of fields explicit. Writing $A_I=r_I A^{\rm R}_I$ with $r_w=z/(z-1)$, $r_{\bar w}=z/(z+1)$ and $r_{\bar z}=z$, compatibility with the BRST differential and constancy of the density of the BV pairing fix the rescalings of the ghost and of the antifields,
\ie
    \mathsf c=z\,\mathsf c^{\rm R} \;,\qquad
    \mathsf B^\vee_{IJ}=r_Ir_J\,z^{-1}\,\mathsf B^{\vee\rm R}_{IJ} \;,\qquad
    \mathsf c^\vee_{w\bar w\bar z}=r_wr_{\bar w}r_{\bar z}\,z^{-2}\,\mathsf c^{\vee\rm R}_{w\bar w\bar z} \;.
    \label{rescaledghosts}
\fe
Accordingly, the endpoint factors of the propagators below are
\ie
    f_{IJ}(z_1,z_2)=r_I(z_1)\,r_J(z_2) \;.
    \label{fIJ}
\fe
 The fields $A^{\rm R}$ here also decay to 0 as $z\rightarrow \infty $.   Now, $\partial_{\bar{z}}$ acting on the simple poles in the gauge fields result in delta functions supported at the locations of these poles, but the corresponding  terms in the action turn out to be zero due to the presence of the zeroes in $\omega$. It turns out that the action can be written in the form
\ie
    2\int \rd^4 x(A^{\rm R}_{\bar{w}}\partial_{\bar{z}}A^{\rm R}_w + (z-1) A^{\rm R}_{\bar{z}}\partial_w A^{\rm R}_{\bar{w}} +(z+1) A^{\rm R}_{w}\partial_{\bar{w}} A^{\rm R}_{\bar{z}}) \;.
\fe
The linearized BRST operator is thus
\ie
    Q = \bar{\partial}_z + (z-1)\partial_w + (z+1)\bar{\partial}_{{w}} \;,
\fe
where $\bar\partial_z=\rd \zbar\,\partial_{\zbar}$, $\partial_w=\rd w\,\partial_w$ and $\bar\partial_w=\rd\wbar\,\partial_{\wbar}$.
This operator is elliptic since the two zeroes $z=\pm 1$ of $\omega$ are distinct.

To define the adjoint of $Q$, we choose the Hermitian metric
\ie \label{hermitianmetric}
 \rd s^2=\alpha(z,\bar z)\,\rd w\,\rd\bar w+\rho\,\rd z\,\rd\bar z \;,
 \qquad \alpha=2\mathcal{N} \;,\quad \mathcal{N}=1+|z|^2 \;.
\fe
The degree-dependent adjoint $Q^\dagger$ and the Laplacians
$\Delta_p=(QQ^\dagger+Q^\dagger Q)|_{\Omega^p}$ are derived in
Appendices~\ref{app:Qadjoint} and~\ref{app:Qlaplacian}, whose formulas
we specialize to $\rho=1$. We shall also briefly discuss other smooth choices of $\rho$ below; for non-constant $\rho$, the weights compatible with the BV formalism are those of \cref{app:Qadjoint} divided by $\rho^{1/2}$. 
The choice of $\alpha$ we have made is a convenient one: since $|z-1|^2+|z+1|^2=\alpha$, for $\rho=1$ the principal metric of $\widehat\Delta_p=2\Delta_p$ is
completely flat, which is convenient for our computations:
\ie
 \widehat G=\rd w\,\rd\bar w+\rd z\,\rd\bar z \;.
\label{main:hk:principalmetric}
\fe
In other words, the effective Laplacians take the form $\widehat{G}^{AB}\partial_A\partial_B+\ldots$.

In operator notation, the regulated Green's operator and propagator for
the rescaled complex are
\ie \label{rescap}
 \mathsf G_{R,\de}^{\dl}
 =\int_{\de}^{\dl}\rd\tau\,e^{-\tau\Delta} \;,\qquad
 \mathsf P_{R,\de}^{\dl}=i\pi Q^\dagger\mathsf G_{R,\de}^{\dl} \;.
\fe
Here $\Delta=\bigoplus_p\Delta_p$ on the rescaled complex, and the factor
$\pi$ is the action normalization used in \eqref{P_G}. The normalization of
\eqref{rescap} is fixed by the quadratic part of the BV action
\eqref{thebvaction}: for a quadratic action $S=\frac12\langle X,KX\rangle$ with
weight $e^{iS}$, the propagator is $iK^{-1}$. Here $K=Q/\pi$ (suppressing $\hbar$
and $h$ as explained in \cref{sec:oneloopPCM}), and on the gauge-fixing
Lagrangian $\operatorname{im}Q^\dagger$ the inverse of $Q$ is
$Q^\dagger\Delta^{-1}$, which gives \eqref{rescap}. The agreement of the resulting
one-loop coefficient with the computation of \cref{sec:rg-torsion}, which
does not use propagators, provides an independent check.
The BV pairing \eqref{sympair} converts this operator expression into a bi-form
kernel.

\subsubsection{Propagator in terms of Heat Kernel}

Let $\mathsf K_p(\tau;x_1,x_2)$ be the operator heat kernel of
$e^{-\tau\Delta_p}$ on the rescaled degree-$p$ complex. It is scalar for
$p=0$ and a $3\times3$ matrix in the basis
$(\rd w,\rd\bar w,\rd\bar z)$ for $p=1$. 
The weighted BV/Hodge pairing
converts an operator kernel into the bi-form used in a propagator: we denote by
$\mathsf K_p^{\rm BV}(\tau;x_1,x_2)$ the kernel of $e^{-\tau\Delta_p}$ whose
second endpoint has been converted by the BV pairing, so that it is a bi-form of
degree $p$ at $x_1$ and $3-p$ at $x_2$. 
For the bosonic physical propagator, the component formula is
\ie
 \mathsf P_{IJ}(x_1,x_2)
 ={}&i\pi f_{IJ}(z_1,z_2)
 \bigl[Q^\dagger_{x_1}\mathsf G_2^{\rm BV}(x_1,x_2)\bigr]_{IJ} \;,\\
 \mathsf G_2^{\rm BV}(x_1,x_2)
 ={}&\int_{d^2}^{D^2}\rd\tau\,
                      \mathsf K_2^{\rm BV}(\tau;x_1,x_2) \;,
 \label{P_G}
\fe
where $I,J=w,\bar w,\bar z$, and no sum over $I,J$ is intended in
this component formula. The factor $\pi$ accounts for the overall
coefficient $1/\pi$ of the quadratic part of \eqref{thebvaction} (for the $\mathsf B\mathsf B$ term, the coefficient $1/(2\pi)$ combined with the factor $2$ from varying the quadratic form). The endpoint factors
$f_{IJ}$ implement \eqref{rescaledfields}; they multiply the result of
$Q^\dagger$ and are not included inside its derivative.

For brevity, we define 
\ie
 \mathscr P_1(\tau;x_1,x_2)
 =\sum_{I,J}f_{IJ}(z_1,z_2)
 \bigl[Q^\dagger_{x_1}\mathsf K_2^{\rm BV}(\tau;x_1,x_2)\bigr]_{IJ}
 \rd x_1^I\wedge\rd x_2^J \;.
 \label{main:hk:propagatorintegrand}
\fe
We use a single definition for all propagators in this section, namely
the operator \eqref{rescap}, $\mathsf P=i\pi Q^\dagger\mathsf G$, whose kernel
is converted into a bi-form by the BV pairing. The bosonic propagator is
its component from $\Omega^2$ to $\Omega^1$, with $\mathsf B$ at both
endpoints, and the ghost propagator is its component from $\Omega^1$ to
$\Omega^0$, with $\mathsf c$ at the first endpoint and $\mathsf B^\vee$ at the
second; we use Grassmann antisymmetry when interchanging the two endpoints.
In both cases $Q^\dagger$ acts at the first endpoint $x_1$, which carries the
field of lower form degree, as in \eqref{main:hk:propagatorintegrand}.
Accordingly, $\mathscr P_0(\tau;x_1,x_2)$ denotes the component of
$Q^\dagger_{x_1}\mathsf K_1^{\rm BV}(\tau;x_1,x_2)$, which has degree zero at $x_1$ and degree two at $x_2$.
Using $Q^\dagger_{p+1}\mathsf K_{p+1}=\mathsf K_pQ^\dagger_{p+1}$, the bosonic
and ghost propagators are expressed through the heat kernels in degrees one
and zero, respectively, and in the one-loop amplitude $Q^\dagger$ is combined
with the vertex insertion as in \eqref{Ffull:vertexoperator}; this is why both
derivatives in \eqref{main:hk:bubble} act on the first argument of the
kernels. Thus
\begin{equation}
 \mathsf P_{\mathsf B\mathsf B}(x_1,x_2)
 =i\pi\int_{d^2}^{D^2}\rd\tau\,
                  \mathscr P_1(\tau;x_1,x_2) \;,
 \label{P_BB}
\end{equation}
while
\begin{equation}
 \mathsf P_{\mathsf c\mathsf B^\vee}(x_1,x_2)
 =i\pi\int_{d^2}^{D^2}\rd\tau\,
                  \mathscr P_0(\tau;x_1,x_2) \;.
 \label{main:hk:ghostpropagator}
\end{equation}
The proper-time cutoffs $d^2,D^2$ have dimensions of length squared.
Here the heat kernels are those of the rescaled fields.
The factors $f_{IJ}$ restore the physical field components and
multiply the differentiated kernels after $Q^\dagger$ has acted.

\subsection{One-Loop Computation}\label{sec:oneloopPCM}

We shall first consider the bosonic contribution to the one-loop effective action, arising from the first term in the parentheses in \eqref{diffbosghos}.
This involves the $\mathsf B\widetilde A\mathsf B$ interaction.
The following diagram displays one possible component assignment for the one-loop Feynman diagram.
\definecolor{figcyan}{RGB}{0,255,255}
\begin{figure}[htbp]
    \centering
    \begin{tikzpicture}[scale=1.0]


        \draw[decorate, decoration={snake, amplitude=2.5pt, segment length=6.9pt},
            line width=1.3pt]
            (-2.25,0) -- (-0.9,0);

        \draw[decorate, decoration={snake, amplitude=2.5pt, segment length=6.9pt},
            line width=1.3pt]
            (-0.9,0) .. controls (-0.35,0.85) and (0.35,0.85) .. (0.9,0);

        \draw[decorate, decoration={snake, amplitude=2.5pt, segment length=6.9pt},
            line width=1.3pt]
            (-0.9,0) .. controls (-0.35,-0.85) and (0.35,-0.85) .. (0.9,0);

        \draw[decorate, decoration={snake, amplitude=2.5pt, segment length=6.9pt},
            line width=1.3pt]
            (0.9,0) -- (2.25,0);

        \node at (-2.48,0) {$\widetilde A_w$};
        \node at (2.48,0) {$\widetilde A_{\bar w}$};

        \node[above] at (-0.7,0.5) {$\sfB_{\bar z}$};
        \node[below] at (-0.7,-0.5) {$\sfB_{\bar w}$};
        \node[above] at (0.7,0.5) {$\sfB_w$};
        \node[below] at (0.7,-0.5) {$\sfB_{\bar z}$};

    \end{tikzpicture}
    \caption{{One component assignment in the bosonic one-loop diagram.}}
    \label{figure2}
\end{figure}
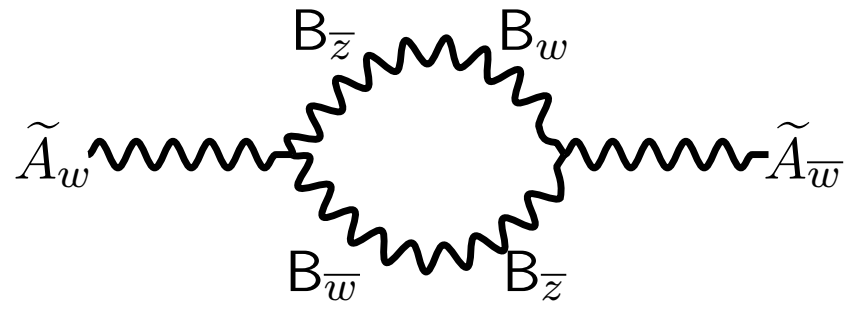
 In fact, there are further assignments allowed by the vertices, which we shall discuss further below.

When evaluating the bosonic one-loop contribution, we need to include a factor of $1/(2\pi)$ for each of the two trivalent vertices
(as is manifest in the action \eqref{thebvaction}). We can also count the power of $\hbar$, with $\hbar$ for each of the two internal propagators and $1/\hbar$ for each of the two vertices,
leading to $\hbar^0$. In \eqref{rescap} and below, we suppress these factors of $\hbar$, as well as the factors of $h$ from $\omega$, which likewise cancel between the propagators and the vertices. We should also include a factor of $i^2$ since we define the exponentiation of the action in the path integral as $e^{iS}$. Together with the factor $1/2!$ from expanding the exponential and the two possible Wick contractions, this gives the prefactor $-(1/2\pi)^2$ in \eqref{APPA}.
Incorporating these numerical factors, we obtain
\ie
    -\left(\frac{1}{2 \pi}\right)^2
    \int_{(\Sigma \times C)^2} & f_{a b c} f_{d e f}\, \varphi(z_1)\rd z_1\, \wedge  \varphi(z_2)\rd z_2 \\
    & \wedge \tA^a(x_1)  \wedge \sfP_{\sfB \sfB}^{b d}(x_1, x_2) \wedge \sfP_{\sfB \sfB}^{c e}(x_2, x_1) \wedge \tA^f(x_2) \;.
    \label{APPA}
\fe
A difference from the ultralocal case is that the interaction vertex here involves the one-form
\ie
   \frac{1}{2\pi} \omega f^{abc} \;,
\fe
where $\omega=\varphi(z)\rd z$.
We have also restricted ourselves to the amplitude involving the $w$ and $\wbar$ components of the classical gauge field. As we shall explain below, the $\zbar$ component of the classical gauge field does not contribute to the 1-loop amplitude.

We can substitute the expression for the propagators as in
\eqref{P_BB}, to have extra factors of $\pi^2$.
Using \eqref{Coxeter_def}, the numerical factors then combine
as
\ie
    -\left(\frac{1}{2 \pi}\right)^2
    (i\pi)^2  f_{a b c} f_{d e f}\delta^{bd}\delta^{ce}
    =\frac{1}{4} \mathsf{h}^{\vee} \delta_{af}
\fe
so that the kernel expression is
\ie
 \frac{\mathsf h^\vee}{4}\delta_{af}
 \int\rd\tau_1\rd\tau_2\int_{(\Sigma\times C)^2}
 &\varphi(z_1)\rd z_1\wedge\varphi(z_2)\rd z_2
 \wedge\widetilde A_w^a(x_1)\rd w_1\\
 &\wedge\mathscr P_1(\tau_1;x_1,x_2)
 \wedge\mathscr P_1(\tau_2;x_2,x_1)
 \wedge\widetilde A_{\bar w}^f(x_2)\rd\bar w_2 \;.
 \label{main:hk:bosonickernels}
\fe
This is the same component amplitude as \eqref{APPA}, for the insertion order with $\widetilde A_w$ at $x_1$ and $\widetilde A_{\bar w}$ at $x_2$; the reverse order gives an equal contribution and is included in $[\cdots]_{ef}$ in \eqref{main:hk:bubble} below.

Given a 4d CS defined with $\omega$ having two zeroes at $z=1$ and $z=-1$, then a typical solution of 4d CS would have the form
\ie
    \label{A_sec_4}
    \tA_{w}&= -\widehat{g} \frac{ \cJ_{w}(w,\bar{w})}{z-1}\widehat{g}^{-1} -\partial_w\widehat{g}\widehat{g}^{-1} \;,
    \\
    \tA_{\bar{w}}&= \widehat{g}\frac{\overline{\cJ}_{\bar{w}}(w,\bar{w})}{z+1}\widehat{g}^{-1}-\partial_{\bar{w}}\widehat{g}\widehat{g}^{-1}\;,
\fe
where the numerator of the first term of each solution involves currents built out of edge modes of the 4d CS theory. In addition, $\widetilde A_{\bar{z}}$ is pure gauge in $\widehat{g}$. These classical gauge fields obey Dirichlet boundary conditions at $z=0$ and $z=\infty$. Here, we have written the classical solutions for the gauge field components along $\Sigma$ as a gauge transformation of a Lax connection. The signs multiplying the currents agree with the PCM Lax connection in the conventions of
\cite{Costello:2019tri, Delduc:2019whp}. Moreover, following \cite{Delduc:2019whp}, we shall choose $\widehat{g}$ to be of archipelago type, meaning that it is the identity outside a disk centered at $z=0$. We denote the radius of the disk by $r_0$; the result below does not depend on $r_0$.

Outside the disk $|z|<r_0$ we have $\widehat g=1$, so that $\widetilde A$ coincides there with the Lax connection $\mathcal L$, with $\mathcal L_w=-\cJ_w/(z-1)$, $\mathcal L_{\bar w}=\overline{\cJ}_{\bar w}/(z+1)$ and $\mathcal L_{\bar z}=0$. Note that $\mathcal L$ itself does not vanish at $z=0$: the Dirichlet condition on $\widetilde A$ at $z=0$ holds only because of a cancellation between the Lax term and $\partial\widehat g\,\widehat g^{-1}$ inside the disk, where the pure-gauge components are large, e.g.\ $\widetilde A_{\bar z}=O(r_0^{-1})$. The contribution of the disk is therefore not suppressed merely by its area. Instead, we remove $\widehat g$ altogether as follows. Since $\widetilde A=\widehat g\cdot\mathcal L$, and since $\operatorname{Ad}_{\widehat g}$ acts pointwise and commutes with the rescalings \eqref{rescaledfields}, the rescaled differentials satisfy
\ie
 Q_{\widetilde A}=\operatorname{Ad}_{\widehat g}\,Q_{\mathcal L}\,\operatorname{Ad}_{\widehat g}^{-1} \;,
\fe
where $Q_{\mathcal L}$ contains the insertions \eqref{main:hk:currentinsertion} on the whole of $C$, including $z=0$. The conjugation preserves the space of fluctuations, since the boundary conditions at $z=0,\pm1,\infty$ are encoded in the rescalings \eqref{rescaledfields}, and $\widehat g$ is smooth and equal to the identity near $z=\pm1$ and $z=\infty$. For $\widehat g$ valued in the compact real form of $G$, $\operatorname{Ad}_{\widehat g}$ is unitary for the Hermitian pairing on $\mathfrak g$, and the Hermitian weights of \cref{app:Qadjoint} are scalar on $\mathfrak g$. Hence, in the background-covariant gauge of \cref{sec:rg-torsion}, $\Delta_p(\widetilde A)=\operatorname{Ad}_{\widehat g}\,\Delta_p(\mathcal L)\,\operatorname{Ad}_{\widehat g}^{-1}$: the two Laplacians are unitarily equivalent, their regularized determinants coincide for any $r_0$, and their local heat-kernel invariants agree pointwise, including inside the disk. Assuming, as usual, that the BV partition function does not depend on the choice of gauge-fixing Lagrangian (within the class preserving the boundary conditions), the same conclusion holds in the gauge used in this section; the computation of \cref{sec:rg-torsion}, which is exact in $\widehat g$, provides a check. We may therefore compute the one-loop amplitude with $\widetilde A$ replaced by $\mathcal L$ on all of $C$, i.e.\ with $\widehat g=1$ everywhere; in particular, the pure-gauge parts of $\widetilde A_w$ and $\widetilde A_{\bar w}$, and $\widetilde A_{\bar z}$, do not contribute. The resulting density (see \eqref{sum_Rc} and \eqref{sec:torsion:density}) is smooth at $z=0$, so no contribution is localized there, and the integral over $C$ equals that over $C^0:=\mathbb{CP}^1\setminus\{0,\infty\}$. Near $z=\infty$ we have $\widehat g=1$, and the density decays at least as $|z|^{-4}$. As explained in \cref{sec:rg-torsion}, $z=\infty$ is at infinite distance for the metric we use, and no additional contribution is localized there.

At each vertex, there is a cancellation between a simple pole in the classical gauge field and the corresponding zero of $\omega$ that appears in the vertex. The remaining fields at the vertex would be either $\mathsf{B}_w$ or $\mathsf{B}_{\bar{w}}$, and $\mathsf{B}_{\bar{z}}$. These fields are replaced by endpoints of propagators in the Feynman diagram, but since the propagator ought to satisfy the boundary conditions of the fields, it turns out that the remaining zeroes and poles of $\omega$ at a vertex are cancelled by the corresponding pole and zero of the propagator for $\mathsf{B}_w$ or $\mathsf{B}_{\bar{w}}$, and the zero in the propagator for $\mathsf{B}_{\bar{z}}$ (which obeys a Dirichlet boundary condition). The latter are contained in  the factors $f_{IJ}(z_1,z_2)$ and $f_{IJ}(z_2,z_1)$.
For example, for the mixed-derivative component assignment shown in \cref{figure2},
the cancellation is explicitly
\ie
    \label{cancellationn}
    &\underbrace{-\frac{1}{z_1-1}}_{\widetilde A_w\text{ classical field}}
    \underbrace{\left(
    \frac{(z_1-1)(z_1+1)}{z_1^2}
    \right)}_{\text{interaction vertex at }x_1}
    \underbrace{
    \frac{\bar z_1+1}{z_1+1}(z_1z_2)
    }_{\mathsf B_{\bar w}\mathsf B_{\bar z}\text{ propagator}}\\
    &\times
    \underbrace{
    \frac{1-\bar z_2}{z_2-1}(z_1z_2)
    }_{\mathsf B_w\mathsf B_{\bar z}\text{ propagator}}
    \underbrace{\left(
    \frac{(z_2-1)(z_2+1)}{z_2^2}
    \right)}_{\text{interaction vertex at }x_2}
    \underbrace{\frac{1}{z_2+1}}_{\widetilde A_{\bar w}\text{ classical field}}
    \\
    &=
    -(\bar z_1+1)(1-\bar z_2) \;.
\fe
Here the antiholomorphic factors $\bar z_1+1$ and $1-\bar z_2$ come from the coefficients of $Q^\dagger$ acting on the propagators; we have omitted the accompanying inverse-metric factors $1/\mathcal{N}$ and factors of $2$, which are included in the vertex operators below.

As shown in Appendix~\ref{app:Qlaplacian}, the one-form Laplacian
contains off-diagonal first-order terms coupling
$\mathsf B_w,\mathsf B_{\bar w},\mathsf B_{\bar z}$ through $w,\bar w$ derivatives with
$z,\bar z$-dependent coefficients. These terms prevent factorization
of its heat kernel into kernels on $\Sigma$ and $C^0$.
We therefore retain the full $3\times3$ matrix heat kernel on
$\Sigma\times C^0$, whose short-time expansion is derived in
Appendix~\ref{shorttimeheat}. The scalar Laplacian splits into
commuting operators on $\Sigma$ and $C^0$, so its heat kernel
does factorize.
Note that we use the normalized Laplacians $\widehat\Delta_p=2\Delta_p$,
so that
\ie
 Q^\dagger\Delta_p^{-1}
 =2Q^\dagger\widehat\Delta_p^{-1} \;.
 \label{main:hk:normalization}
\fe
The factor of $2$ is included in the vertex
operators derived in Appendix~\ref{shorttimeheat}.

The off-diagonal terms in \eqref{oneform_laplacian} mix
$\mathsf B_w,\mathsf B_{\bar w},\mathsf B_{\bar z}$, so its eigenforms can have several
nonzero components. The spectral representation of the heat kernel
\cite[Eq.~(2.20)]{Vassilevich:2003xt} therefore contains products
of different components of these eigenforms. Converting the operator heat kernel to its bi-form representative
using the BV pairing \eqref{sympair} and the weighted inner product
\eqref{weighted_inner_product}, and then applying $Q^\dagger$
as in \eqref{P_G}, these mixed terms can produce a nonzero
$\mathsf P_{\bar z\bar z}$. Thus, in addition to the typical component assignment shown in Figure~\ref{figure2}, the bosonic loop includes
 the pairing
$\mathsf P_{\bar w w}\mathsf P_{\bar z\bar z}$.

There is also a second one-loop contribution arising from the ghost
sector. This comes from the second term in \eqref{thebvaction}, which
contains the interaction
\ie
    \omega\wedge
    \operatorname{Tr}
    \left(
    \mathsf{B}^{\vee}\wedge[\widetilde A,\mathsf{c}]
    \right) \;.
\fe
It gives a loop with two $\mathsf{c}\mathsf{B}^{\vee}$ propagators,
corresponding to the second term in \eqref{diffbosghos}. The prefactor $2(1/2\pi)^2$ in \eqref{APPA2} below arises from the ghost vertex coefficient $1/\pi$ of \eqref{thebvaction} for each vertex, the factor $i^2$, the factor $1/2!$ from expanding the exponential, the unique contraction, and the closed-loop minus sign discussed below. The relative
minus sign between the bosonic and ghost loops follows from the
Grassmann odd nature of the ghost and antifield: contracting
$(\mathsf{B}^{\vee}\wedge\widetilde A\mathsf c)(x_1)
(\mathsf{B}^{\vee}\wedge\widetilde A\mathsf c)(x_2)$ into a closed
loop produces an additional minus sign. The ghost-antifield amplitude is
\ie
    \mathcal M_{\rm gh}
    ={}&2\left(\frac{1}{2\pi}\right)^2
    \int_{(\Sigma\times C)^2}
    f_{abc}f_{def}\,
    \varphi(z_1)\rd z_1\wedge\varphi(z_2)\rd z_2
    \\
    &\wedge\tA^a(x_1)
    \wedge\sfP_{\sfc\sfB^{\vee}}^{bd}(x_1,x_2)
    \\
    &\wedge\sfP_{\sfc\sfB^{\vee}}^{ce}(x_2,x_1)
    \wedge\tA^f(x_2) \;.
    \label{APPA2}
\fe

The same cancellation of meromorphic factors as in
\eqref{cancellationn} occurs for the ghost loop: the relevant antifield
components have the same zeroes and poles as the corresponding bosonic
components, while the ghost has the same zero at $z=0$ as
$\mathsf B_{\bar z}$.

We now extract the local term quadratic in the currents. We write $J_w=\cJ_w$ and
$J_{\bar w}=\overline{\cJ}_{\bar w}$, and pick solutions of the PCM such that they are constant on
$\Sigma$ and valued in a Cartan subalgebra.\footnote{The PCM equations
are the conservation equation
$\partial_w\mathcal J_{\bar w}+\partial_{\bar w}\mathcal J_w=0$
and the Maurer--Cartan equation
$\partial_w\mathcal J_{\bar w}-\partial_{\bar w}\mathcal J_w
 +[\mathcal J_w,\mathcal J_{\bar w}]=0$.
Constant commuting currents satisfy both equations. As explained in \cref{shorttimeheat}, such backgrounds suffice to determine the coefficient of the logarithmically divergent term $\operatorname{Tr}(J_wJ_{\bar w})$.}
The passage from the physical propagators to the rescaled complex is
particularly simple for the background, which by the argument above we may
take to be the Lax connection $\mathcal L$ on all of $C$. Its deformation
of $Q$ is
\ie
 \mathcal A
 ={}&(z-1)\operatorname{ad}_{\widetilde A_w}\,\rd w\wedge
    +(z+1)\operatorname{ad}_{\widetilde A_{\bar w}}\,
                                                  \rd\bar w\wedge\\
 ={}&e\,\rd w\wedge+f\,\rd\bar w\wedge \;,\qquad
 e=-\operatorname{ad}_{J_w} \;,\quad
 f=\operatorname{ad}_{J_{\bar w}} \;.
 \label{main:hk:currentinsertion}
\fe
Thus the Lax denominators have cancelled in the background insertion.

We encode the wedge products at each background vertex and the
coefficients of $2Q^\dagger$ in $\mathcal V_p^\mu$, so that each
vertex together with the adjacent differentiated kernel is written as
\ie
 \mathcal V_p^\mu(x)\,
 \partial_{x^\mu}K_p(t;x,y) \;,\qquad p=0,1 \;.
 \label{main:hk:vertexoperator}
\fe
The coefficients $\mathcal V_0^\mu$ are scalars, while
$\mathcal V_1^\mu$ are matrices acting on the one-form components.
Their explicit expressions and derivation are given in
Appendix~\ref{shorttimeheat}.

We use real coordinate measure
$\rd^4x=\rd\xi^1\rd\xi^2\rd^2z$, with
$\rd^2z=(i/2)\rd z\wedge\rd\bar z$. With $\rho=1$, the principal
metric has volume measure $\rd^4x$.
We write $K_p(t;x,y)$ for the heat kernel of
$\widehat\Delta_p$, normalized with respect to the coordinate
measure $\rd^4y$.
After removing the colour and current factor
$\mathsf h^\vee\operatorname{Tr}(J_wJ_{\bar w})$, the two separation
integrals are precisely
\ie
 \mathcal I_p(t,s;z)=\int_{\Sigma \times C^0}\rd^4x_2
 \Bigl[\operatorname{tr}_{\rm form}\Bigl(
 &\mathcal V_p^\mu(x_1)\partial_{1\mu}K_p(t;x_1,x_2)\\
 &\mathcal V_p^\nu(x_2)\partial_{2\nu}K_p(s;x_2,x_1)
 \Bigr)\Bigr]_{ef} \;,\qquad x_1=(w,z) \;.
 \label{main:hk:bubble}
\fe
Here $[\cdots]_{ef}$ extracts the coefficient bilinear in $e,f$,
including both insertion orders. Both derivatives act on the first
argument of their respective kernels. The trace includes all allowed
component contractions and the position-dependent vertex factors.

Appendix~\ref{shorttimeheat} evaluates \eqref{main:hk:bubble} by
expanding the two differentiated kernels and the second vertex about
$x_1$, multiplying the expansions, and integrating the separation over a copy of $\Sigma \times C^0$.
Let us make the change of variables $T=t+s$ and $u=t/T$. Applying the short-time asymptotics derived in
Appendix~\ref{shorttimeheat} to the full two-kernel integrals, and integrating over one copy of $\Sigma \times C^0$, gives
\ie
 \mathcal I_1(t,s;z)
 =\frac1{(4\pi T)^2}\Bigg\{&
 \frac{4(1-\bar z^2)}{T\mathcal{N}^2}\\
 &+\frac8{\mathcal{N}^4}\Big[(1+u)(1-\bar z^2)
       -2(1-u)^2(2\bar z^2+|z|^2-1)\Big]+O(T)\Bigg\} \;,
 \label{firstexpansion}
\fe
and
\ie
 \mathcal I_0(t,s;z)
 =\frac1{(4\pi T)^2}\Bigg\{&
 \frac{2(1-\bar z^2)}{T\mathcal{N}^2}\\
 &+\frac4{\mathcal{N}^4}\Big[\bar z^2-1
       +2u(1-u)(2\bar z^2+|z|^2-1)\Big]+O(T)\Bigg\} \;.
 \label{secondexpansion}
\fe

In this full matrix-trace convention the bosonic and ghost statistical
weights are $1/4$ and $-1/2$, respectively; the latter is the quadratic
expansion coefficient for the ghost-antifield pair, including the
closed-loop minus sign. These weights arise from the  prefactors
already included in \eqref{APPA} and \eqref{APPA2} and the propagators \eqref{P_BB} and \eqref{main:hk:ghostpropagator}.
The complete quadratic contribution is obtained by integrating the
combination below over $t,s,x_1$ and restoring the single factor
$\mathsf h^\vee\operatorname{Tr}(J_wJ_{\bar w})$.
The leading terms proportional to $1/T$ cancel between the boson and ghost loops, leaving
\ie
 \frac14\mathcal I_1-\frac12\mathcal I_0
 =\frac1{(4\pi T)^2}\left\{
 \frac2{\mathcal{N}^4}\Big[(4-u)-3(2-u)\bar z^2
                         -2(1-u)|z|^2\Big]+O(T)\right\} \;.
 \label{sum_Rc}
\fe
Since $\rd t\,\rd s=T\,\rd T\,\rd u$, the displayed terms give the
logarithmic divergence.

The remaining contribution is evaluated using
\ie
 \int_{\mathbb C}\frac{\rd^2z}{\mathcal{N}^4}=\frac\pi3 \;,\qquad
 \int_{\mathbb C}\frac{|z|^2\rd^2z}{\mathcal{N}^4}=\frac\pi6 \;,\qquad
 \int_{\mathbb C}\frac{\bar z^2\rd^2z}{\mathcal{N}^4}=0 \;.
 \label{main:hk:weightedFS}
\fe
The last equality follows by angular integration. Thus, for every $u$,
\ie
 \int_{C^0}\rd^2z\,
 \frac2{\mathcal{N}^4}\Big[(4-u)-3(2-u)\bar z^2-2(1-u)|z|^2\Big]
 =2\pi \;.
\fe
The
spatially integrated expression in \eqref{sum_Rc} is therefore
$1/(8\pi T^2)$ at logarithmic order.
For cutoffs $d^2\leq t,s\leq D^2$ on the proper times $t,s$ of $\widehat\Delta=2\Delta$ (which are related to the proper times $\tau$ of $\Delta$ in \eqref{P_BB} by $t=\tau/2$; this rescaling of the cutoffs does not affect the logarithmic coefficient), its proper-time integral is
\ie
 \frac1{8\pi}\int_{d^2}^{D^2}\rd t\int_{d^2}^{D^2}\rd s\,
 \frac1{(t+s)^2}
 &=\frac1{8\pi}\log\frac{(D^2+d^2)^2}{4D^2d^2}\\
 &=\frac1{4\pi}\log\frac\Lambda\mu+O(1) \;,
 \qquad \Lambda=d^{-1} \;,\quad\mu=D^{-1} \;.
 \label{main:hk:timeintegral}
\fe

The same result can be checked by integrating the bosonic and ghost
logarithmic terms separately. After the $u$ and weighted $C^0$ integrals,
and including the statistical weights $1/4$ and $-1/2$, their contributions
are respectively $11/18$ and $7/18$ in the common normalization of
\eqref{main:hk:timeintegral}:
\ie
 \frac1{4\pi}
 \left(\underbrace{\frac{11}{18}}_{\mathrm{bosonic}}
       +\underbrace{\frac{7}{18}}_{\mathrm{ghost}}\right)
 \log\frac\Lambda\mu
 =\frac1{4\pi}\log\frac\Lambda\mu \;.
 \label{main:hk:loopFSsplit}
\fe
The split between bosonic and ghost contributions depends on the choice of gauge, and only the sum is meaningful. In \cref{sec:rg-torsion} we shall describe a background-covariant gauge, and in this case the individual contributions are different (for instance, splitting the torsion into the factors $(\det'\Delta_1/\det'\Delta_0)^{-1/4}$ and $(\det'\Delta_0)^{1/2}$, one finds the contributions $17/18$ and $1/18$, respectively). We have not found a direct relation between the two gauges at the level of the integrands, and the agreement of the totals reflects the gauge independence discussed in \cref{sec:rg-torsion}.

We thus find
\ie
 \left.\log|Z_{\mathrm{1-loop}}|\right|_{\log,\mathrm{bulk}}
 =\frac{\mathsf h^\vee}{4\pi}\log\frac\Lambda\mu
   \int_\Sigma\rd\xi^1\rd\xi^2\,
          \operatorname{Tr}(J_wJ_{\bar w}) \;.
 \label{main:hk:completebulk}
\fe
Here the subscript ``bulk'' indicates that this is the integral of a local density over $C^0$; as explained above and in \cref{sec:rg-torsion}, there are no further contributions localized at $z=0$ or $z=\infty$.
In the measure convention
$\rd^2w=\rd w\wedge\rd\bar w=-2i\,\rd\xi^1\rd\xi^2$,
the same real logarithmic contribution is
\ie
 \left.\log|Z_{\mathrm{1-loop}}|\right|_{\log,\mathrm{bulk}}
 =\frac{i\mathsf h^\vee}{8\pi}\log\frac\Lambda\mu
   \int_\Sigma\rd^2w\,\operatorname{Tr}(J_wJ_{\bar w}) \;.
 \label{nonuresult}
\fe

The classical PCM action obtained from 4d Chern--Simons theory with the present conventions is
\ie
 S_{\mathrm{PCM}}
 =-\frac{H}{4\pi}\int_\Sigma\rd^2w\,
                     \operatorname{Tr}(J_wJ_{\bar w}) \;,\qquad
 H=\frac{8\pi i h}{\hbar} \;.
 \label{main:hk:PCMnormalization}
\fe
Since the effective action is $S_{\rm eff}=S_{\rm PCM}-i\log Z_{\rm 1-loop}$, \eqref{nonuresult} implies that the resulting bulk running is
\ie
 H(\mu)=H_{\mathrm{bare}}
       -\frac{\mathsf h^\vee}{2}\log\frac\Lambda\mu \;,
 \qquad
 \beta_H^{\mathrm{bulk}}=\frac{\rd H}{\rd\log\mu}
                       =\frac{\mathsf h^\vee}{2} \;.
 \label{main:hk:beta}
\fe
We have thus derived the RG flow of the PCM directly from 4d Chern--Simons theory.

The calculation above used $\rho=1$, but this is merely part of a convenient choice of gauge. The independence of the result of this choice is discussed at the end of \cref{sec:rg-torsion}.

The localization from a 4d effective action to a 2d effective action that we observe here is conceptually similar to the mechanism at the classical level whereby integrable field theories are derived in the BV formalism from 4d Chern--Simons theory presented in \cite{Benini:2026osk}, where the 2d classical action is obtained by integrating over the Fubini--Study volume form on $C$.

Although we considered two simple zeroes and double poles at $z=0$ and $z=\infty$, we could consider more poles and zeroes, as well as poles and zeroes of higher order.
We expect that the derivation of the RG flow presented above for the principal chiral model should generalize, although this remains to be checked in detail. In particular, the rescalings \eqref{rescaledfields}, the endpoint factors $f_{IJ}$ and the cancellations \eqref{cancellationn} at the vertices depend on the number and orders of the zeroes and poles, and the gauge-fixing metric has to be chosen accordingly.

The computation of the one-loop RG flow has parallels with the computation of the framing anomaly that arises when $\Sigma$ is curved, which was computed explicitly using the BV formalism by Khan \cite{Khan:2022vrx}. In \cite{Khan:2022vrx}, the Riemann surface $C$ was taken to be $\mathbb{C}$, but $\Sigma$ was assumed to have nontrivial curvature. Also, the background gauge field was assumed to be of the form $\widetilde A=\widetilde A(z,\bar{z})\textrm{d}\bar{z}$. Moreover, the factorization and composition properties of the heat kernel are utilized in computing the framing anomaly, leading to dependence of the one-loop effective action on the Ricci scalar of $\Sigma$.

\subsection{Including the Wess--Zumino Term}\label{wzterm}

The derivation of the RG flow of the PCM presented above can be generalized to include a Wess--Zumino (WZ) term. The meromorphic 1-form in this case is 
\ie\label{homegapcmwzterm}
    \omega = h  \frac{(z-1)(z+1)}{\left(z-\frac{k}{h}\right)^2}\rd z \;,
\fe
and from the corresponding 4d Chern--Simons theory we can derive the classical PCM+WZ action
\ie
S_{\mathrm{PCM}+\mathrm{WZ}}[g]=-\frac{H}{4 \pi} \int_{\Sigma} d w \wedge d \bar{w} \operatorname{Tr}\left(J_w J_{\bar{w}}\right)-\frac{K}{4 \pi} I_{\mathrm{WZ}}[g],
\fe
where 
\ie 
H=\frac{8 \pi i h}{\hbar}, \quad K=\frac{8 \pi i k}{\hbar} ,
\fe
and 
\ie 
I_{\mathrm{WZ}}[g]=-\frac{1}{3} \int_{\Sigma\times [0,r_0]} \operatorname{Tr}\left[\left(\widehat{g}^{-1} d \widehat{g}\right)^{\wedge 3}\right].
\fe

In accordance with the Dirichlet boundary conditions on the gauge field at $z=\frac{k}{h}$, the rescalings of \eqref{rescaledfields} are generalized to
\ie \label{rescaledfieldsWZ}
    A_w& = A^{\rm R}_w\frac{z-\frac{k}{h}}{z-1}\\
    A_{\bar{w}}& = A^{\rm R}_{\bar{w}}\frac{z-\frac{k}{h}}{z+1}\\
    A_{\bar{z}}& = A^{\rm R}_{\bar{z}}\left(z-\frac{k}{h}\right) \;.
\fe
Writing $A_I=r_I A^{\rm R}_I$ with $r_w=\left(z-\frac{k}{h}\right)/(z-1)$, $r_{\bar w}=\left(z-\frac{k}{h}\right)/(z+1)$ and $r_{\bar z}=z-\frac{k}{h}$, we can fix the rescalings of the ghost and of the antifields as before,
\ie
    \mathsf c=\left(z-\frac{k}{h}\right)\,\mathsf c^{\rm R} \;,\qquad
    \mathsf B^\vee_{IJ}=r_Ir_J\,\left(z-\frac{k}{h}\right)^{-1}\,\mathsf B^{\vee\rm R}_{IJ} \;,\qquad
    \mathsf c^\vee_{w\bar w\bar z}=r_wr_{\bar w}r_{\bar z}\,\left(z-\frac{k}{h}\right)^{-2}\,\mathsf c^{\vee\rm R}_{w\bar w\bar z} \;.
    \label{rescaledghostsWZ}
\fe
With these rescalings, the cancellation at the vertices described in \eqref{cancellationn} generalizes straighforwardly.

In addition, the currents in the classical background field \eqref{A_sec_4} are multiplied in this case by expressions involving the ratio $k/h$, that is, 
\ie
    \label{A_sec_4WZ}
    \tA_{w}&= -\widehat{g} \left(1-\frac{k}{h}\right)\frac{ \cJ_{w}(w,\bar{w})}{z-1}\widehat{g}^{-1} -\partial_w\widehat{g}\widehat{g}^{-1} \;,
    \\
    \tA_{\bar{w}}&= \widehat{g} \left(1+\frac{k}{h}\right)\frac{\overline{\cJ}_{\bar{w}}(w,\bar{w})}{z+1}\widehat{g}^{-1}-\partial_{\bar{w}}\widehat{g}\widehat{g}^{-1}\;.
\fe

Let us set $p=k/h=K/H$ and assume $p\neq\pm1$. In the BV pairing, the
two factors of $z-p$ supplied by \cref{rescaledfieldsWZ,rescaledghostsWZ}
cancel the double pole of $\omega$. Since $\partial_{\bar z}(z-p)=0$,
the free rescaled BRST differential is the same as for the PCM.
We shall use the gauge \eqref{hermitianmetric} with $\rho=1$.

We take $\widehat g$ to be of archipelago type on a disc of radius $r_0$ centred
at $p$ that avoids the zeroes of $\omega$, with $\widehat g|_{z=p}=g$.
The conjugation argument of \cref{sec:oneloopPCM} then allows us to
use the Lax connection $\mathcal L$ in the one-loop calculation. Since the rescaled fields are regular at $p$,
the density in \eqref{sum_Rc} is smooth there and integrable at
infinity. Its integral over $C_p^0=\mathbb{CP}^1\setminus\{p,\infty\}$
therefore has the same value as in the PCM, using
\eqref{main:hk:weightedFS}.

Now, \eqref{A_sec_4WZ} replaces the insertions
$e$ and $f$ in \eqref{main:hk:currentinsertion} by $(1-p)e$ and
$(1+p)f$, respectively. As in the case with no Wess--Zumino term, we can once again restrict ourselves to commuting, constant currents for simplicity. The bulk logarithm \eqref{nonuresult} hence acquires a factor
of $1-p^2$,
resulting in 
\ie
 \left.\log|Z_{\mathrm{1-loop}}|\right|_{\log,\mathrm{bulk}}
 =\frac{i\mathsf h^\vee}{8\pi} \left(1-\frac{K^2}{H^2}\right)\log\frac\Lambda\mu
   \int_\Sigma\rd^2w\,\operatorname{Tr}(J_wJ_{\bar w}) \;.
\fe
We then obtain the known bulk one-loop running of the PCM+WZ model, 
\ie
 H(\mu)&=H_{\mathrm{bare}}
 -\frac{\mathsf h^\vee}{2}
  \left(1-\frac{K_{\mathrm{bare}}^2}{H_{\mathrm{bare}}^2}\right)
  \log\frac{\Lambda}{\mu},
 \qquad
 K(\mu)=K_{\mathrm{bare}} \;.
\fe
Expressing the beta functions in terms of the running couplings gives
\ie \label{wzbeta}
 \beta_H^{\mathrm{bulk}}
 &=\frac{\rd H}{\rd\log\mu}
 =\frac{\mathsf h^\vee}{2}
  \left(1-\frac{K(\mu)^2}{H(\mu)^2}\right),
 \qquad
 \beta_K=\frac{\rd K}{\rd\log\mu}=0 \;.
\fe
In particular, for in $K\rightarrow H$ limit, we retrieve the known 1-loop scale invariance of the Wess--Zumino--Witten (WZW) model. 
\subsection{Renormalization Group Flow from Analytic Torsion}
\label{sec:rg-torsion}

The logarithmic divergence can also be obtained directly from the Gaussian
BV integral around the classical background. In this description, the
one-loop partition function is expressed through the analytic torsion of
the background-coupled kinetic complex. Its ultraviolet divergence is
then determined by a weighted sum of local heat-kernel coefficients.
This is the relation between quadratic gauge-theory partition functions
and analytic torsion \cite{Ray:1971} described in \cite{Schwarz:1978cn}; see also \cite{Witten:1988hf} for three-dimensional Chern--Simons theory.

We first consider the PCM with \(K=0\) and use the  Hermitian metric \eqref{hermitianmetric}, with $\alpha=2\mathcal{N}$ and auxiliary metric $\rho=1$, as in the diagrammatic
calculation above. The computations below are local in the affine
$z$-coordinate, on $C^0=\mathbb{CP}^1\setminus\{0,\infty\}$, with the background $\widehat g$ removed by the conjugation argument of \cref{sec:oneloopPCM}. With this choice, the
background-coupled Laplacians below are smooth at $z=0$ and at the zeroes
$z=\pm1$, so that the integrals over $C^0$ below are in fact integrals over
$\mathbb C$.

As $z\to\infty$, the invariant $\operatorname{tr}E$ decays as $|z|^{-2}$
(so do the components of $E$ in an orthonormal frame), and the density of
$a_4^{(1)}-3a_4^{(0)}$ below decays as $|z|^{-4}$. The noncompact end at
$z=\infty$ gives no further contribution, for the following reason. For the
metric $\rd z\,\rd\bar z$ on $C$, $z=\infty$ is at infinite distance, so that
there is no boundary at which boundary heat-kernel coefficients could arise; the
boundary conditions at the pole of $\omega$ are encoded in the rescalings
\eqref{rescaledfields} and \eqref{rescaledghosts} and in the $L^2$ condition on the
rescaled fields. Moreover, the principal metric \eqref{main:hk:principalmetric} is
flat, and the lower-order coefficients of the Laplacians
(\cref{app:Qlaplacian,app:bg}) are rational functions of $z,\bar z$ with
denominators given by powers of $\mathcal{N}$, which are bounded together with all
their derivatives. For such operators of bounded geometry, the short-time
expansion of the heat kernel on the diagonal holds uniformly on $C$, so the
logarithmic coefficient of the background-dependent part of the regularized
trace is the integral of the local density, which converges by the decay just
mentioned. Finally, the edge mode $\widehat g$ is the identity near $z=\infty$, so that no background-dependent data is localized there.

We also use a background-covariant version of the gauge we used in previous sections. This differs from the gauge of \cref{sec:oneloopPCM}, where the gauge-fixing operator was kept fixed as the background varied; for instance, the Laplacians below contain terms linear in the background that are absent from the bubble diagrams of \cref{sec:oneloopPCM}. Both gauges belong to the following class. For Hermitian weights satisfying $\sch_I\sch_{I^c}=\mathrm{const}$, see \eqref{sec:torsion:weights}, the subspace $\operatorname{im}Q^\dagger_{A'}$ is Lagrangian with respect to the BV pairing for any background $A'$ with $Q_{A'}^2=0$ (the gauge of \cref{sec:oneloopPCM} corresponds to $A'=0$, and the one here to $A'=\widetilde A$). For the constant commuting currents used below, $Q_{\vartheta J}^2\propto \vartheta^2[e,f]=0$, so the family $A'=\vartheta J$, $0\le \vartheta\le1$, interpolates between the two within this class, and within the space of fluctuations obeying the boundary conditions at the zeroes and poles of $\omega$. Since the BV partition function is invariant under such continuous deformations of the gauge-fixing Lagrangian, the coefficient of the logarithm should agree, and we shall see that it does. The same argument applies to the choice of $\alpha$ in \eqref{hermitianmetric}: the weights of \cref{app:Qadjoint} satisfy $\sch_I\sch_{I^c}=8\rho$ for any positive $\alpha$, so that the coefficient does not depend on $\alpha$; the choice $\alpha=2\mathcal{N}$ only serves to make the principal metric flat. 
We write
\begin{equation}
 a=z-1 \;,\qquad b=z+1 \;,\qquad s=z+\bar z \;,
 \qquad |a|^2=\mathcal{N}-s \;,\qquad |b|^2=\mathcal{N}+s \;.
 \label{sec:torsion:notation}
\end{equation}
Coupling the rescaled differential to the PCM background gives
\begin{equation}
 Q_J=dw\wedge D_w+d\bar w\wedge D_{\bar w}
          +d\bar z\wedge\partial_{\bar z} \;,
 \qquad
 \begin{aligned}
 D_w&=a\partial_w+e \;,\\
 D_{\bar w}&=b\partial_{\bar w}+f \;,
 \end{aligned}
 \label{sec:torsion:q}
\end{equation}
where $e=-\operatorname{ad}_{J_w}$ and $f=\operatorname{ad}_{J_{\bar w}}$ are as in \eqref{main:hk:currentinsertion}.
Notice that the simple poles at the locations of zeroes of $\omega$ have been cancelled by identical factors that remain in $\omega$ after rescaling the fields in the interaction terms in the BV action, as we did previously for the kinetic terms.
As in \cref{sec:oneloopPCM}, we pick solutions of the PCM such that $J_w,J_{\bar w}$ are constant and  lie in a common Cartan
subalgebra. Hence $[e,f]=0$ and $Q_J^2=0$.

In the basis of \cref{app:Lie}, the positive Hermitian pairing is
$h_{\mathfrak g}(X,Y)=\sum_a\overline{X^a}Y^a$.
With the compact reality condition
$J_w^\dagger=-J_{\bar w}$ on the Lie algebra matrices, the induced
endomorphisms obey $e^\dagger=f$. On a root space $\mathfrak g_\gamma$,
these endomorphisms act as the numbers
\begin{equation}
 e_\gamma=-\gamma(J_w) \;,\qquad
 f_\gamma=\gamma(J_{\bar w})=\overline{e_\gamma} \;.
 \label{sec:torsion:roots}
\end{equation}
We shall work on one root space and omit $\gamma$ from $e,f$ until the end.
In particular, complex conjugation of $eu$ then gives
$\overline{eu}=\bar e\,\bar u=f\bar u$. 
Restoring the root sum gives, in the Lie algebra normalization of
\cref{app:Lie},
\begin{equation}
 \sum_\gamma e_\gamma f_\gamma
 =-\operatorname{tr}_{\mathrm{adj}}
          (\operatorname{ad}_{J_w}\operatorname{ad}_{J_{\bar w}})
 =\mathsf h^\vee\operatorname{Tr}(J_wJ_{\bar w}) \;.
 \label{sec:torsion:rootsum}
\end{equation}

The background-dependent Laplacians are
$\Delta_p(J)=Q_JQ^\dagger_{J,p}+Q^\dagger_{J,p+1}Q_J$.
Their explicit expressions and Laplace-type data in degrees zero and one
are given in \eqref{app:bg:lap0}--\eqref{app:bg:mixing} and
\eqref{app:bg:scalarconnection}--\eqref{app:bg:freecurvatures}.

After rescaling the fields as in \eqref{rescaledfields}, the BV pairing has constant density:
\begin{equation}
 \langle X,Y\rangle_{\rm BV}
 =h\int_{\Sigma\times C^0}dz\wedge
                    \operatorname{Tr}(X^{\rm R}\wedge Y^{\rm R}) \;.
 \label{sec:torsion:pairing}
\end{equation}
The Hermitian weights in \eqref{app:bg:weights} satisfy
\begin{equation}
 \sch_I\sch_{I^c}=8 \;.
 \label{sec:torsion:weights}
\end{equation}
After a constant normalization of \eqref{sec:torsion:pairing}, this gives
an antilinear isometry
$\mathcal S_p:\Omega^p\longrightarrow\Omega^{3-p}$,
defined componentwise by
\begin{equation}
 (\mathcal S_pX)_{I^c}
 =\eta(I,I^c)\frac{\sch_I}{\sqrt8}\,\sigma(X_I) \;.
 \label{sec:torsion:dualitymap}
\end{equation}
Here $\eta(I,I^c)$ is the sign of the ordering
$dx^I\wedge dx^{I^c}$ relative to
$dw\wedge d\bar w\wedge d\bar z$, and $\sigma$ is complex conjugation
of the coefficients in the compact real Lie algebra basis.
Since $\sigma$ preserves the Hermitian norm on the Lie algebra and
$|\eta(I,I^c)|=1$, the contribution of the complementary component
to the squared norm is
\begin{equation}
 \sch_{I^c}\bigl|(\mathcal S_pX)_{I^c}\bigr|^2
 =\frac{\sch_I\sch_{I^c}}{8}\,\sch_I|X_I|^2
 =\sch_I|X_I|^2 \;,
\end{equation}
where the last equality uses \eqref{sec:torsion:weights}.
Summing over $I$ and integrating against the coordinate measure gives
$\|\mathcal S_pX\|_{3-p}^2=\|X\|_p^2$.
Cyclicity of the BV pairing implies that $\mathcal S$ exchanges
$Q_J$ and $Q_J^\dagger$ up to signs depending on the form
degree.\footnote{Write $Q_p=Q_J|_{\Omega^p}$, with $Q_p^\dagger$
lowering degree from $p$ to $p-1$. The normalized BV pairing is
represented by $(\mathcal S_pX,Y)_{3-p}$. Thus the cyclicity identity
$B_{\rm BV}(QX,Y)=(-1)^{p+1}B_{\rm BV}(X,QY)$ for $X\in\Omega^p$
gives
\[
 \mathcal S_{p+1}Q_p=(-1)^{p+1}Q^\dagger_{3-p}\mathcal S_p \;.
\]
Since $\mathcal S_{3-p}\mathcal S_p=1$ by
\eqref{sec:torsion:weights} and the wedge signs, the same identity
gives $\mathcal S_{p-1}Q_p^\dagger=(-1)^pQ_{3-p}\mathcal S_p$.}
The signs cancel in each product in the Laplacian, giving
\begin{equation}
 \mathcal S_p\Delta_p(J)
             =\Delta_{3-p}(J)\mathcal S_p \;.
 \label{sec:torsion:duality}
\end{equation}
In using this identity for determinants, we take the operator domains
in complementary degrees to be mapped into one another by $\mathcal S$.
This is a condition on the ghost and antifield domains as well as on the
gauge fields. Since $\sigma$ maps $\mathfrak g_\gamma$ to $\mathfrak g_{-\gamma}$,
\eqref{sec:torsion:duality} relates $\Delta_p$ on $\mathfrak g_\gamma$ to
$\Delta_{3-p}$ on $\mathfrak g_{-\gamma}$. This suffices for our purposes, since
the quadratic-background terms depend on the root only through $e_\gamma f_\gamma$,
which is invariant under $\gamma\to-\gamma$. 

We now evaluate the Gaussian integral. Write
$Q_{J,p}=Q_J|_{\Omega^p}:\Omega^p\to\Omega^{p+1}$.
As in Appendix~\ref{app:bg}, the adjoint is indexed by its domain degree:
$Q^\dagger_{J,p}:\Omega^p\to\Omega^{p-1}$ is the adjoint of $Q_{J,p-1}$.
Omitting harmonic modes, the orthogonal decomposition is
\begin{equation}
 \Omega^p=\operatorname{im}Q_{J,p-1}
                \oplus\operatorname{im}Q^\dagger_{J,p+1} \;.
 \label{sec:torsion:hodge}
\end{equation}
Let
\begin{equation}
 \begin{gathered}
 d_p=\det{}'\left(
 Q^\dagger_{J,p+1}Q_{J,p}
 \big|_{\operatorname{im}Q^\dagger_{J,p+1}}\right) \;,
 \qquad p=0,1,2 \;,\\
 d_{-1}=d_3=1 \;.
 \end{gathered}
 \label{sec:torsion:coexact}
\end{equation}
On coexact $p$-forms, $Q^\dagger_{J,p}$ vanishes, so $\Delta_p$
reduces to $Q^\dagger_{J,p+1}Q_{J,p}$. This block contributes $d_p$.
On exact $p$-forms, $Q_{J,p}$ vanishes, so $\Delta_p$ reduces to
$Q_{J,p-1}Q^\dagger_{J,p}$.

To identify the determinant of the exact block, take a coexact
$(p-1)$-form $u$ satisfying
$Q^\dagger_{J,p}Q_{J,p-1}u=\lambda u$ with $\lambda>0$.
Then $v=Q_{J,p-1}u/\sqrt{\lambda}$ is an exact $p$-form with
\[
 \Delta_p v
 =Q_{J,p-1}Q^\dagger_{J,p}v
 =\lambda v \;.
\]
The inverse map is $v\mapsto Q^\dagger_{J,p}v/\sqrt{\lambda}$.
Thus the exact block in degree $p$ has the same positive spectrum,
with multiplicities, as the coexact block in degree $p-1$.
It contributes $d_{p-1}$, giving
\begin{equation}
 \det{}'\Delta_p=d_pd_{p-1} \;.
 \label{sec:torsion:lapdet}
\end{equation}
Since $\mathcal S_{p+1}$ is an isometry, the squared singular values
of the BV kinetic map $\mathcal S_{p+1}Q_{J,p}$ are the eigenvalues of
$Q^\dagger_{J,p+1}Q_{J,p}$ on coexact $p$-forms, so its determinant has
modulus $d_p^{1/2}$.
The gauge-fixing condition restricts the bosonic fluctuation to coexact one-forms. Its first-order Gaussian
integral has modulus $d_1^{-1/4}$. The odd fields $c$ and $B^\vee$
give a Berezin determinant whose modulus is $d_0^{1/2}$.
Thus, up to the normalization independent of the background,
\begin{equation}
 |Z_{\rm 1-loop}|=d_0^{1/2}d_1^{-1/4}
 =\frac{(\det{}'\Delta_0)^{3/4}}
        {(\det{}'\Delta_1)^{1/4}} \;.
 \label{sec:torsion:z}
\end{equation}
The fourth root in the bosonic factor follows because the squared
singular values of the first-order kinetic operator are the eigenvalues
of $Q_J^\dagger Q_J$.

The Gaussian modulus in \eqref{sec:torsion:z} can also be expressed
in terms of the analytic torsion of the background-coupled BV
complex, as in the one-loop expansion of three-dimensional
Chern--Simons theory \cite{Witten:1988hf}. Let $X$ be an eigenmode of
$\Delta_p(J)$ with eigenvalue $\lambda>0$. Since $\lambda$ is real,
\[
 \Delta_{3-p}(J)\mathcal S_pX
 =\mathcal S_p\Delta_p(J)X
 =\lambda\mathcal S_pX \;.
\]
The matching of the operator domains and the identity
$\mathcal S_{3-p}\mathcal S_p=1$ make this an antilinear bijection
between the corresponding eigenspaces. On the full
Lie-algebra-valued complex this gives
\[
 \det{}'\Delta_3=\det{}'\Delta_0 \;,\qquad
 \det{}'\Delta_2=\det{}'\Delta_1 \;.
\]
Substituting these relations into \eqref{sec:torsion:z} gives
\begin{equation}
 \log|Z_{\rm 1-loop}|
 =-\frac14\sum_{p=0}^3(-1)^p p\,\log\det{}'\Delta_p
 =\frac12\log T(Q_J) \;,
 \label{sec:torsion:torsion}
\end{equation}
where our convention for the analytic torsion is
\begin{equation}
 T(Q_J)=\prod_{p=0}^3
       (\det{}'\Delta_p)^{(-1)^{p+1}p/2} \;.
 \label{sec:torsion:tdef}
\end{equation}
Thus the relevant one-loop determinant is the square root of the
analytic torsion of the length-three kinetic complex.

To extract its logarithmic divergence, we use proper-time
regularization,
\begin{equation}
 \log\det{}'\Delta_p
 =-\int_{\Lambda^{-2}}^{\mu^{-2}}\frac{dt}{t}\,
                          \operatorname{Tr}'_{L^2}e^{-t\Delta_p} \;,
 \label{sec:torsion:proper}
\end{equation}
where the lower limit is the ultraviolet cutoff. (In \cref{sec:oneloopPCM} we instead imposed $d^2\le t\le D^2$ on the proper time of $\widehat\Delta=2\Delta$, with $\Lambda=d^{-1}$ and $\mu=D^{-1}$. The proper time of $\Delta_p$ is twice that of $\widehat\Delta_p$, so the two conventions differ by rescaling both $\Lambda$ and $\mu$ by $1/\sqrt2$, which leaves $\Lambda/\mu$, and hence the logarithmic coefficient, unchanged.) In four real
dimensions the local heat expansion has the form
\begin{equation}
 \operatorname{Tr}_{L^2}e^{-t\Delta_p}
 \sim t^{-2}a_0^{(p)}+t^{-1}a_2^{(p)}+a_4^{(p)}+\cdots \;.
 \label{sec:torsion:heat}
\end{equation}
The first two terms give power divergences. The logarithm comes from
the $t^0$ term \cite{Vassilevich:2003xt}, so that
\begin{equation}
 \left.\log\det{}'\Delta_p\right|_{\log}
 =-2a_4^{(p)}\log\frac{\Lambda}{\mu} \;.
 \label{sec:torsion:detlog}
\end{equation}

The relevant smooth bulk heat coefficient in \eqref{sec:torsion:heat} on a single root space is \cite{Vassilevich:2003xt}
\begin{equation}
 a_4^{(p)}=\frac1{(4\pi)^2}\int d\operatorname{vol}_G
 \left[
 \frac12\operatorname{tr}E_p^2
 +\frac1{12}\operatorname{tr}
             (\Omega^{(p)}_{\mu\nu}\Omega^{(p)\mu\nu})
 +\frac16G^{\mu\nu}\partial_\mu\partial_\nu
                              \operatorname{tr}E_p
 \right] \;.
 \label{sec:torsion:a4}
\end{equation}
Here, $G$ is the principal symbol metric defined in \eqref{app:bg:principalmetric}.
The Riemannian curvature terms vanish because $G$ is flat.
The last term is a total derivative; its integral vanishes since
$\operatorname{tr}E=O(|z|^{-2})$, so that the flux through a large circle in
$C$ is $O(R^{-2})$. Moreover, its $ef$ coefficient cancels pointwise in the
torsion combination, see \eqref{sec:torsion:derivativecancel} below.

Combining \eqref{sec:torsion:z} and \eqref{sec:torsion:detlog}, we obtain
\begin{equation}
 \left.\log|Z_{\rm 1-loop}|\right|_{\log,\mathrm{bulk}}
 =\frac12\left(a_4^{(1)}-3a_4^{(0)}\right)
                                      \log\frac{\Lambda}{\mu} \;.
 \label{sec:torsion:weightedheat}
\end{equation}
The common normalization $\widehat\Delta_p=2\Delta_p$ leaves this
four-dimensional heat coefficient unchanged, as explained at the end
of Appendix~\ref{app:bg}.

We evaluate \eqref{sec:torsion:weightedheat} on a single root space,
where $e,f=\bar e$ are numbers. The coefficient of $ef$ determines the
current bilinear by \eqref{sec:torsion:rootsum}. With
$\mathcal{N}=1+|z|^2$ and $s=z+\bar z$, Appendix~\ref{app:bg} gives
\begin{equation}
 \begin{aligned}
 T_0&=[E_0^2]_{ef}=\frac{s^2}{\mathcal{N}^4},&
 T_1&=[\operatorname{tr}E_1^2]_{ef}
                         =\frac{8\mathcal{N}^2-9s^2}{\mathcal{N}^4} \;,\\
 O_0&=[\Omega^{(0)}_{\mu\nu}\Omega^{(0)\mu\nu}]_{ef}
                         =-\frac{2(\mathcal{N}^2+s^2)}{\mathcal{N}^4},&
 O_1&=3O_0 \;.
 \end{aligned}
 \label{sec:torsion:data}
\end{equation}
These are the invariants of the full matrix-valued operators, including
their off-diagonal entries.

Two cancellations simplify the torsion combination. First, the
curvature-squared term cancels since $O_1-3O_0=0$. Second,
\begin{equation}
 [\operatorname{tr}E_1-3E_0]_{ef}=0 \;.
 \label{sec:torsion:derivativecancel}
\end{equation}
Therefore, even if we did not set it to zero, the total derivative in
\eqref{sec:torsion:a4} would cancel pointwise in the quadratic-background
part of $a_4^{(1)}-3a_4^{(0)}$. Since the coefficient $a_2$ is given by
$\operatorname{tr}E$, \eqref{sec:torsion:derivativecancel} also implies that
the power-law divergence quadratic in the background cancels in the torsion
combination, similarly to the cancellation of the $1/T$ terms between
the bosonic and ghost loops in \cref{sec:oneloopPCM}. Each $a_2^{(p)}$ separately
has a quadratic-background part proportional to $s^2/\mathcal{N}^2$, whose integral over
$C$ diverges logarithmically at $z=\infty$; it is only the torsion combination
that is finite.

We use real coordinate measures
$d^2\xi=d\xi^1d\xi^2$ on $\Sigma$ and $d^2x=dx\,dy$ for $z=x+iy$.
Thus the principal metric of \eqref{app:bg:principalmetric} has volume
element $d\operatorname{vol}_G=4d^2\xi\,d^2x$.
Substitution in \eqref{sec:torsion:a4} gives
\begin{equation}
 \left.\log|Z_{\rm 1-loop}|\right|_{ef,\log,\mathrm{bulk}}
 =ef\log\frac{\Lambda}{\mu}
    \int_\Sigma d^2\xi\int_{C^0}d^2x\,\mathcal D(z,\bar z) \;,
 \label{sec:torsion:densityintegral}
\end{equation}
where
\begin{equation}
 \begin{aligned}
 \mathcal D
 &=\frac{4}{2(4\pi)^2}
       \left[\frac12(T_1-3T_0)+\frac1{12}(O_1-3O_0)\right]\\
 &=\frac{2\mathcal{N}^2-3s^2}{4\pi^2\mathcal{N}^4} \;.
 \end{aligned}
 \label{sec:torsion:density}
\end{equation}
The factor $1/2$ in the first line comes from the determinant weights
in \eqref{sec:torsion:weightedheat}, while the factor $4$ comes from
the principal-metric volume element.

The density is smooth at the zeroes $z=\pm1$ and at $z=0$, and decays
as $|z|^{-4}$ at infinity. Its bulk integral is therefore well defined.
Using $z=re^{i\theta}$, the required integrals are
\begin{equation}
 \int_{\mathbb C}\frac{d^2x}{\mathcal{N}^2}=\pi \;,\qquad
 \int_{\mathbb C}\frac{s^2}{\mathcal{N}^4}\,d^2x
 =4\pi\int_0^\infty\frac{r^3\,dr}{(1+r^2)^4}
 =\frac{\pi}{3} \;.
 \label{sec:torsion:integrals}
\end{equation}
 Consequently,
\begin{equation}
 \int_{C^0}d^2x\,\mathcal D(z,\bar z)=\frac1{4\pi} \;.
 \label{sec:torsion:integrateddensity}
\end{equation}
With the normalized Fubini--Study form
\begin{equation}
 \omega_{\rm FS}
 =\frac{i\,dz\wedge d\bar z}{2\pi(1+|z|^2)^2}
 =\frac{d^2x}{\pi \mathcal{N}^2} \;,\qquad
 \int_{\mathbb{CP}^1}\omega_{\rm FS}=1 \;,
 \label{sec:torsion:fs}
\end{equation}
the integrated result equals
$\int_{C^0}\mathcal D\,d^2x=(4\pi)^{-1}\int_{\mathbb{CP}^1}\omega_{\rm FS}$,
although the two densities differ pointwise.

Restoring the root sum using \eqref{sec:torsion:rootsum}, we find
\begin{equation}
 \left.\log|Z_{\rm 1-loop}|\right|_{J\bar J,\log,\mathrm{bulk}}
 =\frac{\mathsf{h}^\vee}{4\pi}\log\frac{\Lambda}{\mu}
       \int_\Sigma d^2\xi\,\operatorname{Tr}(J_wJ_{\bar w}) \;.
 \label{sec:torsion:logz}
\end{equation}
This is the real-coordinate form of the current-bilinear logarithm.
In the complex measure convention,
$d^2w=dw\wedge d\bar w=-2i\,d^2\xi$, the same expression is
\begin{equation}
 \left.\log|Z_{\rm 1-loop}|\right|_{J\bar J,\log,\mathrm{bulk}}
 =\frac{i \mathsf{h}^\vee}{8\pi}\log\frac{\Lambda}{\mu}
       \int_\Sigma d^2w\,\operatorname{Tr}(J_wJ_{\bar w}) \;,
 \label{sec:torsion:complexlog}
\end{equation}
where we have written the parity-even contribution determined by
the Gaussian modulus. The phase of $Z_{\rm 1-loop}$ does not contribute to
this term: a local term of dimension two that is quadratic in the currents and
odd under the orientation reversal of $\Sigma$ would be proportional to
$\epsilon^{\mu\nu}\operatorname{Tr}(J_\mu J_\nu)$, which vanishes identically.
Moreover, a Wess--Zumino term, which is cubic in the currents, is not generated:
for $z_0=-z_1$ the configuration is invariant under $z\to-z$ combined with
$w\leftrightarrow\bar w$, which maps $\omega$ to $-\omega$ and reverses the
orientation of $\Sigma$, and hence preserves the action, while it reverses the
sign of the Wess--Zumino term.

The classical Euclidean PCM action obtained from our classical 4d CS action is 
\begin{equation}
 S_E=\frac{H_{\rm bare}(\Lambda)}{2\pi}
             \int_\Sigma d^2\xi\,\operatorname{Tr}(J_wJ_{\bar w}) \;.
 \label{sec:torsion:classical}
\end{equation}
Since the one-loop effective action is
$S_{E,\mathrm{eff}}=S_E-\log Z_{\rm 1-loop}$, the bulk logarithm is
absorbed by
\begin{equation}
 H(\mu)=H_{\rm bare}(\Lambda)
                     -\frac{\mathsf{h}^\vee}{2}\log\frac{\Lambda}{\mu} \;.
 \label{sec:torsion:running}
\end{equation}
At fixed bare coupling, this gives
\begin{equation}
 \beta_H^{\rm bulk}
 :=\frac{dH}{d\log\mu}\bigg|_{\rm bulk}
 =\frac{\mathsf{h}^\vee}{2} \;.
 \label{sec:torsion:beta}
\end{equation}
The normalization of $\mathsf{h}^\vee$ is the one in \cref{app:Lie}. Here we read the beta function in Euclidean signature, with $S_E=-iS_{\rm PCM}$ and $H>0$, so that the Euclidean action is positive since $\mathsf h^\vee\operatorname{Tr}(J_wJ_{\bar w})=\sum_\gamma|e_\gamma|^2\geq0$; in terms of the 4d CS data, this corresponds to $i h/\hbar$ being real and positive, which can be arranged by the choice of the phase of the parameter $h$ in $\omega$. 

Finally, we discuss the dependence of the coefficient on the auxiliary factor
$\rho$, which we have set to $1$ both here and in \cref{sec:oneloopPCM}. For non-constant $\rho$, the Hermitian weights of
\cref{app:Qadjoint} satisfy $\sch_I\sch_{I^c}=8\rho$ rather than the constant
value in \eqref{sec:torsion:weights}. Then $\operatorname{im}Q^\dagger$ is
no longer Lagrangian with respect to the BV pairing
\eqref{sec:torsion:pairing}, and the corresponding gauge is not an
admissible BV gauge, so that independence of $\rho$ is not guaranteed for
this family. The admissible family is obtained by rescaling all the weights,
$\sch_p\to\rho^{-1/2}\sch_p$, which restores $\sch_I\sch_{I^c}=8$. For this family the
principal metric becomes $\rd w\,\rd\bar w+\rho\,\rd z\,\rd\bar z$, and the quadratic-background
part of every heat coefficient $a_4^{(p)}$ acquires the additional term
$\tfrac16[\operatorname{tr}E_p]_{ef}\,\partial_z\partial_{\bar z}\log\rho$,
including the contribution of the scalar curvature of the principal metric.
By \eqref{sec:torsion:derivativecancel}, this term cancels pointwise in
$a_4^{(1)}-3a_4^{(0)}$. Hence the density \eqref{sec:torsion:density} and the
coefficient \eqref{sec:torsion:beta} are independent of $\rho$ for any smooth
positive $\rho$ on $\mathbb C$ (here densities are taken with respect to the
coordinate measure $4\,d^2\xi\,d^2x$). The same holds pointwise for the bulk
density with $\rho=|\omega|^2/|h|^2$ regularized as
$\rho_\epsilon=(|z-1|^2+\epsilon^2)(|z+1|^2+\epsilon^2)/|z|^4$, for which
$z=0$ becomes a noncompact end that we treat in the same way as $z=\infty$.

\subsubsection{Including the Wess--Zumino Term}

The computation of the one-loop RG flow via analytic torsion can be generalized to the PCM with a Wess--Zumino term. Let $p=k/h=K/H$ as in
\cref{wzterm}. The background-dependent differential $Q_J$ now depends on $p$. In particular, it can be shown that  \eqref{sec:torsion:dualitymap}-\eqref{sec:torsion:duality} extend to the $p \neq 0$ case.  The endomorphisms and heat-kernel invariants for the case of a nontrivial Wess--Zumino term are derived in \cref{app:bg:wz}. The scalar and one-form endomorphisms are
\ie
 E_0&=-\frac{1}{\mathcal N^2}+q,\qquad
 E_1=\operatorname{diag}(d_1,d_2,d_3)+q\mathbf 1_3+\mathcal U,\\
 q&=-\frac{ef}{2\mathcal N^2}
        \big[(1-p^2)s^2-4ps+4p^2\mathcal N\big].
\fe
Here $d_i$ are given in \eqref{app:bg:di} and $\mathcal U$ is
the off-diagonal matrix in \eqref{app:bg:wz:endomorphisms}.
Writing $r=|z|$, the analogue of \eqref{sec:torsion:data} is
\ie
 T_0(p)
 &=\frac{(1-p^2)s^2-4ps+4p^2\mathcal N}{\mathcal N^4},\\
 T_1(p)
 &=\frac{1}{\mathcal N^4}\big[
 (1-p^2)(8\mathcal N^2-9s^2)
 +4p(7-2r^2)s+4p^2\mathcal N(4r^2-1)\big],\\
 O_0(p)
 &=-\frac{|a|^2|a-pb|^2+|b|^2|b+pa|^2}{\mathcal N^4},
 \qquad O_1(p)=3O_0(p).
 \label{sec:torsion:wzdata}
\fe
The curvature-square and total-derivative terms cancel in the
current-dependent part of \eqref{sec:torsion:weightedheat}.
Consequently, the analogue of \eqref{sec:torsion:density} is
\ie
 \mathcal D_p
 &=\frac{T_1(p)-3T_0(p)}{16\pi^2}\\
 &=(1-p^2)\mathcal D
   +\frac{p(5-r^2)s+2p^2(r^4-1)}{2\pi^2\mathcal N^4}.
 \label{sec:torsion:wzdensity}
\fe
The density is smooth at $z=p$ and integrable at infinity, so
the integral over $C_p^0$ equals the integral over $\mathbb C$.
The term proportional to $s=2r\cos\theta$ has zero angular
integral. The substitution $r\mapsto1/r$ exchanges the radial
integrals of $r^4$ and $1$ against $r\,\rd r/(1+r^2)^4$, implying that the integral of the term proportional to $(r^4-1)$ vanishes as well.
Using \eqref{sec:torsion:integrateddensity}, we obtain
\ie
 \int_{C_p^0}\rd^2x\,\mathcal D_p
 =\frac{1-p^2}{4\pi}.
 \label{sec:torsion:wzintegrateddensity}
\fe
After the root sum \eqref{sec:torsion:rootsum}, the bulk logarithm
is therefore $1-p^2$ times \eqref{sec:torsion:complexlog}, in
agreement with the one-loop diagrammatic calculation of \cref{wzterm}, therefore leading us once again to the beta functions of the PCM + WZ model given in \eqref{wzbeta}.

\subsection{Comments on Elliptic Integrable Field Theories}
\label{sec:comments}

We have shown that the renormalization group flow of the PCM arises from the integral over the (punctured) Riemann
surface $\mathbb{CP}^1$ of a local density determined by the heat kernels and the gauge-fixing metric. A natural question to ask is whether an analogue of this result can be derived for elliptic integrable field theories where the Riemann surface $C$ is some elliptic curve. In this subsection we make some speculative remarks on this question. The general short-time expansion of the heat kernel on $C$ involves not just its curvature, but also the first Chern class of the bundle $E$, which is the coefficient bundle defined over $C$ associated with the de Rham-Dolbeault complex of fields.
In general, for a Dolbeault operator
\ie
    \bar{\partial}_E: \Omega^{0,0}(C, E) \rightarrow \Omega^{0,1}(C, E) \;,
\fe
the heat kernel is a section of
\ie
    \operatorname{Hom}\left(E_y \otimes \Lambda^{0, q} T_y^* C, E_x \otimes \Lambda^{0, q} T_x^* C\right) \;.
\fe
On the diagonal $x=y$, it becomes an element of
\ie
    \operatorname{End}\left(E_x \otimes \Lambda^{0, q} T_x^* C\right) \;.
\fe
With this definition, we expect the diagonal expansion to take the form (for the PCM, the heat kernel does not factorize into kernels on $\Sigma$ and $C$, see \cref{sec:oneloopPCM}, so that this expansion should be understood as a heuristic guide)
\ie \label{firstexpansion22}
    \textrm{d}z_1 \wedge \sfK_{\Omega^1(C)}(z_1,z_1;\tau) = \frac{\mathbf{1}\mathrm{dvol}_C}{4 \pi \tau}-\frac{F_E + \mathbf{1} R_C(z_1) \mathrm{dvol}_C}{12 \pi}+O(\tau)\;.
\fe
Here, $F_E=F_{V} \mathbf{1}_{\mathfrak{g}}+F_{\mathfrak{g}_P}$, when $E=V\otimes \mathfrak{g}_P$, where $\mathfrak{g}_P$ is the associated adjoint bundle of the principal $G$-bundle. In the simple case where $F_{\mathfrak{g}_P}=0$, we then expect the counterterm to have the form
\ie \label{ell_result}
    \left.\log|Z_{\mathrm{1-loop}}|\right|_{\log} =
    \frac{i}{8 \pi}
    \sfh^{\vee}
    \deg V
    \log\!\frac{\Lambda}{\mu}\int_\Sigma\rd^2w\,
    \operatorname{Tr}\left(J_{w}J_{\bar w}\right)\;,
\fe
where $\deg V$ is the degree of $V$, i.e.\ the integral over $C$ of its first Chern class, written in the same normalization as \eqref{nonuresult}, which corresponds to $\deg V=1$. We conjecture that this gives the RG flow of elliptic IFTs predicted by the 4d CS theory. We have not identified the bundle $V$ with $\deg V=1$ for the PCM; doing so would provide a nontrivial check of \eqref{ell_result} in the rational case. A derivation along the lines of \cref{sec:rg-torsion} is left for future work.


\noindent
\textbf{Acknowledgements:} The contents of this work were announced at the IGST 2026 workshop at DESY, Hamburg, and we thank the audience for valuable feedback.
We would like to thank Matthias Blau, Domenico Orlando and Susanne Reffert for helpful discussions.
During the final stages of this project, instances of ChatGPT and Claude were used to aid with coding for computer algebra computations and improving the presentation of the manuscript.
The research of M.A.\ was supported in part by the NCCR SwissMAP of the Swiss National Science Foundation. 
The research of M.Y.\
was supported in part by the World Premier International Research Center Initiative (WPI), MEXT, Japan; by the JSPS KAKENHI Grant No.\ 23K25865; by JST, Japan (CREST Grant No.\ JPMJCR26XA, Moonshot R\&D Grant No.\ JPMJMS2061).

\appendix
\crefalias{section}{appendix}

\section{Lie Algebra Conventions}\label{app:Lie}
In the main text, we use the following Lie algebra conventions:
\begin{itemize}
    \item Lie algebra generators $\{ T^a\}$ are taken to be anti-Hermitian, and satisfy the commutation relations
    \ie
        \relax [T^a,T^b]=f^{ab}{}_{c} T^c
    \fe
    for real structure constants $f^{ab}{}_{c}$.
    
    \item The normalization of the invariant bilinear product $\textrm{Tr}$ on the Lie algebra is chosen such that
    \ie \label{innprod}
        \textrm{Tr}(T^aT^b)= \delta^{ab}
    \fe
    Since the bilinear form is the identity in this basis, adjoint indices are raised and lowered with $\delta^{ab}$, and we do not distinguish between upper and lower indices, e.g., $T_a=T^a$ and $f_{abc}=f^{ab}{}_c$.
    \item The structure constants satisfy
    \ie f^{abc} f_{ab}{}^d = \mathsf{h}^{\vee}\delta^{cd} \;,
    \fe
    where $\mathsf{h}^{\vee}$ is the dual Coxeter number in this normalization, which is twice the dual Coxeter number in the standard normalization. For $\mathfrak{su}(N)$, we have $\mathsf{h}^{\vee}=2N$.
    \item Measures: in \cref{sec:ultralocal} and in the final formulas of \cref{sec:nonultralocal}, $\rd^2w=\rd w\wedge\rd\bar w=-2i\,\rd^2\xi$ denotes the complex two-form on $\Sigma$, where $\rd^2\xi=\rd\xi^1\rd\xi^2$ (also written $\rd^2x$ in \cref{sec:ultralocal}) is the real measure. On $C$, $\rd^2z=\tfrac i2\rd z\wedge\rd\bar z$ (written $\rd^2x$ with $z=x+iy$ in \cref{sec:rg-torsion}) is the real measure. In \cref{app:2dPCM}, $\rd^2w=\rd w\wedge\rd\bar w$ as in \cref{sec:ultralocal}. In \cref{app:Qadjoint,app:Qlaplacian,app:bg}, $\rd^2w$ and $\rd^2z$ both denote real measures.
\end{itemize}
As an example, for $\mathfrak{g}=\mathfrak{su}(2)$, we take $T_a= \frac{-i}{\sqrt{2}}\sigma_a$ where $\sigma_a$ are the standard Pauli matrices satisfying $[\sigma_a,\sigma_b]=2i\epsilon_{ab}{}^c \sigma_c$, implying that the real structure constants $f_{abc}=\sqrt{2}\epsilon_{abc}$. In particular $\textrm{Tr}$ ought to be defined as the negative of the matrix trace, $\textrm{tr}$, to satisfy \eqref{innprod}. The structure constants satisfy $f^{abc} f_{ab}{}^d=4 \delta^{cd}$, which agrees with the dual Coxeter number defined as $\mathsf{h}^{\vee}=2N$ for $\mathfrak{su}(N)$.

\section{Conventions for Elliptic IFTs}\label{ellipticcon}

Let us summarize our conventions for elliptic IFTs.
We shall focus on $G=PGL(N,\mathbb{C})$; as explained in \cite{Costello:2017dso}, this is one of the few cases where we obtain a rigid bundle, by taking a topologically nontrivial $PGL(N,\mathbb C)$-bundle on the elliptic curve (whose degree is coprime to $N$), and hence a non-dynamical solution of the Yang--Baxter equation.
A convenient basis for its Lie algebra is the Belavin basis that involves the clock and shift matrices
\ie
    \mathscr A \equiv\left(\begin{array}{ccccc}1 & 0 & & & \\ 0 & \Omega & \ddots & & \\ & \ddots & \ddots & \ddots & \\ & & \ddots & \Omega^{N-2} & 0 \\ & & & 0 & \Omega^{N-1}\end{array}\right) \;, \quad \mathscr B \equiv\left(\begin{array}{cccc}0 & 1 & & \\ 0 & 0 & \ddots & \\ & \ddots & \ddots & \ddots \\ & & \ddots & 0 \\ 1 & & & 0 \end{array}\right) \;,\fe
where $\Omega=\exp({2\pi i}/{N})$,
satisfying the relations
\ie
    \mathscr A^N=\mathscr B^N=1 \;, \quad \mathscr B\mathscr A=\Omega \mathscr A\mathscr B \;.
\fe
These matrices allow us to define the $\mathfrak{sl}(N,\mathbb{C})$ generators
\ie
    T_{\lambda}\equiv \frac{1}{\sqrt{N}} \mathscr B^{-\lambda_2} \mathscr A^{\lambda_1}
\fe
where $\lambda \in \mathbb{A}:=(\mathbb{Z}_N \times \mathbb{Z}_N) \backslash\{(0,0)\}$, 
satisfying
\ie
    \left[T_\mu, T_\nu \right]=f_{\mu, \nu} T_{\mu+\nu} \;, \quad f_{\mu, \nu}=\frac{1}{\sqrt{N}}\left(\Omega^{\mu_1 \nu_2}-\Omega^{\nu_1 \mu_2}\right) \;,
\fe
and
\ie \textrm{Tr} (T_\mu T_\nu )=-\delta_{\mu+\nu, 0} \Omega^{-\nu_1 \nu_2} \;.
\fe
Unlike in \cref{app:Lie}, this basis is not orthonormal, and we distinguish upper and lower indices. We denote by $T^\alpha=-\Omega^{\alpha_1\alpha_2}T_{-\alpha}$ the dual basis, $\textrm{Tr}(T^\alpha T_\beta)=\delta^\alpha_\beta$, and define $f^{\alpha,\beta}$ by $[T^\alpha,T^\beta]=f^{\alpha,\beta}T^{\alpha+\beta}$. The quadratic Casimir is $\mathsf t=\sum_\alpha T^\alpha\otimes T_\alpha$, and $\mathsf h^\vee$ is defined as in \cref{app:Lie} by $\sum_\alpha[T^\alpha,[T_\alpha,X]]=-\mathsf h^\vee X$, which is basis independent. The Lie algebra of $PGL(N,\mathbb C)$ is $\mathfrak{sl}(N,\mathbb C)$, the complexification of the compact form used in \cref{sec:ultralocal}.

In this basis, the elliptic classical $r$-matrix can be written as
\ie
    r(z)=\sum_{\alpha \in \mathbb{A}} r^\alpha(z) T^\alpha \otimes T_\alpha \;,
\fe
and the group-theory factor in \eqref{ellampl} in this basis is given in \cref{sec:ultralocal}.

The identity \eqref{rmatrixidentity} follows from the classical Yang--Baxter equation
\ie
    \relax[r_{12}(u),r_{13}(u+v)]+[r_{12}(u),r_{23}(v)]+[r_{13}(u+v),r_{23}(v)]=0 \;,
\fe
where $r(z)=\mathsf t/z+O(1)$ near $z=0$. Consider the limit $v\to0$. The terms of order $1/v$ are proportional to $[r_{12}(u)+r_{13}(u),\mathsf t_{23}]$, which vanishes since $\mathsf t_{23}$ commutes with the diagonal action on the second and third factors. At order $v^0$ we find
\ie
    \relax[r_{12}(u),r_{13}(u)]+[\partial_ur_{13}(u),\mathsf t_{23}]+[r_{12}(u)+r_{13}(u),r^{\rm reg}_{23}(0)]=0 \;,
\fe
where $r^{\rm reg}(0)=\sum_\alpha c_\alpha T^\alpha\otimes T_\alpha$ is the regular part of $r$ at the origin. We now apply the map $X\otimes Y\otimes Z\mapsto X\otimes[Y,Z]$, which commutes with the diagonal adjoint action on the last two factors. The first term gives $\sum_\gamma\big(\sum_{\alpha+\beta=\gamma}f_{\alpha,\beta}f^{\alpha,\beta}r^\alpha r^\beta\big)T^\gamma\otimes T_\gamma$, and the second gives $\mathsf h^\vee\sum_\gamma\partial_ur^\gamma\,T^\gamma\otimes T_\gamma$. The third term is mapped to the commutator of $r(u)$ with $\sum_\alpha c_\alpha[T^\alpha,T_\alpha]$, which vanishes since $[T_{-\alpha},T_\alpha]=0$ by $f_{-\alpha,\alpha}=0$. This proves \eqref{rmatrixidentity}. In the rational limit $r=\mathsf t/z$, it reduces to $\mathsf h^\vee\partial_z(1/z)=-\mathsf h^\vee/z^2$. As a further check, for $\mathfrak{sl}(2)$ in the orthonormal basis of \cref{app:Lie} with $r=\sum_a w_a T_a\otimes T_a$ and $(w_1,w_2,w_3)=(1/\mathrm{sn},\mathrm{dn}/\mathrm{sn},\mathrm{cn}/\mathrm{sn})$, the identity becomes $w_1'=-w_2w_3$ and cyclic permutations, which are the derivative formulas of the Jacobi elliptic functions.

\section{2d Derivation of RG Flow of PCM}
\label{app:2dPCM}

In this appendix, we shall review the 2d derivation of the RG flow of the PCM, and also describe how the derivation can be performed by expressing propagators in terms of heat kernels.

Let us consider the PCM with the Lagrangian
\begin{equation}
    L=\frac{H}{2} \operatorname{Tr}\left(J_{+} J_{-}\right) \;,
\end{equation}
where $H$ is the coupling constant, and the conserved current is defined by
$J_{ \pm}(g)=g^{-1} \partial_{ \pm} g$. Here, we work with lightcone coordinates $\xi^{\pm}=\frac{1}{2}(\xi^0\pm \xi^1)$.
We use the normalization
\ie
    S_{\rm PCM}
    =
    \frac{1}{4\pi}
    \int \rd^2\xi\,L \;.
\fe
We write $g$ as the product of a classical solution $\bar{g}$ and fluctuation field $e^{\phi}$. The current can be expanded perturbatively around its classical value as
\begin{equation}
    \begin{aligned}
        J_{ \pm}\left(\bar{g} e^\phi\right) &=  \left(e^{-\phi} \bar{g}^{-1}\right) \partial_{ \pm}\left(\bar{g} e^\phi\right) \\
        &=  J^{\rm cl}_{ \pm}+\left[J^{\rm cl}_{ \pm}, \phi\right]+\frac{1}{2}\left[\left[J^{\rm cl}_{ \pm}, \phi\right], \phi\right] \\
        & \quad +\partial_{ \pm} \phi-\frac{1}{2}\left[\phi, \partial_{ \pm} \phi\right]+O(\phi^3) \;.
    \end{aligned}
\end{equation}
Substituting this expansion into the Lagrangian we obtain, at quadratic order in the fields,
\begin{equation}
    \frac{H}{2} \operatorname{Tr}\left(\partial_{+} \phi \partial_{-} \phi+\frac{1}{2}\left[J^{\rm cl}_{+}, \phi\right] \partial_{-} \phi+\frac{1}{2} \partial_{+} \phi\left[J^{\rm cl}_{-}, \phi\right]\right) \;.
\end{equation}
The corresponding equation of motion is
\begin{equation}
    2 \partial_+\partial_- \phi+\left[J^{\rm cl}_{+}, \partial_{-} \phi\right]+\left[J^{\rm cl}_{-}, \partial_{+} \phi\right]=0 \;.
\end{equation}
Now, observe that the equation of motion can be equivalently written as
\begin{equation}
    \partial_+(D_-\phi)+\partial_-(D_+\phi)=0 \;,
\end{equation}
where $D_{\pm}\phi=\partial_{\pm}\phi+[J^{\rm cl}_{\pm},\phi]$,
using the conservation condition $\partial_+ J^{\rm cl}_-+\partial_- J^{\rm cl}_+=0$. To linear order, $D_{\pm}\phi$ is the fluctuation of the current $J_{\pm}$, which we shall denote $a_{\pm}$. Hence, the equation of motion can be written as
\begin{equation}
    \partial_+a_- +\partial_-a_+=0 \;.
\end{equation}
In addition, the field $a$ also satisfies the linearized Bianchi identity
\begin{equation}\label{linbianch}
    \partial_+ a_{-}-\partial_{-} a_++\left[J^{\rm cl}_{+},a_{-}\right]+\left[a_{+}, J^{\rm cl}_-\right]=0 \;.
\end{equation}

Now, the path integral of the PCM can be written as
\ie
    \int D \phi \exp \left(i\int \textrm{d}^2 \xi \textrm{Tr}\left(\phi O \phi  \right)\right) \;,
\fe
where
\ie
    O\phi =-\frac{H}{2} (2\partial_+\partial_- \phi + [J^{\rm cl}_+,\partial_-\phi ] +[J^{\rm cl}_-,\partial_+\phi ]  ) \;.
\fe
In particular, the one-loop effective action is
\ie
    \widehat S=\frac{i}{2} \log (\det  O) \;.
\fe
This one-loop effective action can equivalently be written as
\ie
    \widehat S
    =
    -\frac{i}{2}\log Z_{\rm aux} \;,
\fe
where
\ie
    Z_{\rm aux}
    =
    \int Du\,D\phi\,
    \exp\left(
    i\int \rd^2\xi\,
    \Tr(uO\phi)
    \right) \;.
\fe
Using the field $a$ instead of $\phi$ incurs a Jacobian that does not contribute to the logarithmic divergence of the current bilinear, see \cite[Sec.~2.1]{Levine:2022hpv}. However, we ought to impose the linearized Bianchi identity \eqref{linbianch}  at the path integral level. The one-loop effective action can thus be expressed as
\ie
    \widehat S=-\frac{i}{2}\log \int & Du Dv Da \exp \left(i\int \textrm{d}^2\xi\, \Tr \left[u (\partial_+a_- +\partial_-a_+  )\right]\right)\\& \times \exp \left(i\int \textrm{d}^2 \xi\, \Tr \left[ v (\partial_+a_- - \partial_-a_+ + [J^{\rm cl}_+,a_- ] +[a_+,J^{\rm cl}_- ] )\right]\right) \;,
\fe
where we have absorbed $\frac{H}{2}$ into the field $u$. Now, performing a further field redefinition $x=u+v$ and $y=u-v$, we arrive at
\ie
    \widehat S=-\frac{i}{2} \log & \int {D} x {D} y {D} a\\
    & \times \exp i \int \textrm{d}^2\xi  \operatorname{Tr}\left[x \partial_{+} a_{-}+y \partial_{-} a_{+}+\frac{1}{2}(x-y)\left(\left[J^{\rm cl}_{+}, a_{-}\right]+\left[a_{+}, J^{\rm cl}_{-}\right]\right)\right] \;.
\fe

In what follows, we shall work with Euclidean signature on the worldsheet by performing the Wick rotation $\xi^0 =-i \xi^2$, under which $\d\xi^0\,\d\xi^1=-i\,\d\xi^1\,\d\xi^2$ and $\operatorname{Tr}(J_+J_-)=-4\operatorname{Tr}(J_wJ_{\bar w})$, so that $e^{iS_{\rm PCM}}=e^{-S_E}$ with $S_E=\frac{H}{2\pi}\int\d^2\xi\,\operatorname{Tr}(J_wJ_{\bar w})\geq0$ for $H>0$ (recall that $\operatorname{Tr}=-\operatorname{tr}$ is positive on the compact real form).
We shall define the complex coordinates $w=\xi^1 + i\xi^2$ and $\bar{w}=\xi^1 -i \xi^2$. In particular we have the relations
\ie
    &\partial_{+}=2\partial_{\bar{w}}\;,\quad
    &\partial_{-}=-2\partial_{w}
\fe
and
\ie
    &a_{+}=2 a_{\bar{w}}\;, \quad
    &a_{-}=-2a_{w} \;.
\fe
The one-loop effective action can be written in Euclidean signature as
\ie
    \widehat S=-\frac{i}{2} \log & \int {D} x {D} y {D} a\\
    & \times \exp \int \frac{\textrm{d}^2 w}{2i}  \operatorname{Tr}\Big[4 x \partial_{\bar{w}} a_{w}+4 y \partial_{w} a_{\bar{w}} \\
    & \qquad +2(x-y)\left(\left[J^{\rm cl}_{\bar{w}}, a_{w}\right]+\left[a_{\bar{w}}, J^{\rm cl}_{w}\right]\right)\Big] \;,
\fe
where $\textrm{d}^2w =\textrm{d}w \wedge \textrm{d}\bar{w}$. To further simplify this expression, we can absorb a constant factor into the Lagrange-multiplier fields $x$ and $y$, resulting in
\ie
    \widehat S=-\frac{i}{2} \log & \int {D} x {D} y {D} a\\
    & \times \exp (-1) \int {\textrm{d}^2 w} \operatorname{Tr}\Big[ x \partial_{\bar{w}} a_{w}+ y \partial_{w} a_{\bar{w}}\\
    &\qquad+\frac{1}{2}(x-y)\left(\left[J^{\rm cl}_{\bar{w}}, a_{w}\right]+\left[a_{\bar{w}}, J^{\rm cl}_{w}\right]\right)\Big] \;.
\fe

To evaluate this effective action, we need the free field propagators
\ie
    \langle a_w^b(w) x^c(w') \rangle = - \frac{1}{2\pi i}\frac{1}{w-w'} \delta^{bc} \;,
\fe
and
\ie
    \langle a_{\bar{w}}^b(w) y^c(w') \rangle = - \frac{1}{2\pi i}\frac{1}{\bar{w}-\bar{w}'} \delta^{bc} \;,
\fe
where $b$ and $c$ are Lie algebra indices.
Expanding the interaction terms in the exponential of the path integral
and retaining the connected contribution quadratic in the background
currents, we obtain the following contribution to
$\frac{1}{2}\log Z_{\rm aux}$:
\ie
    \frac{1}{8}\int \textrm{d}^2 w\int \textrm{d}^2 w' f^{abc}J^{\rm cl}_{w,c}(w)f^{def}J^{\rm cl}_{\bar{w},e}(w') \frac{1}{(2\pi i)^2} \frac{1}{|w-w'|^2}\delta_{af}\delta_{bd} \;.
\fe
We can then shift variables as $w \rightarrow w+w'$, and evaluate the $w$-plane integral using polar coordinates defined as $w=re^{i\theta}$, whereby $\textrm{d}^2w = -i2 r \textrm{d}r\textrm{d}\theta$. The shifted classical current $J^{\rm cl}_{w,c}(w+w')$ can be Taylor expanded around $w'$, but the $w$-dependent terms that result either vanish or do not contribute to UV divergences.
The radial integral can be evaluated as
\ie
    \int^{\frac{1}{\mu}}_{\frac{1}{\Lambda}} \frac{1}{r}\rd r =\log \frac{\Lambda}{\mu} \;,
\fe
where $\mu$ and $\Lambda$ are the IR and UV cutoffs, respectively.
Using the identity
$f^{abc}f_{ab}{}^{e}=\mathsf{h}^{\vee}\delta^{ce}$,
we obtain
\ie
    \left.
    \frac{1}{2}\log Z_{\rm aux}
    \right|_{\rm div}
    =
    -\frac{\mathsf{h}^{\vee}}{8\pi i}
    \log\left(\frac{\Lambda}{\mu}\right)
    \int \rd^2w'\,
    J^{\rm cl}_{w,c}(w')J^{{\rm cl}\,c}_{\bar w}(w') \;.
\fe
Using
$\widehat S=-i\,\frac{1}{2}\log Z_{\rm aux}$,
the logarithmically divergent part of the one-loop effective action is
therefore
\ie
    \widehat S_{\rm div}^{(1)}
    =
    \frac{\mathsf{h}^{\vee}}{8\pi}
    \log\left(\frac{\Lambda}{\mu}\right)
    \int \rd^2w'\,
    J^{\rm cl}_{w,c}(w')J^{{\rm cl}\,c}_{\bar w}(w') \;.
    \label{2d_divergent_effective_action}
\fe

Combining this with the classical action
\ie
    S_{\rm PCM}
    =
    -\frac{H}{4\pi}
    \int \rd^2w'\,
    J^{\rm cl}_{w,c}(w')J^{{\rm cl}\,c}_{\bar w}(w') \;,
\fe
we obtain
\ie
    H(\mu)
    =
    H-\frac{\mathsf{h}^{\vee}}{2}
    \log\left(\frac{\Lambda}{\mu}\right) \;,
\fe
and therefore
\ie \label{beth}
    \beta_H
    =
    \frac{\rd H}{\rd\log\mu}
    =
    \frac{\mathsf{h}^{\vee}}{2} \;.
\fe

Instead of imposing cutoffs when evaluating the $r$ integral, an alternative approach is to regularize the propagators using heat kernels, before proceeding to evaluate the Feynman amplitude. The regularized propagators take the form
\ie
    \langle a_w^b(w) x^c(w') \rangle_{\textrm{reg}} = \delta^{bc}\frac{2}{i} \partial_w \int^{\frac{1}{\mu^2}}_{\frac{1}{\Lambda^2} } \frac{1}{4\pi \tau } e^{-\frac{|w-w'|^2}{4\tau }}\textrm{d}\tau
\fe
and
\ie
    \langle a_{\bar{w}}^b(w) y^c(w') \rangle_{\textrm{reg}}  =\delta^{bc} \frac{2}{i} \partial_{\bar{w}} \int^{\frac{1}{\mu^2}}_{\frac{1}{\Lambda^2} } \frac{1}{4\pi \tau } e^{-\frac{|w-w'|^2}{4\tau }}\textrm{d}\tau \;.
\fe
The derivation of the beta function \eqref{beth} can be equivalently performed with these regularized propagators. 

\section{Adjoint of the Rescaled BRST Differential}
\label{app:Qadjoint}

Appendices~\ref{app:Qadjoint}--\ref{app:bg} contain the technical details of \cref{sec:nonultralocal}:
\begin{itemize}
\item \cref{app:Qadjoint} derives the adjoint $Q^\dagger$ and the Hermitian weights, which define the gauge-fixing condition and the propagators of \cref{heatk} and satisfy the condition \eqref{sec:torsion:weights} used in \cref{sec:rg-torsion};
\item \cref{app:Qlaplacian} derives the Laplacians $\Delta_p$, whose principal metric is \eqref{main:hk:principalmetric};
\item \cref{shorttimeheat} derives the short-time expansions \eqref{firstexpansion} and \eqref{secondexpansion}, which determine the boson and ghost bubbles in \cref{sec:oneloopPCM};
\item \cref{app:bg} derives the background-coupled Laplacians and the heat-kernel data \eqref{sec:torsion:data}, which determine the density \eqref{sec:torsion:density} in \cref{sec:rg-torsion}.
\end{itemize}
Appendices~\ref{app:Qadjoint} and~\ref{app:Qlaplacian} are thus used in both computations of the one-loop RG flow, \cref{shorttimeheat} only in \cref{sec:oneloopPCM}, and \cref{app:bg} only in \cref{sec:rg-torsion}.

In this appendix, we derive the adjoint of the rescaled BRST differential 
that appears in the discussion of the PCM in the main text. We denote it as $Q^{\dagger}$ to emphasize its property of being the adjoint of the BRST differential, $Q$. It replaces the operator $\delta$ used for the flat case in \cref{sec:nonultralocal}, and is the gauge-fixing operator used for the PCM in the main text.

We write
\ie
    Q
    =
    a\,\rd w\,\partial_w
    +
    b\,\rd\bar w\,\partial_{\bar w}
    +
    \rd\bar z\,\partial_{\bar z} \;,
    \qquad
    a=z-1 \;,
    \qquad
    b=z+1 \;.
\fe
It is useful to introduce
\ie
    \mathcal{N}=1+z\bar z \;,
\fe
so that
\ie
    |a|^2+|b|^2=2\mathcal{N} \;.
\fe

We choose the Hermitian metric
\ie
    \label{warped_metric}
    \rd s^2
    =
    \alpha(z,\bar z)\,\rd w\,\rd\bar w
    +
    \rho(z,\bar z)\,\rd z\,\rd\bar z \;,
\fe
where
\ie
 \alpha=2\mathcal{N} \;,\qquad \rho>0 \;.
 \label{alpha_rho_def}
\fe
We retain a general smooth positive auxiliary factor $\rho(z,\bar z)$
in the formulas of this appendix and Appendix~\ref{app:Qlaplacian}.
The one-loop calculation in the main text and
Appendix~\ref{shorttimeheat} uses $\rho=1$. For non-constant $\rho$, the
Hermitian weights below satisfy $\sch_I\sch_{I^c}=8\rho$, and the weights compatible
with the BV pairing are obtained by dividing them by $\rho^{1/2}$; see the
discussion at the end of \cref{sec:rg-torsion}.

We use the complex-coordinate convention
\ie
    w=x+i y \;.
\fe
Thus
\ie
    g_{w\bar w}=g_{\bar w w}=\frac{\alpha}{2} \;,
    \qquad
    g^{w\bar w}=g^{\bar w w}=\frac{2}{\alpha} \;,
\fe
and similarly
\ie
    g_{z\bar z}=g_{\bar z z}=\frac{\rho}{2} \;,
    \qquad
    g^{z\bar z}=g^{\bar z z}=\frac{2}{\rho} \;.
\fe
The corresponding volume forms are
\ie
    \rd\mathrm{vol}_{\Sigma}
    =
    \frac{i\alpha}{2}\,\rd w\wedge\rd\bar w \;,
    \qquad
    \rd\mathrm{vol}_{C}
    =
    \frac{i\rho}{2}\,\rd z\wedge\rd\bar z \;,
\fe
up to the choice of orientation. For $\rho=1$, the choice $\alpha=2\mathcal{N}$ makes the
principal second-order parts of the Laplacian along  $\Sigma$ and $C$ 
have the same overall normalization of $-2$.

The four-dimensional volume density is therefore
\ie
    \rd\mathrm{vol}_4
    =
    \alpha\rho\,\rd^2w\,\rd^2z \;.
\fe
We shall write all inner products with respect to the fixed coordinate
measure
\ie
    \rd\mu_0
    =
    \rd^2w\,\rd^2z \;,
\fe
where, in this appendix and in \cref{app:Qlaplacian,app:bg}, $\rd^2w$ and $\rd^2z$ denote real coordinate measures (in \cref{sec:ultralocal}, $\rd^2w$ instead denotes the two-form $\rd w\wedge\rd\bar w$). We
absorb all metric dependence into Hermitian weights $\sch_p$ acting on
$p$-forms:
\ie
    \label{weighted_inner_product}
    \langle X,Y\rangle_p
    =
    \int \rd\mu_0\,X^\dagger \sch_pY \;.
\fe

With this convention,
\ie
    |\rd w|^2
    =
    |\rd\bar w|^2
    =
    \frac{2}{\alpha} \;,
    \qquad
    |\rd z|^2
    =
    |\rd\bar z|^2
    =
    \frac{2}{\rho} \;.
\fe
The weight $\sch_p$ is obtained by multiplying the volume density by the
inverse metric factor associated with every one-form index.

For a scalar $f$,
\ie
    \|f\|^2
    =
    \int\rd\mu_0\,\alpha\rho\,|f|^2 \;,
\fe
and hence
\ie
    \sch_0=\alpha\rho \;.
\fe

For a one-form
\ie
    B
    =
    B_w\,\rd w
    +
    B_{\bar w}\,\rd\bar w
    +
    B_{\bar z}\,\rd\bar z \;,
\fe
we have
\ie
    |B|^2
    =
    \frac{2}{\alpha}|B_w|^2
    +
    \frac{2}{\alpha}|B_{\bar w}|^2
    +
    \frac{2}{\rho}|B_{\bar z}|^2 \;.
\fe
Multiplying by the volume density gives
\ie
    \|B\|^2
    =
    \int\rd\mu_0
    \left(
    2\rho|B_w|^2
    +
    2\rho|B_{\bar w}|^2
    +
    2\alpha|B_{\bar z}|^2
    \right) \;.
\fe
Thus, in the basis
\ie
    (\rd w,\rd\bar w,\rd\bar z) \;,
\fe
we obtain
\ie
    \sch_1
    =
    2\,\operatorname{diag}
    (\rho,\rho,\alpha) \;.
\fe

For a two-form
\ie
    F
    =
    F_{w\bar w}\,\rd w\wedge\rd\bar w
    +
    F_{w\bar z}\,\rd w\wedge\rd\bar z
    +
    F_{\bar w\bar z}\,\rd\bar w\wedge\rd\bar z \;,
\fe
the norms of the basis forms are
\ie
    |\rd w\wedge\rd\bar w|^2
    &=
    \frac{4}{\alpha^2} \;,
    \quad
    |\rd w\wedge\rd\bar z|^2
    &=
    \frac{4}{\alpha\rho} \;,
    \quad
    |\rd\bar w\wedge\rd\bar z|^2
    &=
    \frac{4}{\alpha\rho} \;.
\fe
It follows that
\ie
    \|F\|^2
    =
    \int\rd\mu_0
    \left(
    \frac{4\rho}{\alpha}|F_{w\bar w}|^2
    +
    4|F_{w\bar z}|^2
    +
    4|F_{\bar w\bar z}|^2
    \right) \;,
\fe
so that, in the basis
\ie
    (
    \rd w\wedge\rd\bar w,
    \rd w\wedge\rd\bar z,
    \rd\bar w\wedge\rd\bar z
    ) \;,
\fe
we have
\ie
    \sch_2
    =
    4\,\operatorname{diag}
    \left(
    \frac{\rho}{\alpha},1,1
    \right) \;.
\fe

Finally, the degree-three component has basis
\ie
    \Theta
    =
    \rd w\wedge\rd\bar w\wedge\rd\bar z \;.
\fe
Its pointwise norm is
\ie
    |\Theta|^2
    =
    \frac{8}{\alpha^2\rho} \;,
\fe
and therefore
\ie
    \|G\Theta\|^2
    =
    \int\rd\mu_0\,\frac{8}{\alpha}|G|^2 \;.
\fe
Hence
\ie
    \sch_3=\frac{8}{\alpha} \;.
\fe

We now derive the adjoint of $Q$. Let $Q^\flat$ denote the adjoint with
respect to the unweighted coordinate pairing
\ie
    \langle X,Y\rangle_{\rm flat}
    =
    \int\rd\mu_0\,X^\dagger Y \;.
\fe
We define contraction operators $\iota_w$, $\iota_{\bar w}$ and
$\iota_{\bar z}$ by
\ie
    \iota_w(\rd w)=1 \;,
    \qquad
    \iota_{\bar w}(\rd\bar w)=1 \;,
    \qquad
    \iota_{\bar z}(\rd\bar z)=1 \;,
\fe
with contraction against the remaining basis one-forms equal to zero,
and extend them to higher forms by the graded Leibniz rule.
Integration by parts then gives
\ie
    \label{Qflat}
    Q^\flat
    =
    -\bar a\,\partial_{\bar w}\iota_w
    -\bar b\,\partial_w\iota_{\bar w}
    -\partial_z\iota_{\bar z} \;.
\fe

The metric adjoint
\ie
    Q_p^\dagger:
    \Omega^p
    \longrightarrow
    \Omega^{p-1}
\fe
is defined by
\ie
    \label{adjoint_def}
    \langle QX,Y\rangle_p
    =
    \langle X,Q_p^\dagger Y\rangle_{p-1} \;.
\fe
Using \eqref{weighted_inner_product}, we have
\ie
    \begin{aligned}
        \langle QX,Y\rangle_p
        &=
        \langle QX,\sch_pY\rangle_{\rm flat}
        \\
        &=
        \langle X,Q^\flat(\sch_pY)\rangle_{\rm flat}
        \\
        &=
        \left\langle
        X,
        \sch_{p-1}^{-1}Q^\flat(\sch_pY)
        \right\rangle_{p-1} \;.
    \end{aligned}
\fe
Consequently,
\ie
    Q_p^\dagger
    =
    \sch_{p-1}^{-1}Q^\flat \sch_p \;.
    \label{weighted_adjoint}
\fe
Importantly, $Q^\flat$ acts on the product $\sch_pY$. Hence derivatives
of the metric weights generate connection terms in $Q^\dagger$.

Let us now evaluate \eqref{weighted_adjoint} explicitly.
For the one-form
\ie
    B
    =
    B_w\,\rd w
    +
    B_{\bar w}\,\rd\bar w
    +
    B_{\bar z}\,\rd\bar z \;,
\fe
we first have
\ie
    \sch_1B
    =
    2\rho B_w\,\rd w
    +
    2\rho B_{\bar w}\,\rd\bar w
    +
    2\alpha B_{\bar z}\,\rd\bar z \;.
\fe
Applying \eqref{Qflat} and multiplying by $\sch_0^{-1}$ gives
\ie
    Q_1^\dagger B
    =
    -\frac{2\bar a}{\alpha}\partial_{\bar w}B_w
    -\frac{2\bar b}{\alpha}\partial_wB_{\bar w}
    -\frac{2}{\alpha\rho}
    \partial_z(\alpha B_{\bar z}) \;.
\fe
Since
\ie
    \partial_z\log\alpha
    =
    \partial_z\log \mathcal{N}
    =
    \frac{\bar z}{\mathcal{N}} \;,
\fe
this becomes
\ie
    Q_1^\dagger B
    =
    -\frac{\bar z-1}{\mathcal{N}}\partial_{\bar w}B_w
    -\frac{\bar z+1}{\mathcal{N}}\partial_wB_{\bar w}
    -2\rho^{-1}
    \left(
    \partial_z+\frac{\bar z}{\mathcal{N}}
    \right)B_{\bar z} \;.
    \label{Q1dagger}
\fe

Next consider the two-form
\ie
    F
    =
    F_{w\bar w}\,\rd w\wedge\rd\bar w
    +
    F_{w\bar z}\,\rd w\wedge\rd\bar z
    +
    F_{\bar w\bar z}\,\rd\bar w\wedge\rd\bar z \;.
\fe
Using
\ie
    \sch_2F
    =
    4\frac{\rho}{\alpha}
    F_{w\bar w}\,\rd w\wedge\rd\bar w
    +
    4F_{w\bar z}\,\rd w\wedge\rd\bar z
    +
    4F_{\bar w\bar z}\,\rd\bar w\wedge\rd\bar z
\fe
and applying \eqref{weighted_adjoint}, we obtain
\ie
    (Q_2^\dagger F)_w
    =
    \frac{2\bar b}{\alpha}
    \partial_wF_{w\bar w}
    +
    2\rho^{-1}\partial_zF_{w\bar z} \;,
    \label{Q2dagger1}
\fe
\ie
    (Q_2^\dagger F)_{\bar w}
    =
    -\frac{2\bar a}{\alpha}
    \partial_{\bar w}F_{w\bar w}
    +
    2\rho^{-1}\partial_zF_{\bar w\bar z} \;,
    \label{Q2dagger2}
\fe
and
\ie
    (Q_2^\dagger F)_{\bar z}
    =
    -\frac{2\bar a}{\alpha}
    \partial_{\bar w}F_{w\bar z}
    -\frac{2\bar b}{\alpha}
    \partial_wF_{\bar w\bar z} \;.
    \label{Q2dagger3}
\fe
Since $\alpha=2\mathcal{N}$, the coefficients $2\bar a/\alpha$ and $2\bar b/\alpha$ equal $\bar a/\mathcal{N}=(\bar z-1)/\mathcal{N}$ and $\bar b/\mathcal{N}=(\bar z+1)/\mathcal{N}$, as in \eqref{Q1dagger} and \eqref{Q3dagger}.

Finally, consider the three-form
\ie
    G\Theta
    =
    G\,\rd w\wedge\rd\bar w\wedge\rd\bar z \;.
\fe
Since
\ie
    \sch_3(G\Theta)
    =
    \frac{8G}{\alpha}
    \rd w\wedge\rd\bar w\wedge\rd\bar z \;,
\fe
we find
\ie
    \begin{aligned}
        Q_3^\dagger(G\Theta)
        ={}&
        -2\rho^{-1}
        \left(
        \partial_z-\frac{\bar z}{\mathcal{N}}
        \right)G\,
        \rd w\wedge\rd\bar w
        \\
        &+
        \frac{\bar z+1}{\mathcal{N}}
        \partial_wG\,
        \rd w\wedge\rd\bar z
        \\
        &-
        \frac{\bar z-1}{\mathcal{N}}
        \partial_{\bar w}G\,
        \rd\bar w\wedge\rd\bar z \;.
    \end{aligned}
    \label{Q3dagger}
\fe

The factors $1 / \mathcal{N}=2 / \alpha$ in the mixed components of $Q^{\dagger}$ are inverse-metric factors, separate from the meromorphic field rescalings. For example, the degree-two adjoint contains
\begin{equation}
\left(Q_2^{\dagger}\right)_{\bar{z}, w \bar{z}}=-\frac{\bar{a}}{\mathcal{N}} \partial_{\bar{w}} \;, \quad\left(Q_2^{\dagger}\right)_{\bar{z}, \bar{w} \bar{z}}=-\frac{\bar{b}}{\mathcal{N}} \partial_w \;.
\end{equation}
These factors enter the vertex operators of the one-loop amplitude, see \eqref{Ffull:V0} and \eqref{Ffull:V1} in \cref{shorttimeheat}.

\section{Laplacians Associated with the Rescaled BRST Differential}
\label{app:Qlaplacian}

We now derive the Laplacians associated with the differential $Q$ and
the adjoints obtained in Appendix~\ref{app:Qadjoint}. The Laplacian on
degree $p$ is
\ie
    \Delta_p
    =
    Q_{p+1}^\dagger Q
    +
    QQ_p^\dagger \;,
\fe
with the appropriate term omitted at the ends of the complex.

We first consider a scalar function $f$. We have
\ie
    Qf
    =
    a\,\partial_wf\,\rd w
    +
    b\,\partial_{\bar w}f\,\rd\bar w
    +
    \partial_{\bar z}f\,\rd\bar z \;.
\fe
Applying \eqref{Q1dagger},
\ie
    \begin{aligned}
        \Delta_0 f
        ={}&
        -\frac{2|a|^2}{\alpha}
        \partial_{\bar w}\partial_w f
        -\frac{2|b|^2}{\alpha}
        \partial_w\partial_{\bar w}f
        -
        \frac{2}{\alpha\rho}
        \partial_z
        \left(
        \alpha\partial_{\bar z}f
        \right) \;.
    \end{aligned}
\fe
Using
$
    |a|^2+|b|^2
    =
    2\mathcal{N}
    =
    \alpha
    $, 
we obtain
\ie
    \Delta_0
    =
    -2\partial_w\partial_{\bar w}
    -
    2\rho^{-1}
    \left(
    \partial_z+\frac{\bar z}{\mathcal{N}}
    \right)
    \partial_{\bar z} \;.
    \label{scalar_laplacian}
\fe

The two principal second-order terms in \eqref{scalar_laplacian} have
the same normalization. For direct comparison with the ordinary
real Laplacian and with the standard heat kernel used in the main
text, it is convenient to introduce
\ie
    \widehat{\Delta}_p=2\Delta_p \;.
    \label{normalized_laplacian}
\fe
In particular,
\ie
    \widehat{\Delta}_{\Sigma}
    =
    -4\partial_w\partial_{\bar w}
    =
    -\partial_x^2-\partial_y^2 \;.
\fe
If $\widehat G_p=\widehat{\Delta}_p^{-1}$ and
$G_p=\Delta_p^{-1}$, then
\ie
    G_p=2\widehat G_p \;.
    \label{Green_rescaling}
\fe
Hence, when the propagator is written using the standard heat kernel
of $\widehat{\Delta}_p$, each internal propagator carries one overall
factor of $2$. In \cref{shorttimeheat}, these factors of $2$ are included in the
vertex operators.

We next derive the Laplacian on one-forms. For
\ie
    B
    =
    B_w\,\rd w
    +
    B_{\bar w}\,\rd\bar w
    +
    B_{\bar z}\,\rd\bar z \;,
\fe
the two-form $QB$ has components
\ie
    (QB)_{w\bar w}
    &=
    a\,\partial_wB_{\bar w}
    -
    b\,\partial_{\bar w}B_w \;,
    \\
    (QB)_{w\bar z}
    &=
    a\,\partial_wB_{\bar z}
    -
    \partial_{\bar z}B_w \;,
    \\
    (QB)_{\bar w\bar z}
    &=
    b\,\partial_{\bar w}B_{\bar z}
    -
    \partial_{\bar z}B_{\bar w} \;.
\fe
The one-form Laplacian is
\ie
    \Delta_1B
    =
    QQ_1^\dagger B
    +
    Q_2^\dagger QB \;.
\fe
Substituting \eqref{Q1dagger} and
\eqref{Q2dagger1}--\eqref{Q2dagger3}, and simplifying using
$|a|^2+|b|^2=\alpha$, gives
\ie
    \Delta_1
    =
    \begin{pmatrix}
        \mathcal{H}
        &
        0
        &
        \dfrac{2(\bar z+1)}{\mathcal{N}\rho}\partial_w
        \\[3mm]
        0
        &
        \mathcal{H}
        &
        -\dfrac{2(\bar z-1)}{\mathcal{N}\rho}\partial_{\bar w}
        \\[3mm]
        -\dfrac{z+1}{\mathcal{N}^2}\partial_{\bar w}
        &
        \dfrac{z-1}{\mathcal{N}^2}\partial_w
        &
        \mathcal{H}_{\bar z}
    \end{pmatrix} \;,
    \label{oneform_laplacian}
\fe
where
\ie
    \mathcal{H}
    =
    -2\partial_w\partial_{\bar w}
    -
    2\rho^{-1}\partial_z\partial_{\bar z} \;,
    \label{Hoperator}
\fe
and
\ie
    \mathcal{H}_{\bar z}
    =
    -2\partial_w\partial_{\bar w}
    -
    2\partial_{\bar z}
    \left[
    \rho^{-1}
    \left(
    \partial_z+\frac{\bar z}{\mathcal{N}}
    \right)
    \right] \;.
    \label{Hzoperator1}
\fe
Equivalently,
\ie
    \mathcal{H}_{\bar z}
    =
    -2\partial_w\partial_{\bar w}
    -
    2\rho^{-1}
    \left(
    \partial_{\bar z}
    -
    \partial_{\bar z}\log\rho
    \right)
    \left(
    \partial_z+\frac{\bar z}{\mathcal{N}}
    \right) \;.
    \label{Hzoperator2}
\fe

The structure of \eqref{oneform_laplacian} will be important below.
The diagonal blocks all have the same flat second-order operator
$-2\partial_w\partial_{\bar w}$ along $\Sigma$. After the common
normalization \eqref{normalized_laplacian}, this becomes the ordinary
real flat Laplacian $-4\partial_w\partial_{\bar w}$. Therefore the
Riemannian curvature of the principal metric has no component along $\Sigma$.
Note, however, that the bundle connection and its curvature do have
components along $\Sigma$: the off-diagonal first-order terms of $\Delta_1$
give the connection components $\mathcal W^{(1)}_w,\mathcal W^{(1)}_{\bar w}$
of \eqref{Ffull:connections}, whose curvature enters $\widehat E_1$ through
\eqref{Ffull:mixingcontribution}.

Along $C$, there are two further types of connection terms. First,
\ie
    \partial_z\log\alpha
    =
    \frac{\bar z}{\mathcal{N}} \;,
\fe
which is induced by the Hermitian weight associated with the warped
metric. Second,
\ie
    \partial_{\bar z}\log\rho \;,
\fe
which appears for the $C$-one-form component and is the usual
connection associated with the cotangent bundle of $C$.

\section{Short-Time  Asymptotics of 1-loop Amplitudes}
\label{shorttimeheat}

In this appendix we derive the two short-time expansions
\eqref{firstexpansion} and \eqref{secondexpansion} used in the main
text, with $\rho=1$ throughout. We expand the background-independent scalar and one-form
heat kernels, keeping the background fields in the interaction vertices.
The adjoint $Q^\dagger$ is the one derived in
Appendix~\ref{app:Qadjoint}, specialized to this choice of metric.

\subsection{Connections and Heat Coefficients}

We use the Laplacians of Appendix~\ref{app:Qlaplacian}, normalized by
$\widehat\Delta_p=2\Delta_p$, and write
\begin{equation}
 \mathcal{N}=1+|z|^2 \;,\qquad \chi=\frac{\bar z}{\mathcal{N}} \;.
 \label{Ffull:notation}
\end{equation}
The metric determined by the principal symbol is now flat:
\begin{equation}
 \widehat G=\rd w\,\rd\bar w+\rd z\,\rd\bar z \;,
 \qquad \rd\operatorname{vol}_{\widehat G}=\rd^4x \;,
 \qquad R=0 \;,
 \label{Ffull:metric}
\end{equation}
where $\rd^4x=\rd\xi^1\rd\xi^2\rd^2z$ and
$\rd^2z=(i/2)\rd z\wedge\rd\bar z$.
This principal metric differs from the auxiliary metric with
$\alpha=2\mathcal{N}$ used to define $Q^\dagger$. The lower-order connection
terms remain nontrivial.

Let $\mathcal W_\mu^{(p)}$ denote the connection of the
Laplace-type operator, so that
\begin{equation}
 \widehat\Delta_p=-\left[
 \widehat G^{\mu\nu}\nabla_\mu\nabla_\nu+\widehat E_p\right] \;,
 \qquad \nabla_\mu=\partial_\mu+\mathcal W_\mu^{(p)} \;.
 \label{Ffull:laplaceform}
\end{equation}
If the operator is written as
$-(\widehat G^{\mu\nu}\partial_\mu\partial_\nu+
\widehat A_p^\mu\partial_\mu+\widehat{\mathcal B}^{(p)})$,
completing the connection square gives
\begin{equation}
 \mathcal W_\nu^{(p)}=\frac12\widehat G_{\nu\mu}\widehat A_p^\mu \;,
 \qquad
 \widehat E_p=\widehat{\mathcal B}^{(p)}
 -\widehat G^{\mu\nu}
 \bigl(\partial_\mu\mathcal W_\nu^{(p)}
       +\mathcal W_\mu^{(p)}\mathcal W_\nu^{(p)}\bigr) \;.
 \label{Ffull:connectionE}
\end{equation}

For scalars the only nonzero connection component is
$\mathcal W_z^{(0)}=\chi$, and
$\widehat{\mathcal B}^{(0)}=0$. Hence
\begin{equation}
 \widehat E_0=-\frac2{\mathcal{N}^2} \;.
 \label{Ffull:E0}
\end{equation}
For one-forms, use the ordered basis
$(\rd w,\rd\bar w,\rd\bar z)$ and let $M_{ij}$ be the matrix
units. Reading all first-order terms in \eqref{oneform_laplacian} gives
\begin{equation}
 \begin{aligned}
 \mathcal W_w^{(1)}
 &=\frac{\bar z-1}{\mathcal{N}}M_{23}
       +\frac{z+1}{2\mathcal{N}^2}M_{31},&
 \mathcal W_{\bar w}^{(1)}
 &=-\frac{\bar z+1}{\mathcal{N}}M_{13}
       -\frac{z-1}{2\mathcal{N}^2}M_{32} \;,\\
 \mathcal W_z^{(1)}&=\chi M_{33},&
 \mathcal W_{\bar z}^{(1)}&=0 \;.
 \end{aligned}
 \label{Ffull:connections}
\end{equation}
The multiplication coefficient of the normalized operator is
$\widehat{\mathcal B}^{(1)}=4(\partial_{\bar z}\chi)M_{33}$.
In particular, the off-diagonal connections contribute
\begin{equation}
 -2\{\mathcal W_w^{(1)},\mathcal W_{\bar w}^{(1)}\}
 =\operatorname{diag}\left(
 \frac{|z+1|^2}{\mathcal{N}^3},\frac{|z-1|^2}{\mathcal{N}^3},\frac2{\mathcal{N}^2}\right) \;.
 \label{Ffull:mixingcontribution}
\end{equation}
Using $\partial_{\bar z}\chi=\mathcal{N}^{-2}$, the third entry is therefore
$(\widehat E_1)_{33}=4/\mathcal{N}^2-2/\mathcal{N}^2+2/\mathcal{N}^2=4/\mathcal{N}^2$.
Thus the full endomorphism is
\begin{equation}
 \widehat E_1=\operatorname{diag}\left(
 \frac{|z+1|^2}{\mathcal{N}^3},\frac{|z-1|^2}{\mathcal{N}^3},\frac4{\mathcal{N}^2}\right) \;.
 \label{Ffull:E1}
\end{equation}

The heat kernel, normalized against the coordinate measure $\rd^4y$ at the second point, has the local expansion
\begin{equation}
 K_p(t;x,y)\sim
 \frac{e^{-|x-y|^2/(4t)}}{(4\pi t)^2}
 \bigl[U_p(x,y)+tU_{1,p}(x,y)+\cdots\bigr] \;,
 \label{Ffull:parametrix}
\end{equation}
where $U_p$ is parallel transport from $y$ to $x$ for
$\mathcal W^{(p)}$. At coincidence $U_p(x,x)=\mathbf1$.
The coefficient underlying $a_2$ in Eq.~(4.27) of
\cite{Vassilevich:2003xt} is
\begin{equation}
 b_{2,p}:=U_{1,p}(x,x)=\widehat E_p+\frac{R}{6}\mathbf1
 =\widehat E_p \;.
 \label{Ffull:diagonal}
\end{equation}
Explicitly,
\begin{equation}
 b_{2,0}=-\frac2{\mathcal{N}^2} \;,\qquad
 b_{2,1}=\operatorname{diag}\left(
 \frac{|z+1|^2}{\mathcal{N}^3},\frac{|z-1|^2}{\mathcal{N}^3},\frac4{\mathcal{N}^2}\right) \;.
 \label{Ffull:b2}
\end{equation}

\subsection{Background Insertions and the Two-kernel Integrals}

After the rescalings and the cancellation of meromorphic factors
described in \eqref{cancellationn}, the background insertion is
\begin{equation}
 \mathcal A=e\,\rd w\wedge+f\,\rd\bar w\wedge \;,
 \label{Ffull:A}
\end{equation}
with $e,f$ as in \eqref{main:hk:currentinsertion}.
As in the main text, solutions of the PCM are chosen such that the currents are constant, in order to extract the local quadratic term without
their derivatives. (This choice is sufficient to determine the coefficient: a logarithmically divergent local term quadratic in the background has dimension two and hence contains no derivatives, so it is a combination of $\operatorname{Tr}(J_wJ_{\bar w})$, $\operatorname{Tr}(J_wJ_w)$ and $\operatorname{Tr}(J_{\bar w}J_{\bar w})$. The last two have spin $\pm2$ and are excluded by the rotation symmetry of $\Sigma$, and the coefficient of the first is an invariant tensor of $\mathfrak g$, hence proportional to $\delta_{ab}$ for simple $\mathfrak g$. It can therefore be computed on constant commuting backgrounds.) On such backgrounds we treat $e,f$ as scalars on each root space.
The notation $[\cdots]_{ef}$ below extracts their bilinear coefficient,
including both insertion orders. The colour sum is restored using
\begin{equation}
 \operatorname{tr}_{\rm ad}(ef)
 =\mathsf h^\vee\operatorname{Tr}(J_wJ_{\bar w}) \;.
\end{equation}

Each normalized propagator contains
$2Q^\dagger\int K(t)\rd t$. Using
$Q_{p+1}^\dagger K_{p+1}=K_pQ_{p+1}^\dagger$ and the cyclic
contraction of the form indices around the loop, the insertions can
be written as
\begin{equation}
 2Q_{p+1}^\dagger\circ\mathcal A
 =\mathcal V_p^\mu\partial_\mu \;.
 \label{Ffull:vertexoperator}
\end{equation}
This is a composition of operators: it sends a field $b$ to
$2Q_{p+1}^\dagger(\mathcal A b)$. Thus its derivatives act on the
propagator in the loop. The constant rescaled background has no
derivative, and no zeroth-order vertex remains.

For the ghost sector,
\begin{equation}
 \mathcal V_0^w=-\frac{2(\bar z+1)f}{\mathcal{N}} \;,\qquad
 \mathcal V_0^{\bar w}=-\frac{2(\bar z-1)e}{\mathcal{N}} \;,\qquad
 \mathcal V_0^z=\mathcal V_0^{\bar z}=0 \;.
 \label{Ffull:V0}
\end{equation}
For one-forms,
\begin{equation}
 \begin{aligned}
 \mathcal V_1^w&=\frac{2(\bar z+1)}{\mathcal{N}}
 \begin{pmatrix}-f&e&0\\0&0&0\\0&0&-f\end{pmatrix},&
 \mathcal V_1^{\bar w}&=\frac{2(\bar z-1)}{\mathcal{N}}
 \begin{pmatrix}0&0&0\\f&-e&0\\0&0&-e\end{pmatrix} \;,\\
 \mathcal V_1^z&=4
 \begin{pmatrix}0&0&e\\0&0&f\\0&0&0\end{pmatrix},&
 \mathcal V_1^{\bar z}&=0 \;.
 \end{aligned}
 \label{Ffull:V1}
\end{equation}
In particular, $\partial_z$ occurs because $Q_2^\dagger$ contains
$2\partial_zF_{w\bar z}$ and
$2\partial_zF_{\bar w\bar z}$, even though the background
has only $w$ and $\bar w$ components.

With $\rho=1$, the geometric measure of $\widehat G$ and the coordinate measure coincide, so that we need not distinguish the corresponding normalizations of $K_p$.
The separation integrals are precisely
\begin{equation}
 \begin{aligned}
 \mathcal I_p(t,s;z)
 =\int\rd^4x_2\Big[\operatorname{tr}_{\rm form}\bigl(
 &\mathcal V_p^\mu(x_1)\partial_{1\mu}K_p(t;x_1,x_2)\\[-2pt]
 &\mathcal V_p^\nu(x_2)\partial_{2\nu}K_p(s;x_2,x_1)
 \bigr)\Big]_{ef} \;,\qquad x_1=(w,z) \;.
 \end{aligned}
 \label{Ffull:bubble}
\end{equation}
Both derivatives act on the first argument of their respective kernels.

The full matrix contractions give the statistical combination
$\tfrac14\mathcal I_1-\tfrac12\mathcal I_0$.
Writing the interactions as
$I_B=\tfrac12 B^T H_BB$ and $I_{\rm gh}=\bar cH_{\rm gh}c$,
the connected second-order Wick contractions have coefficients
$\tfrac12(\tfrac12)^2\times2=\tfrac14$ and $-\tfrac12$,
respectively. These factors include the symmetry factors of the
component amplitudes in the main text, and the extraction of $ef$ already
includes both external orderings.

\subsection{The Two-kernel Separation Integral at Fixed Proper-time Ratio}
\label{Ffull:gaussian-derivation}

The calculation consists of expanding the two differentiated heat kernels
and the second vertex in \eqref{Ffull:bubble}, multiplying these expansions,
and integrating the separation over a copy of $\Sigma \times C^0$. We use $p=1$ for the boson and $p=0$ for
the ghost. All coefficient matrices below are evaluated at $x_1=(w,z)$,
and we write $x_2=x_1+r$.

\paragraph{Expanding the heat kernels.}
Put
\begin{equation}
 T=t+s \;,\qquad u=\frac{t}{T} \;,\qquad v=\frac{ts}{T} \;,\qquad
 G_t(r)=\frac{e^{-|r|^2/(4t)}}{(4\pi t)^2} \;.
 \label{Fdirect:times}
\end{equation}
The Gaussian localizes $r$ to order $\sqrt T$ at fixed $0<u<1$.
For the connection $\nabla_i=\partial_i+\mathcal W_i$, the parallel
transport from $y$ to $x$ is
\begin{equation}
 U(x,y)=\mathcal P\exp\left[-\int_0^1\rd a\,
 (x-y)^i\mathcal W_i\bigl(y+a(x-y)\bigr)\right] \;.
 \label{Ffull:transport-exponential}
\end{equation}
Expanding this expression about $x_1$ gives
\begin{equation}
 \begin{aligned}
 U(x_1,x_2)&=I+\mathcal W_i r^i+\frac12P^+_{ij}r^ir^j
   +\frac16C^+_{ijk}r^ir^jr^k+O(|r|^4) \;,\\
 U(x_2,x_1)&=I-\mathcal W_i r^i+\frac12P^-_{ij}r^ir^j
   +\frac16C^-_{ijk}r^ir^jr^k+O(|r|^4) \;.
 \end{aligned}
 \label{Ffull:transport-expansion}
\end{equation}
where
\begin{equation}
 P^\pm_{ij}=\frac12\left(
 \mathcal W_i\mathcal W_j+\mathcal W_j\mathcal W_i
 \pm\partial_i\mathcal W_j\pm\partial_j\mathcal W_i\right) \;,
 \qquad D_{\mu j}=\partial_\mu\mathcal W_j-P^+_{\mu j} \;.
 \label{Ffull:transport-tensors}
\end{equation}
The cubic coefficients can be written in terms of the same connection as
\begin{equation}
 \begin{aligned}
 C^+_{ijk}&=\operatorname{Sym}_{ijk}\left(
 \partial_iP^+_{jk}+\mathcal W_iP^+_{jk}\right) \;,\\
 C^-_{ijk}&=\operatorname{Sym}_{ijk}\left(
 \partial_iP^-_{jk}-P^-_{jk}\mathcal W_i\right) \;.
 \end{aligned}
 \label{Fdirect:cubic-transport}
\end{equation}
Here $\operatorname{Sym}_{ijk}$ averages over the six permutations of the
indices; the order of each matrix product is retained. The degree label
on $\mathcal W$ is suppressed. We keep these cubic terms to display the
complete quadratic endpoint derivatives below, but they will eventually not contribute to the final integral. The two heat kernels
needed in the loop have the expansions
\begin{equation}
 \begin{aligned}
 \frac{K_p(t;x_1,x_2)}{G_t(r)}
 &=I+\mathcal W_i r^i+\frac12P^+_{ij}r^ir^j
   +\frac16C^+_{ijk}r^ir^jr^k+t b_{2,p}
   +\mathcal R_{p,t} \;,\\
 \frac{K_p(s;x_2,x_1)}{G_s(r)}
 &=I-\mathcal W_i r^i+\frac12P^-_{ij}r^ir^j
   +\frac16C^-_{ijk}r^ir^jr^k+s b_{2,p}
   +\widetilde{\mathcal R}_{p,s} \;.
 \end{aligned}
 \label{Fdirect:two-kernels}
\end{equation}
Here $\mathcal R_{p,t}=O(|r|^4)+O(t|r|)+O(t^2)$, with the analogous
estimate for the reverse kernel. In particular, the separation dependence
of the coefficient multiplying $t$ is included in $\mathcal R_{p,t}$.
In addition, the diagonal coefficients $b_{2,p}$ are given in \eqref{Ffull:b2}.
The one-form expansion is a matrix expansion, using all the connection
components in \eqref{Ffull:connections}. The ghost expansion is scalar,
with only $\mathcal W_z^{(0)}=\bar z/\mathcal{N}$ nonzero.

\paragraph{Differentiating the two kernels.}
 Define
\begin{equation}
 \lambda_w=\frac{r^{\bar w}}4 \;,\quad
 \lambda_{\bar w}=\frac{r^w}4 \;,\quad
 \lambda_z=\frac{r^{\bar z}}4 \;,\quad
 \lambda_{\bar z}=\frac{r^z}4 \;.
\end{equation}
Both derivatives in the loop act on the first argument of their respective
kernel. Since $r=x_2-x_1$, the first acts as $\partial_{x_1^\mu}-\partial_{r^\mu}$
on an expression anchored at $x_1$, whereas the second acts as
$\partial_{r^\nu}$. Consequently,
\begin{equation}
 \begin{aligned}
 \partial_{1\mu}U(x_1,x_2)
 &=-\mathcal W_\mu+D_{\mu j}r^j
   +\frac12\left(\partial_\mu P^+_{ij}-C^+_{\mu ij}\right)r^ir^j
   +O(|r|^3) \;,\\
 \partial_{2\nu}U(x_2,x_1)
 &=-\mathcal W_\nu+P^-_{\nu j}r^j
   +\frac12C^-_{\nu ij}r^ir^j+O(|r|^3) \;.
 \end{aligned}
 \label{Ffull:transport-endpoint-derivatives}
\end{equation}

The differentials of the heat kernels entering the one-loop amplitude are thus
\begin{equation}
 \begin{aligned}
 \frac{\partial_{1\mu}K_p(t;x_1,x_2)}{G_t}
 ={}&\frac{\lambda_\mu}{t}
 \left(I+\mathcal W_i r^i+\frac12P^+_{ij}r^ir^j+t b_{2,p}\right)
 -\mathcal W_\mu+D_{\mu j}r^j\\
 &+\frac{\lambda_\mu}{6t}C^+_{ijk}r^ir^jr^k
 +\frac12\left(\partial_\mu P^+_{ij}-C^+_{\mu ij}\right)r^ir^j
 +t\partial_\mu b_{2,p}\\
 &+\frac16(\partial_\mu C^+_{ijk})r^ir^jr^k
 +\frac{\lambda_\mu}{t}\mathcal R_{p,t}
 +\partial_{1\mu}\mathcal R_{p,t} \;,\\[2pt]
 \frac{\partial_{2\nu}K_p(s;x_2,x_1)}{G_s}
 ={}&-\frac{\lambda_\nu}{s}
 \left(I-\mathcal W_i r^i+\frac12P^-_{ij}r^ir^j+s b_{2,p}\right)
 -\mathcal W_\nu+P^-_{\nu j}r^j\\
 &-\frac{\lambda_\nu}{6s}C^-_{ijk}r^ir^jr^k
 +\frac12C^-_{\nu ij}r^ir^j
 -\frac{\lambda_\nu}{s}\widetilde{\mathcal R}_{p,s}
 +\partial_{2\nu}\widetilde{\mathcal R}_{p,s} \;.
 \end{aligned}
 \label{Ffull:differentiated-parametrices}
\end{equation}
We can now substitute these expressions into the one-loop integral and multiply them, and eventually proceed to integrate over $\Sigma \times C^0$.

The cubic transport terms in \eqref{Ffull:differentiated-parametrices},
together with the quadratic transport derivatives, are of order $T$ under $r\sim\sqrt T$ at fixed
$u$. Their leading products with the $T^{-1/2}$ derivative of the other
kernel are of order $\sqrt T$ and odd in $r$, so their Gaussian integrals
vanish. The remainders $\mathcal R_{p,t}$ and $\widetilde{\mathcal R}_{p,s}$, which arise from the expansion of the
vertex factors and the parallel transports, likewise first appear at order
$\sqrt T$ with odd parity in $r$, so their Gaussian integrals also vanish. 
The first even corrections are of order $T$ in the
Gaussian-averaged product, beyond the logarithmic coefficient that is our goal.

To multiply the heat kernel derivative factors, we group the terms that can contribute as
\begin{equation}
 \begin{aligned}
 \frac{\partial_{1\mu}K_p}{G_t}
 &=H_{\mu,-1}+H_{\mu,0}+H_{\mu,1}+O(T) \;,\\
 \frac{\partial_{2\nu}K_p}{G_s}
 &=\widetilde H_{\nu,-1}+\widetilde H_{\nu,0}
  +\widetilde H_{\nu,1}+O(T) \;.
 \end{aligned}
 \label{Fdirect:H-expansion}
\end{equation}
where the subscript $k$ denotes order $T^{k/2}$, and
\begin{equation}
 \begin{aligned}
 H_{\mu,-1}&=\frac{\lambda_\mu}{t}I \;,\\
 H_{\mu,0}&=\frac{\lambda_\mu}{t}\mathcal W_i r^i-\mathcal W_\mu \;,\\
 H_{\mu,1}&=\frac{\lambda_\mu}{t}
 \left(\frac12P^+_{ij}r^ir^j+t b_{2,p}\right)+D_{\mu j}r^j \;,\\
 \widetilde H_{\nu,-1}&=-\frac{\lambda_\nu}{s}I \;,\\
 \widetilde H_{\nu,0}&=\frac{\lambda_\nu}{s}\mathcal W_i r^i-\mathcal W_\nu \;,\\
 \widetilde H_{\nu,1}&=-\frac{\lambda_\nu}{s}
 \left(\frac12P^-_{ij}r^ir^j+s b_{2,p}\right)+P^-_{\nu j}r^j \;.
 \end{aligned}
 \label{Fdirect:H-pieces}
\end{equation}
These symbols are matrices for $p=1$ and scalar functions for $p=0$.

\paragraph{Expanding the vertex and multiplying.}
Before proceeding to integrate over $\Sigma \times C^0$, we need to also expand the second vertex located at $x_2$.
The first vertex is already evaluated at $x_1$. For the second one write
\begin{equation}
 \begin{aligned}
 \mathcal V_p^\nu(x_2)
 &=\mathcal V_p^\nu+\mathcal V_{p,[1]}^\nu+\mathcal V_{p,[2]}^\nu
   +O(|r|^3) \;,\\
 \mathcal V_{p,[1]}^\nu&=r^i\partial_i\mathcal V_p^\nu \;,\qquad
 \mathcal V_{p,[2]}^\nu=\frac12r^ir^j\partial_i\partial_j\mathcal V_p^\nu \;.
 \end{aligned}
 \label{Ffull:vertex-separation}
\end{equation}
In particular, this expands the factors $(\bar z\pm1)/\mathcal{N}$, including their
denominators.

The product in \eqref{Ffull:bubble} is now a multiplication of
the three expansions we have derived, with the unexpanded first vertex on the left.
Write $\mathfrak t(M)=[\operatorname{tr}_{\rm form}M]_{ef}$ as before;
derivative indices are summed. The leading term in the short-time expansion is
\begin{equation}
 \mathcal P_{p,-2}
 =\mathfrak t\left(
 \mathcal V_p^\mu H_{\mu,-1}\mathcal V_p^\nu\widetilde H_{\nu,-1}\right) \;.
\end{equation}
At order $T^{-1/2}$ the three products are
\begin{equation}
 \mathcal P_{p,-1}=\mathfrak t\left[\mathcal V_p^\mu\left(
 H_{\mu,0}\mathcal V_p^\nu\widetilde H_{\nu,-1}
 +H_{\mu,-1}\mathcal V_p^\nu\widetilde H_{\nu,0}
 +H_{\mu,-1}\mathcal V_{p,[1]}^\nu\widetilde H_{\nu,-1}
 \right)\right] \;.
 \label{Fdirect:odd-product}
\end{equation}
Each is odd in $r$, so this order integrates to zero separately in each
sector. At order $T^0$, there are exactly six products:
\begin{equation}
 \begin{aligned}
 \mathcal P_{p,0}=\mathfrak t\Bigl[\mathcal V_p^\mu\Bigl(
 &H_{\mu,1}\mathcal V_p^\nu\widetilde H_{\nu,-1}
 +H_{\mu,0}\mathcal V_p^\nu\widetilde H_{\nu,0}
 +H_{\mu,-1}\mathcal V_p^\nu\widetilde H_{\nu,1}\\
 &+H_{\mu,0}\mathcal V_{p,[1]}^\nu\widetilde H_{\nu,-1}
 +H_{\mu,-1}\mathcal V_{p,[1]}^\nu\widetilde H_{\nu,0}
 +H_{\mu,-1}\mathcal V_{p,[2]}^\nu\widetilde H_{\nu,-1}
 \Bigr)\Bigr] \;.
 \end{aligned}
 \label{Fdirect:six-products}
\end{equation}
The first three keep the vertex fixed and correct one or both kernel
derivatives. The next two use the linear vertex expansion together with
one order-$T^0$ kernel term. The last uses the quadratic vertex expansion
and the two leading kernel derivatives.

\paragraph{Integrating the separation.}
The product of Gaussians and their derivatives satisfies
\begin{equation}
 G_tG_s=G_T(0)G_v \;,\qquad
 \partial_{1\mu}G_t=\frac{\lambda_\mu}{t}G_t \;,\qquad
 \partial_{2\nu}G_s=-\frac{\lambda_\nu}{s}G_s \;.
 \label{Ffull:gaussian-product-fixedu}
\end{equation}
Thus the remaining separation integral is an ordinary Gaussian average,
\begin{equation}
 \langle F\rangle_v=\int\rd^4r\,G_v(r)F(r) \;.
\end{equation}
Let $\mathsf J_{ij}=1$ for the conjugate pairs $(w,\bar w)$,
$(\bar w,w)$, $(z,\bar z)$, $(\bar z,z)$, and zero otherwise.
Then, its moments are
\begin{equation}
 \begin{aligned}
 \langle r^ir^j\rangle_v&=4v\mathsf J_{ij},&
 \langle\lambda_\mu r^i\rangle_v&=v\delta_\mu^i,&
 \langle\lambda_\mu\lambda_\nu\rangle_v&=\frac v4\mathsf J_{\mu\nu} \;,\\
 \langle\lambda_\mu\lambda_\nu r^ir^j\rangle_v
 &=v^2\left(\mathsf J_{\mu\nu}\mathsf J_{ij}
 +\delta_\mu^i\delta_\nu^j+\delta_\mu^j\delta_\nu^i\right) \;.
 \end{aligned}
 \label{Ffull:gaussian-moments-fixedu}
\end{equation}
For example, the terms in the first and third products of
\eqref{Fdirect:six-products} that contain $b_{2,p}$ give
\begin{equation}
 -\frac14\sum_{\mu,\nu}\mathsf J_{\mu\nu}\,
 \mathfrak t\left[\mathcal V_p^\mu\left(
 u b_{2,p}\mathcal V_p^\nu
 +(1-u)\mathcal V_p^\nu b_{2,p}\right)\right] \;.
 \label{Fdirect:heat-term}
\end{equation}
The last product, containing the quadratic vertex, gives
\begin{equation}
 -\frac{u(1-u)}2\sum_{\mu,\nu,i,j}
 \left(\mathsf J_{\mu\nu}\mathsf J_{ij}
 +\delta_\mu^i\delta_\nu^j+\delta_\mu^j\delta_\nu^i\right)
 \mathfrak t\left(\mathcal V_p^\mu\partial_i\partial_j\mathcal V_p^\nu\right) \;.
 \label{Fdirect:quadratic-vertex-average}
\end{equation}
Substituting the one-form vertex matrices into this expression yields
$16u(1-u)(2\bar z^2+|z|^2-1)/\mathcal{N}^4$. The other five products are integrated
using the same second and fourth moments. All components of the one-form
matrices are multiplied before taking the trace. More explicitly, for the
bosonic sector the six Gaussian averages, in the order in
\eqref{Fdirect:six-products}, give
\begin{equation}
 \begin{aligned}
 \mathcal L_1=\frac8{\mathcal{N}^4}\Bigl[\,
 &u(1+u)(1-\bar z^2)
 -2u(1-u)(3\bar z^2+|z|^2-2)\\
 &-(1-u)^2(4\bar z^2+|z|^2-3)
 -\bar z u(1-u)(z+\bar z)\\
 &+\bar z(1-u)(2u-1)(z+\bar z)
 +2u(1-u)(2\bar z^2+|z|^2-1)\,\Bigr] \;.
 \end{aligned}
 \label{Fdirect:six-bosonic-averages}
\end{equation}

The result takes the form
\begin{equation}
 \mathcal I_p(t,s;z)=G_T(0)\left[-\frac{S_p}{4T}
 +\mathcal L_p(u;z)+O(T)\right] \;,\qquad
 \mathcal L_p=\langle\mathcal P_{p,0}\rangle_v \;,
 \label{Fdirect:bubble-expansion}
\end{equation}
where
\begin{equation}
 S_p=\sum_{\mu,\nu}\mathsf J_{\mu\nu}
 \mathfrak t(\mathcal V_p^\mu\mathcal V_p^\nu) \;,\qquad
 S_0=\frac{8(\bar z^2-1)}{\mathcal{N}^2} \;,\qquad S_1=2S_0 \;.
 \label{Ffull:leading-contractions}
\end{equation}
The leading $1/T$ terms therefore cancel in
$\mathcal I_1/4-\mathcal I_0/2$. Applying the moments to the six products
gives, after simplifying \eqref{Fdirect:six-bosonic-averages},
\begin{equation}
 \begin{aligned}
 \mathcal L_1(u;z)&=\frac8{\mathcal{N}^4}\left[
 (1+u)(1-\bar z^2)-2(1-u)^2(2\bar z^2+|z|^2-1)\right] \;,\\
 \mathcal L_0(u;z)&=\frac4{\mathcal{N}^4}\left[
 \bar z^2-1+2u(1-u)(2\bar z^2+|z|^2-1)\right] \;.
 \end{aligned}
 \label{Ffull:fixedu-flat-coefficients}
\end{equation}

We have derived the boson and ghost contributions in tandem, but the ghost contribution is in fact simpler. This can be attributed to the fact that unlike the 1-form Laplacian, the scalar Laplacian decomposes cleanly into Laplacians for $\Sigma$ and $C^0$
\ie 
\widehat{\Delta}_0=\underbrace{-4 \partial_w \partial_{\bar{w}}}_{\Delta_{\Sigma}}+\underbrace{-4\left(\partial_z+\frac{\bar{z}}{\mathcal{N}}\right) \partial_{\bar{z}}}_{\Delta_C} \;,
\fe
and therefore the corresponding heat kernel is factorizable, that is,
\ie 
K_0\left(t ; x_1, x_2\right)=\underbrace{\frac{1}{4 \pi t} e^{-\left|w_1-w_2\right|^2 /(4 t)}}_{G_t^{\Sigma}\left(w_1, w_2\right)} k_0^C\left(t ; z_1, z_2\right) \;.
\fe
The degree-zero vertex operators in \eqref{Ffull:V0} differentiate
only in $w,\bar w$. Since the kernel $k_0^C$ on the spectral-parameter curve is independent of
these coordinates, these derivatives act only on $G_t^\Sigma$.
Equivalently, $\mathcal W_w^{(0)}=\mathcal W_{\bar w}^{(0)}=0$,
so there are no transport derivatives in these directions.
The surviving corrections are the heat coefficient that comes entirely from the short-time asymptotics of the scalar heat kernel along $C^0$, and the second
derivative of the second vertex:
\begin{equation}
 \mathcal L_0=-\frac{b_{2,0}}4 S_0
 -u(1-u)\sum_{\mu,\nu}\mathsf J_{\mu\nu}
 \bigl[\mathcal V_0^\mu\partial_z\partial_{\bar z}
                  \mathcal V_0^\nu\bigr]_{ef} \;.
 \label{Fdirect:ghost-reduction}
\end{equation}
For the vertex factors in \eqref{Ffull:V0},
\begin{equation}
 \begin{aligned}
 \partial_z\partial_{\bar z}\frac{\bar z\pm1}{\mathcal{N}}
 &=\frac{-2\bar z\pm(|z|^2-1)}{\mathcal{N}^3} \;,\\
 \sum_{\mu,\nu}\mathsf J_{\mu\nu}
 \bigl[\mathcal V_0^\mu\partial_z\partial_{\bar z}
                  \mathcal V_0^\nu\bigr]_{ef}
 &=-\frac{8(2\bar z^2+|z|^2-1)}{\mathcal{N}^4} \;.
 \end{aligned}
\end{equation}
Thus \eqref{Fdirect:ghost-reduction} gives the two terms in
$\mathcal L_0$ directly.

Together with \eqref{Ffull:leading-contractions}, the coefficients in
\eqref{Ffull:fixedu-flat-coefficients} give exactly
\eqref{firstexpansion} and \eqref{secondexpansion}. Their weighted
combination is \eqref{sum_Rc}; the remaining integrals over the spectral-parameter
curve and the proper times are evaluated in \cref{sec:oneloopPCM}.

\section{Background-Coupled Laplacians}
\label{app:bg}

In this appendix, we extend the computations of Appendices~\ref{app:Qadjoint} and~\ref{app:Qlaplacian} to the
rescaled BRST differential coupled to a constant commuting PCM background.
We derive the scalar and one-form Laplacians, and collect the corresponding
connections, endomorphisms and curvatures which enter the local heat-kernel
coefficient. We shall use the background-dependent adjoint throughout. Thus
both the differential and its adjoint contain the background.

As in \cref{app:Qadjoint}, we use the coordinate measure $d\mu_0=d^2w\,d^2z$
and write
\begin{equation}
 \langle X,Y\rangle_p=\int d\mu_0\,X^\dagger \sch_pY \;.
 \label{app:bg:pairing}
\end{equation}
In the ordered bases
\begin{equation}
 \begin{aligned}
 \Omega^1:\;&(dw,d\bar w,d\bar z) \;,\\
 \Omega^2:\;&(dw\wedge d\bar w,dw\wedge d\bar z,
                         d\bar w\wedge d\bar z) \;,
 \end{aligned}
 \label{app:bg:bases}
\end{equation}
the weights are
\begin{equation}
 \sch_0=2\mathcal{N} \;,\qquad \sch_1=2\operatorname{diag}(1,1,2\mathcal{N}) \;,
 \qquad \sch_2=\operatorname{diag}(2/\mathcal{N},4,4) \;,\qquad \sch_3=4/\mathcal{N} \;.
 \label{app:bg:weights}
\end{equation}
It is useful to introduce the flat-measure adjoints
\begin{equation}
 R_w=-\bar a\partial_{\bar w}+f \;,\qquad
 R_{\bar w}=-\bar b\partial_w+e \;.
 \label{app:bg:r}
\end{equation}
Integration by parts, with vanishing boundary terms, gives
\begin{equation}
 Q_J^\flat=R_w\iota_w+R_{\bar w}\iota_{\bar w}
                         -\partial_z\iota_{\bar z} \;.
 \label{app:bg:flatadj}
\end{equation}
Consequently, the weighted adjoint on degree $p$ is
\begin{equation}
 Q^\dagger_{J,p}=\sch_{p-1}^{-1}Q_J^\flat \sch_p:
                 \Omega^p\longrightarrow\Omega^{p-1} \;.
 \label{app:bg:weightedadj}
\end{equation}
The derivative in $Q_J^\flat$ acts on the weight as well as on the field.
On a one-form $B=B_wdw+B_{\bar w}d\bar w+B_{\bar z}d\bar z$, this gives
\begin{equation}
 Q^\dagger_{J,1}B
 =\frac{R_w}{\mathcal{N}}B_w+\frac{R_{\bar w}}{\mathcal{N}}B_{\bar w}
       -2\left(\partial_z+\frac{\bar z}{\mathcal{N}}\right)B_{\bar z} \;.
 \label{app:bg:adj1}
\end{equation}
For a two-form $F$ with components ordered as in
\eqref{app:bg:bases}, we similarly obtain
\begin{equation}
 \begin{aligned}
 (Q^\dagger_{J,2}F)_w
    &=-\frac{R_{\bar w}}{\mathcal{N}} F_{w\bar w}+2\partial_zF_{w\bar z} \;,\\
 (Q^\dagger_{J,2}F)_{\bar w}
    &=\frac{R_w}{\mathcal{N}} F_{w\bar w}+2\partial_zF_{\bar w\bar z} \;,\\
 (Q^\dagger_{J,2}F)_{\bar z}
    &=\frac{R_w}{\mathcal{N}} F_{w\bar z}+\frac{R_{\bar w}}{\mathcal{N}} F_{\bar w\bar z} \;.
 \end{aligned}
 \label{app:bg:adj2}
\end{equation}
For example, the term $-\partial_{\bar z}B_w$ in
$(Q_JB)_{w\bar z}$ contributes
\begin{equation}
 -\int d\mu_0\,4(\partial_z\bar B_w)F_{w\bar z}
 =\int d\mu_0\,4\bar B_w\partial_zF_{w\bar z} \;.
 \label{app:bg:ibp}
\end{equation}
Dividing by the one-form weight $(H_1)_{ww}=2$ gives the
$2\partial_zF_{w\bar z}$ term in \eqref{app:bg:adj2}.

We now compute the Laplacians
\begin{equation}
 \Delta_p(J)=Q^\dagger_{J,p+1}Q_J+Q_JQ^\dagger_{J,p} \;.
 \label{app:bg:lapdef}
\end{equation}
For a scalar $u$, only the first term contributes, and
\begin{equation}
 \Delta_0(J)u
 =\frac{R_wD_w+R_{\bar w}D_{\bar w}}{\mathcal{N}} u
       -2\left(\partial_z+\frac{\bar z}{\mathcal{N}}\right)\partial_{\bar z}u \;.
 \label{app:bg:scalarcomp}
\end{equation}
Using $|a|^2+|b|^2=2\mathcal{N}$, define
\begin{equation}
 \begin{aligned}
 \mathcal H_J={}&-2\partial_w\partial_{\bar w}
                    -2\partial_z\partial_{\bar z}\\
 &+\frac{f(a-\bar b)}{\mathcal{N}}\partial_w
       +\frac{e(b-\bar a)}{\mathcal{N}}\partial_{\bar w}
       +\frac{2ef}{\mathcal{N}} \;.
 \end{aligned}
 \label{app:bg:h}
\end{equation}
Then
\begin{equation}
 \Delta_0(J)=\mathcal H_J-\frac{2\bar z}{\mathcal{N}}\partial_{\bar z} \;.
 \label{app:bg:lap0}
\end{equation}
In particular, the background contributes both first-order terms and the
zeroth-order term $2ef/\mathcal{N}$.

For a one-form, the two-form $Q_JB$ has components
\begin{equation}
 \begin{aligned}
 (Q_JB)_{w\bar w}&=D_wB_{\bar w}-D_{\bar w}B_w \;,\\
 (Q_JB)_{w\bar z}&=D_wB_{\bar z}-\partial_{\bar z}B_w \;,\\
 (Q_JB)_{\bar w\bar z}&=D_{\bar w}B_{\bar z}
                                  -\partial_{\bar z}B_{\bar w} \;.
 \end{aligned}
 \label{app:bg:q1}
\end{equation}
Both terms in $\Delta_1(J)B=Q_JQ^\dagger_{J,1}B+
Q^\dagger_{J,2}Q_JB$ must be retained. Substituting
\eqref{app:bg:adj1}, \eqref{app:bg:adj2} and \eqref{app:bg:q1}, we find
\begin{equation}
 \Delta_1(J)=
 \begin{pmatrix}
 \mathcal H_J&0&X_w\\
 0&\mathcal H_J&X_{\bar w}\\
 Y_w&Y_{\bar w}&
 \mathcal H_J-\dfrac{2\bar z}{\mathcal{N}}\partial_{\bar z}-\dfrac2{\mathcal{N}^2}
 \end{pmatrix} \;,
 \label{app:bg:lap1}
\end{equation}
where
\begin{equation}
 \begin{aligned}
 X_w&=\frac{2\bar b}{\mathcal{N}}\partial_w-\frac{2\bar z e}{\mathcal{N}} \;,
 &X_{\bar w}&=-\frac{2\bar a}{\mathcal{N}}\partial_{\bar w}
                                      -\frac{2\bar z f}{\mathcal{N}} \;,\\
 Y_w&=-\frac b{\mathcal{N}^2}\partial_{\bar w}-\frac{zf}{\mathcal{N}^2} \;,
 &Y_{\bar w}&=\frac a{\mathcal{N}^2}\partial_w-\frac{ze}{\mathcal{N}^2} \;.
 \end{aligned}
 \label{app:bg:mixing}
\end{equation}
Every derivative in these expressions acts on the field to its right.
The coefficients must also be differentiated when the operators are composed.
For example,
\begin{equation}
 \begin{aligned}
 (\Delta_1(J))_{w\bar z}
 &=-2D_w\left(\partial_z+\frac{\bar z}{\mathcal{N}}\right)
                                      +2\partial_zD_w\\
 &=2[\partial_z,D_w]-\frac{2\bar z}{\mathcal{N}}D_w
   =\frac{2\bar b}{\mathcal{N}}\partial_w-\frac{2\bar z e}{\mathcal{N}} \;,
 \end{aligned}
 \label{app:bg:entry13}
\end{equation}
and
\begin{equation}
 (\Delta_1(J))_{\bar zw}
 =\left[\partial_{\bar z},\frac{R_w}{\mathcal{N}}\right]
 =-\frac b{\mathcal{N}^2}\partial_{\bar w}-\frac{zf}{\mathcal{N}^2} \;.
 \label{app:bg:entry31}
\end{equation}
Setting $e=f=0$ recovers the operators of \cref{app:Qlaplacian} at $\rho=1$.

We next put the operators in Laplace form. To distinguish the connection
from the meromorphic differential $\omega$ of the action, we denote the
connection one-form in degree $p$ by
$\boldsymbol\omega^{(p)}=\omega_\mu^{(p)}dx^\mu$. Our convention is
\begin{equation}
 \begin{aligned}
 \Delta_p(J)
 &=-\left(G^{\mu\nu}\partial_\mu\partial_\nu
                      +A^{(p)\mu}\partial_\mu+\mathcal B^{(p)}\right)\\
 &=-\left(G^{\mu\nu}\nabla_\mu^{(p)}\nabla_\nu^{(p)}
                                      +E_p\right) \;,
 \qquad \nabla_\mu^{(p)}=\partial_\mu+\omega_\mu^{(p)} \;.
 \end{aligned}
 \label{app:bg:laplaceform}
\end{equation}
The metric determined by the principal symbol is
\begin{equation}
 G=2\,dw\,d\bar w+2\,dz\,d\bar z \;,\qquad
 G_{w\bar w}=G_{\bar ww}=G_{z\bar z}=G_{\bar zz}=1 \;,
 \qquad d\operatorname{vol}_G=4d\mu_0 \;.
 \label{app:bg:principalmetric}
\end{equation}
It is this metric, rather than the auxiliary metric
\eqref{hermitianmetric}, which enters \eqref{app:bg:laplaceform}.
Its Christoffel symbols vanish in these coordinates. Consequently,
\begin{equation}
 \omega_\nu^{(p)}=\frac12G_{\nu\mu}A^{(p)\mu} \;,\qquad
 E_p=\mathcal B^{(p)}
      -G^{\mu\nu}\left(\partial_\mu\omega_\nu^{(p)}
                            +\omega_\mu^{(p)}\omega_\nu^{(p)}\right) \;.
 \label{app:bg:connectionE}
\end{equation}
In particular, $\omega_w^{(p)}=\tfrac12A^{(p)\bar w}$ and
$\omega_z^{(p)}=\tfrac12A^{(p)\bar z}$. The overall minus sign in
\eqref{app:bg:laplaceform} is included in the definitions of
$A^{(p)\mu}$ and $\mathcal B^{(p)}$.

Define
\begin{equation}
 \vartheta_w=-\frac{e(b-\bar a)}{2\mathcal{N}} \;,\qquad
 \vartheta_{\bar w}=-\frac{f(a-\bar b)}{2\mathcal{N}} \;,\qquad
 \vartheta=\vartheta_wdw+\vartheta_{\bar w}d\bar w \;,\qquad
 \kappa=-\frac{s^2}{2\mathcal{N}^2} \;.
 \label{app:bg:thetakappa}
\end{equation}
Reading off the scalar operator gives
\begin{equation}
 \mathcal B^{(0)}=-\frac{2ef}{\mathcal{N}} \;,\qquad
 \boldsymbol\omega^{(0)}=\vartheta+\frac{\bar z}{\mathcal{N}}dz \;.
 \label{app:bg:scalarconnection}
\end{equation}
Since the coefficients are independent of $w,\bar w$, we obtain
\begin{equation}
 \begin{aligned}
 E_0
 &=-\frac{2ef}{\mathcal{N}}-\partial_{\bar z}\left(\frac{\bar z}{\mathcal{N}}\right)
                                    -2\vartheta_w\vartheta_{\bar w}\\
 &=-\frac1{\mathcal{N}^2}+\kappa\,ef \;.
 \end{aligned}
 \label{app:bg:e0}
\end{equation}

For the one-form operator, the multiplication coefficient is
\begin{equation}
 \mathcal B^{(1)}=
 \begin{pmatrix}
 -\dfrac{2ef}{\mathcal{N}}&0&\dfrac{2\bar z e}{\mathcal{N}}\\[2mm]
 0&-\dfrac{2ef}{\mathcal{N}}&\dfrac{2\bar z f}{\mathcal{N}}\\[2mm]
 \dfrac{zf}{\mathcal{N}^2}&\dfrac{ze}{\mathcal{N}^2}&-\dfrac{2ef}{\mathcal{N}}+\dfrac2{\mathcal{N}^2}
 \end{pmatrix} \;.
 \label{app:bg:b1}
\end{equation}
The first-order coefficients give
\begin{equation}
 \begin{aligned}
 \omega_w^{(1)}&=
 \begin{pmatrix}
 \vartheta_w&0&0\\
 0&\vartheta_w&\bar a/\mathcal{N}\\
 b/(2\mathcal{N}^2)&0&\vartheta_w
 \end{pmatrix} \;,\\
 \omega_{\bar w}^{(1)}&=
 \begin{pmatrix}
 \vartheta_{\bar w}&0&-\bar b/\mathcal{N}\\
 0&\vartheta_{\bar w}&0\\
 0&-a/(2\mathcal{N}^2)&\vartheta_{\bar w}
 \end{pmatrix} \;,
 \end{aligned}
 \label{app:bg:connection1w}
\end{equation}
and
\begin{equation}
 \omega_z^{(1)}=\operatorname{diag}(0,0,\bar z/\mathcal{N}) \;,
 \qquad \omega_{\bar z}^{(1)}=0 \;.
 \label{app:bg:connection1z}
\end{equation}
In particular, all background dependence of the connection is scalar in
the form indices:
\begin{equation}
 \boldsymbol\omega^{(1)}
 =\boldsymbol\omega^{(1)}\big|_{J=0}+\vartheta\,\mathbf1_3 \;.
 \label{app:bg:scalarshift}
\end{equation}
Using \eqref{app:bg:connectionE}, the endomorphism reduces to
\begin{equation}
 E_1=\mathcal B^{(1)}-\partial_{\bar z}\omega_z^{(1)}
                       -\{\omega_w^{(1)},\omega_{\bar w}^{(1)}\} \;.
 \label{app:bg:e1definition}
\end{equation}
Here $\{X,Y\}=XY+YX$. The result is
\begin{equation}
 E_1=
 \begin{pmatrix}
 \dfrac{|b|^2}{2\mathcal{N}^3}&0&
             \dfrac{e\bar a(2\mathcal{N}+s)}{\mathcal{N}^2}\\[3mm]
 0&\dfrac{|a|^2}{2\mathcal{N}^3}&
             \dfrac{f\bar b(2\mathcal{N}-s)}{\mathcal{N}^2}\\[3mm]
 \dfrac{fa(2\mathcal{N}+s)}{2\mathcal{N}^3}&
 \dfrac{eb(2\mathcal{N}-s)}{2\mathcal{N}^3}&\dfrac2{\mathcal{N}^2}
 \end{pmatrix}
 +\kappa\,ef\,\mathbf1_3 \;.
 \label{app:bg:e1}
\end{equation}
For example,
\begin{equation}
 (E_1)_{w\bar z}
 =\frac{2\bar z e}{\mathcal{N}}+\frac{2\bar b}{\mathcal{N}}\vartheta_w
 =\frac{e\bar a(2\mathcal{N}+s)}{\mathcal{N}^2} \;.
 \label{app:bg:e1example}
\end{equation}
Thus the off-diagonal connection contributes to $E_1$, including to its
diagonal entries at zero background.

We now compute the curvatures. Our convention is
\begin{equation}
 \Omega^{(p)}_{\mu\nu}
 =\partial_\mu\omega_\nu^{(p)}-\partial_\nu\omega_\mu^{(p)}
                   +[\omega_\mu^{(p)},\omega_\nu^{(p)}] \;,
 \qquad
 \boldsymbol\Omega^{(p)}
 =\frac12\Omega^{(p)}_{\mu\nu}dx^\mu\wedge dx^\nu \;.
 \label{app:bg:curvaturedef}
\end{equation}
Let $\mathcal F=d\vartheta$. Its nonzero independent components are
\begin{equation}
 \begin{aligned}
 \mathcal F_{zw}&=-\frac{e\bar a^2}{2\mathcal{N}^2} \;,
 &\mathcal F_{\bar zw}&=\frac{eb^2}{2\mathcal{N}^2} \;,\\
 \mathcal F_{z\bar w}&=-\frac{f\bar b^2}{2\mathcal{N}^2} \;,
 &\mathcal F_{\bar z\bar w}&=\frac{fa^2}{2\mathcal{N}^2} \;,
 \end{aligned}
 \qquad
 \mathcal F_{w\bar w}=\mathcal F_{z\bar z}=0 \;.
 \label{app:bg:f}
\end{equation}
For the scalar operator, the complete curvature is
\begin{equation}
 \boldsymbol\Omega^{(0)}
 =-\frac1{\mathcal{N}^2}\,dz\wedge d\bar z+\mathcal F \;.
 \label{app:bg:omega0}
\end{equation}
For the one-form operator, \eqref{app:bg:scalarshift} and the commutativity
of the background give
\begin{equation}
 \Omega^{(1)}_{\mu\nu}
 =\left.\Omega^{(1)}_{\mu\nu}\right|_{J=0}
                              +\mathcal F_{\mu\nu}\mathbf1_3 \;.
 \label{app:bg:omega1}
\end{equation}
There is no quadratic-background term in the curvature.

To display its background-independent part, let $M_{ij}$ be the matrix with a single $1$ in
row $i$, column $j$, in the one-form basis \eqref{app:bg:bases}. Thus
\begin{equation}
 (M_{ij})_{kl}=\delta_{ik}\delta_{jl} \;,\qquad
 M_{ij}M_{kl}=\delta_{jk}M_{il} \;.
 \label{app:bg:units}
\end{equation}
All six independent components are
\begin{equation}
 \begin{aligned}
 \left.\Omega^{(1)}_{w\bar w}\right|_{J=0}
 &=\frac1{2\mathcal{N}^3}
       \operatorname{diag}(|b|^2,-|a|^2,|a|^2-|b|^2) \;,\\
 \left.\Omega^{(1)}_{z\bar z}\right|_{J=0}
 &=-\frac1{\mathcal{N}^2}M_{33} \;,\\
 \left.\Omega^{(1)}_{zw}\right|_{J=0}
 &=-\frac{2\bar z\bar a}{\mathcal{N}^2}M_{23}
                            -\frac{\bar a}{2\mathcal{N}^3}M_{31} \;,\\
 \left.\Omega^{(1)}_{\bar zw}\right|_{J=0}
 &=\frac b{\mathcal{N}^2}M_{23}-\frac{zb}{\mathcal{N}^3}M_{31} \;,\\
 \left.\Omega^{(1)}_{z\bar w}\right|_{J=0}
 &=\frac{2\bar z\bar b}{\mathcal{N}^2}M_{13}
                            -\frac{\bar b}{2\mathcal{N}^3}M_{32} \;,\\
 \left.\Omega^{(1)}_{\bar z\bar w}\right|_{J=0}
 &=\frac a{\mathcal{N}^2}M_{13}+\frac{za}{\mathcal{N}^3}M_{32} \;.
 \end{aligned}
 \label{app:bg:freecurvatures}
\end{equation}
The remaining components follow by antisymmetry. For instance, the third
line is obtained from
\begin{equation}
 \left.\Omega^{(1)}_{zw}\right|_{J=0}
 =\partial_z\left(\frac{\bar a}{\mathcal{N}}M_{23}
                         +\frac b{2\mathcal{N}^2}M_{31}\right)
  +\left[\frac{\bar z}{\mathcal{N}}M_{33},
          \frac{\bar a}{\mathcal{N}}M_{23}+\frac b{2\mathcal{N}^2}M_{31}\right] \;.
 \label{app:bg:curvatureexample}
\end{equation}

We finally collect the traces which enter the heat coefficient.
Here $\operatorname{tr}$ denotes the trace over form components on one
root space, and $[\cdot]_{ef}$ denotes the coefficient of $ef$.
The scalar endomorphism immediately gives
\begin{equation}
 E_0^2=\frac1{\mathcal{N}^4}+\frac{s^2}{\mathcal{N}^4}ef
                         +\frac{s^4}{4\mathcal{N}^4}(ef)^2 \;.
 \label{app:bg:e0square}
\end{equation}
For the one-form endomorphism, put
\begin{equation}
 d_1=\frac{\mathcal{N}+s}{2\mathcal{N}^3} \;,\qquad
 d_2=\frac{\mathcal{N}-s}{2\mathcal{N}^3} \;,\qquad d_3=\frac2{\mathcal{N}^2} \;.
 \label{app:bg:di}
\end{equation}
Since $(E_1)_{12}=(E_1)_{21}=0$, its ordinary matrix square obeys
\begin{equation}
 \operatorname{tr}E_1^2
 =\sum_{i=1}^3(d_i+\kappa ef)^2
    +2(E_1)_{13}(E_1)_{31}+2(E_1)_{23}(E_1)_{32} \;.
 \label{app:bg:traceexpansion}
\end{equation}
The diagonal and off-diagonal contributions are, respectively,
\begin{equation}
 \sum_i(d_i+\kappa ef)^2
 =\frac{9\mathcal{N}^2+s^2}{2\mathcal{N}^6}-\frac{3s^2}{\mathcal{N}^4}ef
                           +\frac{3s^4}{4\mathcal{N}^4}(ef)^2 \;,
 \label{app:bg:diagtrace}
\end{equation}
and
\begin{equation}
 \begin{aligned}
 &2(E_1)_{13}(E_1)_{31}+2(E_1)_{23}(E_1)_{32}\\
 &\quad=\frac{ef}{\mathcal{N}^5}
       \left[(\mathcal{N}-s)(2\mathcal{N}+s)^2+(\mathcal{N}+s)(2\mathcal{N}-s)^2\right]
       =\frac{8\mathcal{N}^2-6s^2}{\mathcal{N}^4}ef \;.
 \end{aligned}
 \label{app:bg:offdiagtrace}
\end{equation}
Consequently,
\begin{equation}
 \begin{aligned}
 \operatorname{tr}E_1^2
 &=\frac{9\mathcal{N}^2+s^2}{2\mathcal{N}^6}
       +\frac{8\mathcal{N}^2-9s^2}{\mathcal{N}^4}ef\\
 &\quad+\frac{3s^4}{4\mathcal{N}^4}(ef)^2 \;.
 \end{aligned}
 \label{app:bg:e1square}
\end{equation}

Indices on $\Omega$ are raised with the inverse of
\eqref{app:bg:principalmetric}. In complex coordinates the contraction is
\begin{equation}
 \begin{aligned}
 \operatorname{tr}(\Omega_{\mu\nu}\Omega^{\mu\nu})
 ={}&-2\operatorname{tr}
       \left(\Omega_{w\bar w}^{\,2}+\Omega_{z\bar z}^{\,2}\right)\\
 &+4\operatorname{tr}
       \left(\Omega_{zw}\Omega_{\bar z\bar w}
                    +\Omega_{\bar zw}\Omega_{z\bar w}\right) \;.
 \end{aligned}
 \label{app:bg:contraction}
\end{equation}
In particular, \eqref{app:bg:f} gives
\begin{equation}
 \mathcal F_{\mu\nu}\mathcal F^{\mu\nu}
 =-\frac{|a|^4+|b|^4}{\mathcal{N}^4}ef
 =-\frac{2(\mathcal{N}^2+s^2)}{\mathcal{N}^4}ef \;.
 \label{app:bg:fsquare}
\end{equation}
The mixed components of the background-independent one-form curvature are off-diagonal
matrices and have zero trace. Since $\mathcal F_{w\bar w}
=\mathcal F_{z\bar z}=0$, the cross term between the free and
background-dependent curvatures also has zero trace. Thus the complete
contractions are
\begin{equation}
 \Omega^{(0)}_{\mu\nu}\Omega^{(0)\mu\nu}
 =-\frac2{\mathcal{N}^4}-\frac{2(\mathcal{N}^2+s^2)}{\mathcal{N}^4}ef \;,
 \label{app:bg:omega0square}
\end{equation}
and
\begin{equation}
 \operatorname{tr}
   \left(\Omega^{(1)}_{\mu\nu}\Omega^{(1)\mu\nu}\right)
 =\frac{-16\mathcal{N}^3+9\mathcal{N}^2-3s^2}{\mathcal{N}^6}
                   -\frac{6(\mathcal{N}^2+s^2)}{\mathcal{N}^4}ef \;.
 \label{app:bg:omega1square}
\end{equation}
In particular, the quadratic-background invariants are
\begin{equation}
 \begin{aligned}
 T_0&:=[E_0^2]_{ef}=\frac{s^2}{\mathcal{N}^4} \;,
 &T_1&:=[\operatorname{tr}E_1^2]_{ef}
                        =\frac{8\mathcal{N}^2-9s^2}{\mathcal{N}^4} \;,\\
 O_0&:=[\Omega^{(0)}_{\mu\nu}\Omega^{(0)\mu\nu}]_{ef}
                        =-\frac{2(\mathcal{N}^2+s^2)}{\mathcal{N}^4} \;,
 &O_1&:=[\operatorname{tr}
      (\Omega^{(1)}_{\mu\nu}\Omega^{(1)\mu\nu})]_{ef}=3O_0 \;.
 \end{aligned}
 \label{app:bg:invariants}
\end{equation}

All formulas above use the unnormalized $\Delta_p$ of \cref{app:Qlaplacian}.
For $\widehat\Delta_p=2\Delta_p$, the principal metric is
$\widehat G=G/2$, the connection and curvature two-form are unchanged,
and $\widehat E_p=2E_p$. The volume element scales by $1/4$ and
the contracted invariants in \eqref{sec:torsion:a4} scale by $4$.
Thus the four-dimensional bulk coefficient $a_4^{(p)}$ is unchanged by
this common normalization.

\subsection{Extension to the PCM with a Wess--Zumino term}
\label{app:bg:wz}

The computations above can be generalized to the case of the PCM with Wess--Zumino (WZ) term. For the constant commuting background of \cref{wzterm}, set
$p=k/h$ with $p\in\mathbb R$ and $p\ne\pm1$. On one root space,
write $u=(1-p)e$ and $v=(1+p)f$, with $f=\bar e$ as in
\eqref{sec:torsion:roots}. The WZ factors multiply the insertions in
$D_w$ and $D_{\bar w}$, so taking the Hermitian adjoint gives
\ie
 R_w&=-\bar a\partial_{\bar w}+\bar u
     =-\bar a\partial_{\bar w}+(1-p)f,\\
 R_{\bar w}&=-\bar b\partial_w+\bar v
     =-\bar b\partial_w+(1+p)e .
\fe
Generalizing the calculations in \eqref{app:bg:entry13} and
\eqref{app:bg:entry31} to the case of the PCM+WZ model, we find that the
multiplication terms in $(X_w,X_{\bar w})$ involve $(u,v)$,
while those in $(Y_w,Y_{\bar w})$ involve $(\bar u,\bar v)$.

For the scalar Laplacian \eqref{app:bg:scalarcomp}, the
zeroth-order part of $(R_wD_w+R_{\bar w}D_{\bar w})/\mathcal N$ is
$m=(|u|^2+|v|^2)/\mathcal N=2(1+p^2)ef/\mathcal N$.
In the Laplace-form convention \eqref{app:bg:laplaceform}, this gives
$\mathcal B^{(0)}=-m$. The first-order coefficients give
\ie
 \vartheta_w=\frac{\bar a u-b\bar v}{2\mathcal N},
 \qquad
 \vartheta_{\bar w}=\frac{\bar b v-a\bar u}{2\mathcal N}
                  =-\overline{\vartheta_w}.
\fe
Using $|a|^2+|b|^2=2\mathcal N$, the multiplication and
connection-square terms in \eqref{app:bg:e0} combine to give
\ie
 E_0&=-\frac{1}{\mathcal N^2}+q,\\
 q&=-m-2\vartheta_w\vartheta_{\bar w}
   =-\frac{|bu+\bar a\bar v|^2}{2\mathcal N^2}.
\fe
For $p=0$, the $+1$ in $b$ and the $-1$ in $\bar a$ cancel in
$bu+\bar a\bar v=e(b+\bar a)=es$, giving $q=\kappa ef$.
For nonzero $p$, the unequal weights contribute
$p(\bar z-z)e$ from the $z$-dependent terms and $-2pe$ from
the constants. Thus, in the present case, we have 
\ie
 bu+\bar a\bar v
 &=e\big[(1-p)(z+1)+(1+p)(\bar z-1)\big]\\
 &=e\big[s+p(\bar z-z)-2p\big],
\fe
and
\ie
 q&=-\frac{(1-p^2)s^2-4ps+4p^2\mathcal N}{2\mathcal N^2}\,ef .
\fe

In degree one, the same terms $-m$ and
$-2\vartheta_w\vartheta_{\bar w}$ contribute $q\mathbf 1_3$ to the generalization of 
\eqref{app:bg:e1definition}. The unequal weights also affect the
two upper off-diagonal entries of $E_1$:
\ie
 U_w:=(E_1)_{13}
 &=\frac{2}{\mathcal N}(\bar z u+\bar b\vartheta_w)
   =\frac{u\bar a(2\mathcal N+s)+(u-\bar v)|b|^2}{\mathcal N^2},\\
 U_{\bar w}:=(E_1)_{23}
 &=\frac{2}{\mathcal N}(\bar z v-\bar a\vartheta_{\bar w})
   =\frac{v\bar b(2\mathcal N-s)-(v-\bar u)|a|^2}{\mathcal N^2}.
\fe
In each entry, $\mathcal B^{(1)}$ supplies the term linear in the
insertion and the anticommutator in \eqref{app:bg:e1definition}
supplies the term involving $\vartheta$. Their sum produces the
differences $u-\bar v$ and $v-\bar u$, which vanish in
\eqref{app:bg:e1} at $p=0$. The lower off-diagonal entries are
$\overline{U_w}/(2\mathcal N)$ and
$\overline{U_{\bar w}}/(2\mathcal N)$, as required by the weighted
Hermitian pairing \eqref{app:bg:weights}.

The scalar shift in \eqref{app:bg:scalarshift} commutes with the
free connection, so its curvature contribution is
$\mathcal F=\rd\vartheta$ times the identity. With the raising
convention of \eqref{app:bg:contraction}, the replacement for
\eqref{app:bg:fsquare} is
\ie
 \mathcal F_{\mu\nu}\mathcal F^{\mu\nu}
 &=-4\left(|\partial_z\vartheta_w|^2
           +|\partial_{\bar z}\vartheta_w|^2\right)\\
 &=-\frac{ef}{\mathcal N^4}
    \left(|a|^2|a-pb|^2+|b|^2|b+pa|^2\right).
\fe
The mixed free curvature matrices in \eqref{app:bg:freecurvatures}
are traceless, so the cross term with $\mathcal F\mathbf 1_3$
has zero trace. The current-dependent part of the curvature-square
trace has multiplicity one in degree zero and three in degree one.
It therefore cancels in \eqref{sec:torsion:weightedheat}.

With the diagonal entries $d_i$ of \eqref{app:bg:di}, the
endomorphisms take the form
\ie
 E_0&=-\frac{1}{\mathcal N^2}+q,\qquad
 E_1=\operatorname{diag}(d_1,d_2,d_3)+q\mathbf 1_3+\mathcal U,\\
 \mathcal U&=
 \begin{pmatrix}
  0&0&U_w\\
  0&0&U_{\bar w}\\
  \overline{U_w}/(2\mathcal N)&
  \overline{U_{\bar w}}/(2\mathcal N)&0
 \end{pmatrix}.
 \label{app:bg:wz:endomorphisms}
\fe
As in \eqref{app:bg:e0square} and \eqref{app:bg:traceexpansion},
the scalar square and the trace of the one-form square are
\ie
 E_0^2
 &=\frac{1}{\mathcal N^4}-\frac{2q}{\mathcal N^2}+q^2,\\
 \operatorname{tr}E_1^2
 &=\sum_{i=1}^3(d_i+q)^2
   +\frac{|U_w|^2+|U_{\bar w}|^2}{\mathcal N}\\
 &=\frac{9\mathcal N^2+s^2}{2\mathcal N^6}
   +\frac{6q}{\mathcal N^2}+3q^2
   +\frac{|U_w|^2+|U_{\bar w}|^2}{\mathcal N}.
 \label{app:bg:wz:traces}
\fe
The off-diagonal terms pair the $13$ entry with the $31$ entry
and the $23$ entry with the $32$ entry.
Substituting $U_w$ and $U_{\bar w}$ gives
\ie
 \frac{|U_w|^2+|U_{\bar w}|^2}{\mathcal N}
 &=\frac{ef}{\mathcal N^4}\big[
  (1-p^2)(8\mathcal N^2-6s^2)\\
 &\qquad\qquad
  +8p(2-r^2)s+8p^2\mathcal N(2r^2+1)\big],
 \qquad r=|z|.
\fe
Extracting the terms quadratic in the currents, with the
original $ef$ factored out as in \eqref{app:bg:invariants}, gives
\ie
 ef\,T_0(p)&=-\frac{2q}{\mathcal N^2},\\
 ef\,T_1(p)&=\frac{6q}{\mathcal N^2}
              +\frac{|U_w|^2+|U_{\bar w}|^2}{\mathcal N}.
 \label{app:bg:wz:invariants}
\fe
The curvature-square invariants follow from the contraction
of $\mathcal F$,
\ie
 O_0(p)&=[\Omega^{(0)}_{\mu\nu}\Omega^{(0)\mu\nu}]_{ef}
 =-\frac{|a|^2|a-pb|^2+|b|^2|b+pa|^2}{\mathcal N^4},\\
 O_1(p)&=[\operatorname{tr}
      (\Omega^{(1)}_{\mu\nu}\Omega^{(1)\mu\nu})]_{ef}
 =3O_0(p).
\fe
 The $q^2$ terms cancel between $\operatorname{tr}E_1^2$
and $3E_0^2$. In addition,
$\operatorname{tr}E_1-3E_0=6/\mathcal N^2$ is independent of the
currents, so the total-derivative term in \eqref{sec:torsion:a4}
makes no pointwise contribution to the $ef$ coefficient of
$a_4^{(1)}-3a_4^{(0)}$.

\bibliographystyle{ytphys.bst}
\bibliography{4dCS}

\end{document}